\documentclass[aps,prl,twocolumn,superscriptaddress]{revtex4-1}

\usepackage{graphicx}  
\usepackage{amsmath,amssymb,amsfonts}
\usepackage{mathrsfs} 

\usepackage{textcomp}
\usepackage{hyperref}
\usepackage{gensymb}
\usepackage{verbatim}
\usepackage{braket}

\hypersetup{breaklinks=true,colorlinks=true,urlcolor=black}
\usepackage{color}
\usepackage{soul}
\usepackage{gensymb}

\DeclareGraphicsExtensions{.jpg,.pdf,.png,.eps}

\begin{document}


\title{Position operators in terms of converging finite-dimensional matrices: Exploring their interplay with geometry, transport, and gauge theory.}



\author{B.Q.~Song}
\affiliation{Ames Laboratory, Iowa State University, Ames, Iowa 50011, USA}
\affiliation{Department of Physics and Astronomy, Iowa State University, Ames, Iowa 50011, USA}
\author{J.D.H.~Smith}
\affiliation{Ames Laboratory, Iowa State University, Ames, Iowa 50011, USA}
\affiliation{Department of Mathematics, Iowa State University, Ames, Iowa 50011, USA}
\author{J.~Wang}
\affiliation{Ames Laboratory, Iowa State University, Ames, Iowa 50011, USA}
\affiliation{Department of Physics and Astronomy, Iowa State University, Ames, Iowa 50011, USA}


\date{\today}

\begin{abstract}
The position operator $\hat{r}$ appears as $i{\partial_p}$ in wave mechanics, while its matrix form (e.g., under a Bloch basis) is well known diverging in diagonals, causing serious difficulties in basis transformation, observable yielding, etc. The present work arises from a belief that the matrix of a physical operator should not diverge. We aim to find a convergent $r$-matrix (CRM) to improve the existing divergent $r$-matrix (DRM), and investigate its influence at both the conceptual and the application levels. Unlike the spin matrix, which affords a Lie algebra representation as the solution of $[s_i,s_j]={\epsilon}_{i,j,k}s_k$, the $r$-matrix cannot be a solution for $[\hat{r},p]=i\hbar$, namely Weyl algebra. Indeed: (1) matrix representations of Weyl algebras prove not existing; thus, (2) neither a CRM nor a DRM would afford a representation. Instead, the CRM should be viewed as a procedure of encoding $\hat{r}$, an operator of continuous spectrum and infinite dimension, using matrices of arbitrary finite dimensions. Deriving CRM is aligned with the spirit of DRM, while it recognizes that the limited understanding about Weyl algebra has led to the divergence. A key modification is increasing the 1-st Weyl algebra (the familiar substitution $\hat{r}{\rightarrow}i{\partial_p}$) to the $N$-th Weyl algebra. Resolving the divergence makes $r$-matrix rigorously defined, and we are able to show $r$-matrix is distinct from a spin matrix in terms of its defining principles, transformation behavior, and the observable it yields. At the conceptual level, the CRM fills the logical gap between the $r$-matrix and the Berry connection (this unremarked vagueness has caused the diagonal divergence); and helps to show that Bloch space $\mathcal{H}_B$ is incomplete for $\hat{r}$. At the application level, we focus on transport, and discover that the Hermitian matrix is not identical with the associative Hermitian operator, i.e., $r_{m,n}=r_{n,m}^*{\nLeftrightarrow}\hat{r}=\hat{r}^{\dagger}$, which subtly affects the celebrated Berry curvature formula for adiabatic current. We also discuss how such a non-representation CRM can contribute to building a unified transport theory. 
\end{abstract}



\maketitle

\section{1. Introduction}
In the change from classical to quantum mechanics, the position $r$ was promoted to the operator $\hat{r}$, conjugated with momentum $p$ by the non-trivial commutator relation $[\hat{r},p]=i\hbar$ that leads to uncertainty in their values \cite{1,2}. As such, $\hat{r}$ appears as $i{\partial}_p$ acting on wavefunctions coordinated by $p$, with crucial consequences whether ${\langle}\hat{r}{\rangle}$ is explicitly evaluated (e.g., for transport), or if ${\langle}\hat{r}{\rangle}$ is trivially constant (e.g., with an atomic Hamiltonian) but energies/eigenstates are to be solved \cite{1}.

The familiar $i{\partial}_p$ is the form of $\hat{r}$ in wave mechanics, while the matrix form of $\hat{r}$ is rarely seen. This disproportionate observation is due to the lack of convergence of the $r$-matrix: the diagonals diverge in plane waves, Bloch bases, etc. \cite{3,4,5,6,7,8} Given the autonomy of matrix mechanics, and its equivalence with wave mechanics \cite[Ch.~3]{2}, it is weird that the diagonals of a physical operator (i.e., their expectation values) all diverge. This undermines preservation of the spectrum of the matrix; obstructs obtaining an equivalent $r$-matrix by basis transformation; and even casts doubt on the off-diagonal convergence, given the holistic nature of a matrix. We elaborate on the misbehavior of the $r$-matrix in Sec. 2.

In contrast, the spin matrix can readily be found, by solving $[s_i,s_j]={\epsilon}_{i,j,k}s_k$. Pauli matrices provide the simplest solutions, and many others exist, either reducible or irreducible. Formally, representations of Lie algebras are involved \cite{9}. In physics, we consider $\mathfrak{su}(2,\mathbb{C})$ as it stands for spin. Generally, systematic construction and classification of Lie algebras have been achieved for finite (Cartan matrices \cite{9,10}) and infinite dimensions (e.g., Kac-Moody algebras \cite{10}). It is tempting to try the same for the $r$-matrix by solving $[\hat{r},p]=i\hbar$: the operator on the right is now replaced by a complex number, leading to Weyl algebras \cite{10,11}. Unfortunately, this is doomed to fail, because Weyl algebras admit no matrix representations (Sec. 2). Does this mean that an $r$-matrix cannot exist? Given that $[\hat{r},p]=i\hbar$ cannot be solved with matrices, could any matrix be assigned to $\hat{r}$?

A major goal of this work is to derive convergent $r$-matrices (CRM) in arbitrary finite dimensions by introducing the $N$-th Weyl algebra $A_N$ (Sec. 2). It can be viewed as a procedure for encoding $\hat{r}$, an infinite dimensional operator of continuous spectrum, using matrices of finite dimension. Theoretically, it is convenient to  have such a formal conversion for dimensions, and such a matrix description of a differential operator \cite{12}. However, we stress that the $r$-matrices do not yield representations. In particular, we must carefully distinguish the terminologies “matrix” and “matrix representation.'' In fact, the known $r$-matrices \cite{3,4,5,8}, called \emph{divergent} $r$-\emph{matrices} (DRM) in this context, do not yield representations either | a fact which has somehow been concealed by the divergence. The DRMs derive from the 1-st Weyl algebra $A_1$ (just the familiar substitution $\hat{r}{\rightarrow}i{\partial}_p$); we analyze the divergence, and find that expanding $A_1$ to $A_N$ can fix it. Here, $N$ depends on the dimension of the Bloch space. Although $A_1$ is the special case of $A_N$ with $N=1$, the DRMs are \textit{not} 1-D CRMs. In fact, the CRMs converge for arbitrary dimensions, and are thus distinct from the DRMs, rather than including them as a special case (Sec. 4). 

Why are $r$-matrices important? A short answer is that they are involved in both transport \cite{4,5,6,7,8,13,14,15,16,17,18} and topology \cite{19,20,21,22,23,24,25} in crystals (e.g., under Bloch bases). For example, when band electrons are exposed to light \cite{26,27,28,29} and undergo resonant inter-band transition \cite{4,5,7,8,13,18}, the charge center changes, leading to a \emph{shift current} $J_s{\propto}{\gamma}_{m,n}{\cdot}R_{m,n}$ \cite{8,18}. The ${\gamma}_{m,n}$ are hopping rates, and $R_{m,n}$ is the position shift associated with $|\psi_n{\rangle}{\rightarrow}|\psi_m{\rangle}$: 
\begin{equation}
\begin{split}
R_{m,n}(k):=r_{m,m}(k)-r_{n,n}(k)-\mathfrak{X}_{m,n}(k)
\end{split}
\label{eq1}
\end{equation} 
Here, $R_{m,n}(k)$ is the \emph{shift vector}, obtained by subtracting diagonals between bands $m$, $n$; while $\mathfrak{X}_{m,n}(k)$ is a complementary term (involving off-diagonals) to ensure gauge invariance \cite{4,5,18}. As such, $r$-matrices enter through the distance shifted during hopping. 

The $r$-matrices make another entrance in connection with the hopping rates ${\gamma}_{m,n}$. The light field is usually modelled with
\begin{equation}
\begin{split}
\hat{V}_{\text{int}}=eE\hat{r}
\end{split}
\label{eq2}
\end{equation} 
\cite{4}.
The modification of quantum states is encoded in the matrix ${\langle}{\psi}_m|eE\hat{r}|{\psi}_n{\rangle}$, which forms the basic building unit for $n$-th order perturbation \cite{4,5,8}. For instance, consider the linear response 
\begin{equation}
\begin{split}
{\gamma}_{m,n}(k){\propto}\tfrac{1}{h}|{\langle}u_m|\hat{V}_{\text{int}}|u_n{\rangle}|^2(f_n(k)-f_m(k))
\end{split}
\label{eq3}
\end{equation} 
(the Fermi Golden Rule).
Combined with Eq.~\ref{eq1}, the DC component of the second-order $J_s$ response to the external driving with frequency $\omega$ is found to be 
\begin{equation}
\begin{split}
J_s(\omega)={\int}&f_{m,n}{\cdot}(r_{m,m}-r_{n,n}-\mathfrak{X}_{m,n}){\cdot}\\
&r_{m,n}r_{n,m}{\delta}(\omega_{m,n}-\omega)E(\omega)E(-\omega){\cdot}dk,
\end{split}
\label{eq4}
\end{equation} 
\cite{8}, where $f_{m,n} ({\omega}_{m,n})$ is the Fermi distribution (energy) difference between two bands:
\begin{equation}
\begin{split}
f_{m,n}:=f_n(k)-f_m(k),~{\omega}_{m,n}:={\omega}_m(k)-{\omega}_n(k).
\end{split}
\label{eq5}
\end{equation} 
Compared with $J_s{\propto}{\gamma}_{m,n}{\cdot}R_{m,n}$, we recognize 
\begin{equation}
\begin{split}
{\gamma}_{m,n}(k)=f_{m,n}r_{m,n}r_{n,m}{\delta}({\omega}_{m,n}-{\omega})E(\omega)E(-\omega)
\end{split}
\label{eq6}
\end{equation} 
for $m{\neq}n$. Clearly, the hopping rate ${\gamma}_{m,n}(k)$ relies on the $r$-matrix. When higher-order perturbations are counted, ${\gamma}_{m,n}(k)$ should involve higher-order products of $r$-matrices: $r_{m,j}r_{j,l}r_{l,n}r_{n,m}$, etc. \cite{4,5,8,24}.

Simply speaking, the matrix components $r_{m,n}$ originate from band labels $m$, $n$. The $r$-matrix is linked to observables in various forms \cite{15,16,30,31,32,33,34,35,36}. In view of these consequences of the $r$-matrix, the dichotomy between wave and matrix mechanics, and their mutual replaceability, deserve second thought. We should not naively attribute the $r$-matrix solely to matrix mechanics, since evaluation and transformation of the $r$-matrix inevitably involve the nature of $\hat{r}$ as a differential operator, sometimes in implicit ways. Nor should we think that using the differential form $i{\partial}_p$ makes the $r$-matrix redundant. Matrices and differential operators will be extensively discussed in Sec. 5.

We have surpressed the Cartesian indices in Eq.~\ref{eq1}--\ref{eq6}. In general, optical conductivities are tensors. Nevertheless, the simple fact is that the $r$-matrix appears in optical responses. More importantly, its role is clear: since the $r$-matrix takes the form of a connection, a  differential-geometric notion, it opens the door to quantum geometry \cite{30,31,32,33,34,35,36,36a}. A notable success has been the linkage of the Berry connection with the Wannier center \cite{21,25}, leading to quantization of adiabatic charge pumping \cite{20} and the Berry phase theory of polarization \cite{25}. In this vein, given appropriate coupling forms or scenarios (potentially ignoring certain degrees of freedom \cite{33}), more geometric interpretations appear, such as curvature \cite{19}, a quantum metric (as a distance defined between quantum states) \cite{31}, and tangent space \cite{33}. It is fascinating that these geometric notions enter into diverse phenomena which are seemingly irrelevant.

\begin{table*}
\caption{\label{tab:table1} Major results of the present work. In particular, we  highlight 2.3, 3.2, 3.3, 4.2, 4.3, 4.4, 4.5, 4.6, 5.2, 5.3, 5.4, 5.5, 6.1, 6.3, 7.2, and 7.3 as major innovation points that might update or challenge certain pre-existing viewpoints, or have significant impact on subsequent research.\label{tab1}}
\begin{ruledtabular}
\begin{tabular}{l l}
& \textbf{Sec. 2 Position operator $\hat{r}$ and the Weyl algebra.} \\
\hline
& 2.1 Introduce three spaces: (i) $\mathcal{H}$ spanned by $\hat{r}$'s eigenstates, (ii) Bloch space $\mathcal{H}_B$, (iii) quotient space $\mathcal{V}$ of Bloch space. \\
& 2.2 Show the relation between $\hat{r}$ and generators of Weyl algebra.\\
& 2.3 Non-existence of matrix representations of Weyl algebra; need for new principles to determine $r$-matrices.\\
\hline
& \textbf{Sec. 3 Bloch space structure and its quotient space.} \\
\hline
& 3.1 The norm of Bloch space $\mathcal{H}_B$ is divergent; the expectation value ${\langle}\hat{r}{\rangle}$ is divergent in Bloch bases.\\
& 3.2 To avoid ${\int}_{-\infty}^{\infty}dr$, we introduce an isomorphic product space $\mathcal{V}{\otimes}E$ to substitute fo $\mathcal{H}_B$, which is realized by a projection \\
& map $\Pi$: $\mathcal{H}_B{\rightarrow}\mathcal{V}{\otimes}E$. \\
& 3.3 Prove Bloch space $\mathcal{H}_B$ is incomplete for $\hat{r}$ (with counterexamples), i.e., $\mathcal{H}_B{\ncong}\mathcal{H}$. \\
\hline
& \textbf{Sec. 4 Matrices of position operators.} \\
\hline
& 4.1 Derive DRM with $A_1$ (reproduce previous result). \\
& 4.2 Derive converging $r$-matrix (CRM) of arbitrary dimensions with $N$-th Weyl algebra $A_N$. \\
& 4.3 Define $r$-matrix $\mathfrak{r}_{m,n}(k,k')$ and reduced $r$-matrix $r_{m,n}(k)$. \\
& 4.4 Show the space spanned by periodic functions $u_{n,k}(r)$ is isomorphic to $\mathcal{V}$, a quotient space of $\mathcal{H}_B$. Show geometric \\
& quantities, e.g., Berry connection or curvature, are defined on $\mathcal{V}$, not on $\mathcal{H}_B$. \\
& 4.5 Articulate how the divergence in DRM is fixed; demonstrate the relation between DRM and CRM. \\
& 4.6 Show neither DRM nor CRM will satisfy the commutation $[\hat{r},p]=i\hbar$. \\
\hline
& \textbf{Sec. 5 Properties of the $\hat{r}$ operator and $r$-matrix.} \\
\hline
& 5.1 Define ribbon and its transformation in bundle space in analog with basis and its transformation in vector space. \\
& 5.2 Under a unified frame of ribbon, two types of operators are recognized: matrix operator and differential operators. \\
& 5.3 Show $r_{m,n}=r_{n,m}^*{\nLeftrightarrow}\hat{r}=\hat{r}^{\dagger}$; show the well-known Berry curvature expression for polarization is conditionally true. \\
& 5.4 Algebraic rules for matrix of differential operator: complex conjugation, inner product, transformation, etc. \\
& 5.5 Inapplicability of bra/ket designations for denoting $r$-matrix. \\
\hline
& \textbf{Sec. 6 Gauge, ribbon, and basis transformations.} \\
\hline
& 6.1 Show relations between gauge transformation $T_G$, ribbon transformation $T_R$, basis transformation $T_B$; show $T_G$ can be \\ 
& induced by $T_R$. \\
& 6.2 Define gauge invariance, ribbon (transformation) invariance. \\
& 6.3 Procedures of extracting observables for matrix and differential operators, characterized by gauge symmetry principle. \\
\hline
& \textbf{Sec. 7 Discussion and outlook.} \\
\hline
& 7.1 Several spaces related to $r$-matrix \\
& 7.2 Principles for defining $r$-matrix in comparison with those for defining spin matrix and group matrix. \\
& 7.3 Applications of CRM and its implications in building a unified transport mechanism.
\end{tabular}
\end{ruledtabular}
\end{table*}

Although substantial advances have been made in the geometrical interpretation of optical transitions \cite{17,18,30,31,32,33,34,35,36}, we have not yet dealt with the issue of divergence. The $r$-matrix is still based on DRMs, the divergence arising from the diagonal ${\partial}_k(k-k')$ terms. The issue was raised by Blount in the 1950s \cite{3}. Since then, no essential progress has been made on resolving the divergence, or on its origin and significance. Thus, the diagonal terms $r_{m,m}$, $r_{n,n}$ in Eq.~\ref{eq1} are not evaluated with Bloch functions, but with periodic functions $u_{m,k}(r)$, $u_{n,k}(r)$. In arguing for such a substitution, resort is made to a heuristic: “The Bloch wave does not work, so something else should be used.” However, it remains unaddressed whether $u_{n,k}(r)$ is the only possible choice, and whether using $u_{n,k}(r)$ can be attributed to a certain general principle. Moreover, the implications of employing $u_{n,k}(r)$ and integrating it over $k$ to yield ${\langle}\hat{r}{\rangle}$ are disturbing, because when observables are no longer taken from diagonals of a physical operator, not the orthodoxy of matrix mechanics \cite{2}. There has not yet been any comprehensive justification of this procedure, which seems quite remarkable given the maturity of quantum mechanics \cite{2}.

One can ensure that the diagonal entries $r_{n,n}$ appear in quarantined form, just like cutting off the rotten parts of an apple, while it is unclear if converging entries in a diverging matrix are still meaningful, at the very least in the absence of a renormalization protocol. Another concern is that geometry often relies on perturbation series \cite{4,33,35}, which incurs risks: strong interaction or gap closing might undermine the perturbation treatment; the geometric interpretation could be sensitive to the orders of truncation; it is hard to recognize a geometric effect when it is confounded with other effects \cite{35}. Any of these possibilities could diminish the fundamentality and elegance of a geometric formula. Additionally, hopping, which should be a continuous process, is usually interpreted in terms of a pair of states (initial and final), while the geometric interpretation requires the number of intermediate states to be accounted for \cite{33,35}. In short, the present situation is not satisfactory, and worry arises from the logic gap: the $r$-matrix does not stand on a solid foundation, and observable extraction is clearly incompatible with the basic rules for matrices, while observables based on DRMs are continually being proposed \cite{8,17,18,30,31,32,33,34,35,36}.

Our aim in this work is two-fold. At the conceptual level, we resolve the vagueness in using ${\psi}_{n,k}$ or $u_{n,k}$ by introducing the “$r$-matrix” $\mathfrak{r}_{m,n}(k,k')$ and the “reduced $r$-matrix” $r_{m,n}(k)$. Here, $\mathfrak{r}_{m,n}(k,k')$ is evaluated with ${\psi}_{n,k}$, and $r_{m,n}(k)$ is evaluated with $u_{n,k}$. Both are defined in \textit{convergent} fashion, regarded as different facets of the CRM, and their relations are deduced. The vector space spanned by $u_{n,k}$ is recognized, whose dimension and relation with the Bloch space (spanned by ${\psi}_{n,k}$) are clarified. With CRM, the difficulty in using either diagonals or off-diagonals disappears. Moreover, we recall the non-existence of a matrix representation for a Weyl algebra, and show that Bloch waves are incomplete for $\hat{r}$. As a consequence, the principles for deducing the $r$-matrix must be different from, for instance, those for the spin operator. For spin, a complete “total space” serves as a Hilbert space which affords a Lie algebra representation. For $\hat{r}$ and the Weyl algebra, it is a quotient space of a total space (or fiber space of a bundle space in bundle theory \cite{37,38}) which serves as a Hilbert space. The algebraic procedure for this expands $A_1$ to $A_N$.

At the application level, our main focus is on transport, which, by definition, means position change of charge carriers. Thus, it ultimately concerns the expectation value of $\hat{r}$. The analysis of diverse transport mechanisms, such as injection currents \cite{8,26}, shift currents \cite{8,18}, and adiabatic currents \cite{20,25}, currently involves vagueness or arbitrariness in extraction of the observable ${\langle}\hat{r}{\rangle}$. We leave the development of a unified transport theory based on CRM for the future. Here, we concentrate on the issue of observable extraction. Since the definition of the $r$-matrix is subject to different principles, distinct transformation behaviors and gauge issues emerge. The methods for extracting expectation values also vary. All these phenomena suggest that $\hat{r}$ is not the same type of operator as spin. To unify the differeing concepts, we introduce the notion of a \emph{ribbon}. Another focus is on \emph{designation systems}, and we point out the risks in using the bra/ket notation when differential operators are involved. The organization, major results, and innovative features of the paper are summarized in Table~\ref{tab1}.

\section{2. Position operators and Weyl algebras.}

We shall introduce three important vector spaces that will be involved. The first is the space $\mathcal{H}$ spanned by the eigenstates of the position operator $\hat{r}$ (or of $p$ | the two sets of eigenstates are equivalent bases linked by Fourier transformation). The second vector space is the Bloch space $\mathcal{H}_B$ by bases $|\psi_{n,k}{\rangle}$. Bloch space $\mathcal{H}_B$ is isomorphic to the space $\mathcal{H}_W$ spanned by the Wannier functions \cite{25}. The third vector space $\mathcal{V}$ will be defined shortly, as a quotient space of $\mathcal{H}_B$ (i.e., $\mathcal{H}_B$ can be expressed as $\mathcal{V}{\otimes}E$, where $E$ is another vector space). We will show how to bring $\hat{r}$ down from $\mathcal{H}$, on which it is originally defined, to a matrix defined on a finite-dimensional quotient space $\mathcal{V}$.

Let us first introduce $\mathcal{H}$. In general, the identity of a vector space is characterized by: the dimension and the inner product. If and only if both aspects are the same, two vector spaces are considered identical; if there exist invertible (one-to-one) map between two spaces and the inner product remains unchanged after the map (formally, such a map is called an inner-product-preserving or structure-preserving map), the two spaces are said isomorphic.

The dimension of $\mathcal{H}$ is evidently infinite as 
eigenvalue $r$ takes all possible $\mathbb{R}$. To more accurate, the dimension is uncountable infinite, as detailed shortly. The inner product defined for a vector space (often said ``equipped on the space") is formally a map $\mathcal{H}{\times}\mathcal{H}{\rightarrow}\mathbb{C}$, which means the inputs (one the left of $\rightarrow$) are two elements (vectors) in space $\mathcal{H}$ and the output is a complex number $\mathbb{C}$.

On top of inner products, one could say the space $\mathcal{H}$ is complete (such that it is qualified for a Hilbert space), referring to the following fact,
\begin{equation}
\begin{split}
{\langle}r|r'{\rangle}={\delta}_{r,r'}~~r,r'{\in}\mathbb{R}.
\end{split}
\label{eq7}
\end{equation} 
Each $r$ corresponds to a distinct eigenstate, thus these eigenstates are as numerous as real numbers. With more rigor, the \textit{cardinality} (a term characterizing the population of an infinite set) of the eigenstates is equal to that of $\mathbb{R}$. The set $\mathbb{R}$ is known as “uncountably infinite”, by its meaning, unable to list the entries in a one-to-one correspondence with the set of natural numbers $\mathbb{N}$, which is called countable infinite. In other words, $\mathbb{R}$ is ``more" than $\mathbb{N}$, although both are infinite. Therefore, if there is another space whose dimension is countably infinite, it is smaller than $\mathcal{H}$, and thus cannot be isomorphic to $\mathcal{H}$.

When dimensions rise to infinity, some fundamental changes take place. For example, it is possible to write a finite-dimensional operator in matrix form. With respect to eigenstates forming a basis, the matrix is diagonal. Now suppose we try to consider
\begin{equation}
\begin{split}
{\delta}_{r,r'}{\cdot}r_{r,r'}
\to
\begin{pmatrix} r_{1,1} & \cdots & 0 \\ \vdots & r_{i,i} & \vdots \\ 0 & \cdots & r_{N,N} \end{pmatrix}_{N{\rightarrow}{\infty}}.
\end{split}
\label{eq8}
\end{equation}
It might tempting to think that ${\delta}_{r,r'}{\cdot}r_{r,r'}$ could be a matrix, if $r$, $r'$ may be regarded as row/column labels, except that now the matrix becomes infinitely big ($N{\rightarrow}{\infty}$) to host the infinite number of elements. However, this is incorrect. Firstly, the row/column labels take values in the set of (positive) natural numbers $N$, which is countably infinite. This procedure does not apply to uncountably infinite sets such as $\mathbb{R}$. Intuitively speaking, the matrix with countably infinite many rows/columns is still “not big enough.” Secondly, the matrix formalism stipulates that contraction of $i$ should sum over all its possible values: ${\sum}_{j}M_{i,j}{\phi}_{j}$. When this is extended to $r$, it becomes an uncountably infinite sum ${\sum}_{r{\in}{\mathbb{R}}}M_{r,r'}{\phi}_{r'}$, which in general diverges \cite{Note1}.

Another issue regarding the infinite dimension is the lack of converging norms. The norm means the “length” of a vector, i.e., $\|r\Vert:={\langle}r|r{\rangle}$, which should be positive definite, and physically gives the probability density of a particular eigenstate. To normalize the integration over a continuous range, one has to accept $\|r\|$ as an infinite spike, i.e., a $\delta$-function. The diverging norm will make derivatives of states ill-defined and thus, in this space, a Berry connection
\begin{equation}
\begin{split}
{\langle}r|{\partial}_{r}r{\rangle}
\end{split}
\label{eq9}
\end{equation}
is also ill-defined. This is understandable, since an arbitrarily small deviation $r+{\Delta}r$ will make Eq.~\ref{eq7} jump from infinity to zero (i.e., ${\Delta}r=0$, ${\langle}r|r{\rangle}={\infty}$, and ${\Delta}r{\neq}0$, ${\langle}r|r+{\Delta}r{\rangle}=0$), evidently not differentiable. In fact, the following aspects are interrelated: (1) the norm of a space; (2) the dimension of a space; (3) differentiability and derivative of vector states; (4) geometric notions, such as Berry connection and curvatures. To define notions such as Berry curvatures, we must reduce the dimension of the space, and the four aspects above need to be modified in parallel. 

In addition, the average position of an extensive state, e.g., plane waves, will also diverge. This is not a concern for scattering problems, where normalization is not required, and just the relative amplitudes of incoming/outgoing beams are adequate; or again when only localized states are involved and the average position is constantly fixed, such as for atomic Hamiltonians or harmonic oscillators \cite{1}. However, for transport (e.g., shift currents \cite{8,18}), a diverging position could be fatal, destroying any attempt at a meaningful definition of transport.

In the space $\mathcal{H}$, the operator $\hat{r}$ is needed in the commutation relation
\begin{equation}
\begin{split}
[{\hat{r},p}]=i\hbar\,,
\end{split}
\label{eq10}
\end{equation}
but never stands alone. It is always paired with its conjugate, the momentum $p$ \cite{1,2}. By linearity, one may adsorb $i$ into the operator to obtain the alternative convention $[\hat{r},p]=1$. With Eq.~\ref{eq10} as the generator, one obtains an infinite set of operators forming a ring. A \emph{ring} is an algebra equipped with two operations: addition and multiplication \cite{40} (division is not required). Addition must be Abelian and invertible; multiplication is not required to be commutative or invertible. 

A most familiar ring is the set of integer $\mathbb{Z}$. Apparently, one has addition and multiplication defined among integers; most importantly, addition and multiplication of two integers will give another integer - this requirement is known as \textit{closure}. Thus, in this case, the ring is the set of integer numbers combined with operations defined on them (or said equipped on them); thus, it is more than just a set.

In general, the elements in a ring could be anything. Here we concern a ring composed by polynomials and derivatives as below
\begin{equation}
\begin{split}
f_m(r)\frac{{\partial}^m}{{\partial}r^m}\ \mbox{ or }\ f_m(p)\frac{{\partial}^m}{{\partial}p^m}
\end{split}
\label{eq11}
\end{equation}
using the Einstein convention, where $f_m$ is a polynomial serving as the “coefficients” of partial derivatives. One is at liberty to select either $r$ or $p$ as the variable. The ring generated by Eq.~\ref{eq11} is called a \emph{Weyl algebra} \cite{10,11}. In fact, we encounter many Weyl algebras in quantum mechanics. Consider the formalism 
\begin{equation}
\begin{split}
{\int}{\psi}^*(r)(-i)\frac{\partial}{{\partial}r}{\phi}(r)dr
\end{split}
\label{eq12}
\end{equation}
for yielding the expectation value of momentum.
It involves the multiplication of ${\psi}^*(r)(-i)\frac{\partial}{{\partial}r}$ and ${\phi}(r)$, where generic functions ${\psi}^*(r)$ and ${\phi}(r)$ can be approximated by Taylor expansion with polynomials $f_m(r)$ serving the role of coefficients. The integration over $r$ arises from the addition operation equipped on the ring. Hence, defining the position operator, which is a major goal in this work, boils down to its mathematical role in constructing generators of Weyl algebras. (Appx. F) 

In quantum mechanism, we tend to interpret ${\partial}_r{\phi}(r)$ as a derivative ``acting" on a function of $r$. In other words, ${\partial}_r$ is operation, and ${\phi}(r)$ is a function for ${\partial}_r$ to act on; ${\partial}_r$ and ${\phi}(r)$ are not on the equal status. In the ring framework, this is equivalently interpreted as an abstract multiplication of ${\partial}_r$ with ${\phi}(r)$. The ${\partial}_r$ is interpreted with $1{\cdot}{\partial}_r$, and ${\phi}(r)$ is interpreted as  ${\phi}(r){\cdot}{\partial}_{r^0}$, such that the two are on the equal status as elements in a ring. For example, if ${\phi}(r)=r^2$, the result of multiplication is $2r{\partial}_r$, which corresponds in Eq.~\ref{eq11} to $f_m(r)=2r$ and $m=1$. In abstract algebra, such a mutliplication is no different than a ``normal" multiplication like $2{\cdot}4=8$, as long as closure (definition) of the multiplication is respected. The closure will determine the range of the ring. Note that the conjugate pair $\hat{r}$ and $p$ are Weyl algebra generators; the full Weyl algebra contain all possible orders of polynomials for multiplication closure. 

It is instructive to compare the Weyl algebra with Lie algebra; the latter is a vector space $V$ ($s_x$, $s_y$, $s_z$ serve as bases), over $\mathbb{R}$ or $\mathbb{C}$ (real or complex numbers as coefficients to be multiplied with bases $s_i$), equipped with Lie brackets \cite{38}, which is just the commutator ``$[~,~]$". Consider spin operators $\mathfrak{su}(2,\mathbb{C})$ with $[s_i,s_j]={\epsilon}_{i,j,k}s_k$. In intuitive language, the bracket will make two operators (ones plugged in brackets) become a single operator on the right. Formally, the bracket is a binary map: $V{\times}V{\rightarrow}V$. Finding spin representation is just looking for mathematical objects (matrices or any other well-defined terms) that will reproduce the relation described by the brackets. For a Weyl algebra, a crucial difference is Eq.~\ref{eq10} replacing the operator on the right by a complex number, as a map $V{\times}V{\rightarrow}\mathbb{C}$, where $\mathbb{C}$ is a complex number.

This subtle difference leads to significant consequences: finite-dimension matrix representations of Weyl algebras do not exist. In other words, Weyl algebras cannot be represented by matrices. If they could, we would have
\begin{equation}
\begin{split}
\text{Tr}([A,B])=A_{i,l}B_{l,i}-B_{i,l}A_{l,i}=0=\text{Tr}(i{\cdot}\mathbb{I}_N)=Ni\,,
\end{split}
\label{eq13}
\end{equation}
where the only solution would be with trivial zero-dimensional matrices | $N=0$. 

This is why Eq.~\ref{eq11}--\ref{eq12} take the form of polynomials and differential operators rather than matrices. In general, they belong to a Weyl algebra. (Appx. F) The representation is 
\begin{equation}
\begin{split}
\hat{r}{\mapsto}i\frac{\partial}{{\partial}p}\,,
\end{split}
\label{eq14}
\end{equation}
yielding the 1-st Weyl algebra $A_1$. Lie algebras (such as $\mathfrak{sl}(2,\mathbb{C})$) can be realized as subalgebras of $A_1$ \cite{11}.
We can also consider multiple pairs of variables ${\lbrace}r_i,p_i{\rbrace}_N$, yielding the $N$-th Weyl algebra $A_N$. Consequently, the polynomials may involve multiple variables: 
\begin{equation}
\begin{split}
f_m(p)\frac{{\partial}^m}{{\partial}p^m}{\mapsto}f_{m_1,\ldots,m_N}(p_1,\ldots,p_N){\prod}_i^N\frac{{\partial}^{m_i}}{{\partial}p_i^{m_i}}.
\end{split}
\label{eq15}
\end{equation}
The $N$-th Weyl algebra $A_N$ (Eq.~\ref{eq15}) will be used to construct an $r$-matrix in Sec. 4. (Also see Appx. F)

\section{3. Structure of Bloch space and its quotients.}

In this section, we introduce the other two spaces: Bloch space $\mathcal{H}_B$ and its quotient space $\mathcal{V}$ (in some literature, quotient space is also called \textit{factor space}). We will first point out some “bad features” of $\mathcal{H}_B$. Then, we will construct a product space $\mathcal{V}{\otimes}E$, which is isomorphic to $\mathcal{H}_B$ (the definition of isomorphic is given in Sec. 2). The space $\mathcal{V}{\otimes}E$ is well behaved, and the $r$-matrix will be defined on the product space instead of on $\mathcal{H}_B$. Since $\mathcal{V}{\otimes}E{\cong}{\mathcal{H}_B}$, the space $\mathcal{V}$ is also a quotient space of $\mathcal{H}_B$. 

Some features of $\mathcal{H}_B$ are unsuitable for serving as a Hilbert space. We give two examples. Firstly, Hilbert space should be a Banach space (a complete normed vector space) \cite{2,38}. The ``normed" means a vector could be normalized to unity, such that a physical probability could be recognized. The norm is defined as 
\begin{equation}
\begin{split}
\|{\psi}_{n,k}\|
&=\bigg[{\int}|{\psi}_{n,k}(r)|^p\bigg]^{\frac{1}{2}}
=\bigg[{\int}_{-{\infty}}^{\infty}{\psi}_{n,k}^*(r){\psi}_{n,k}(r)dr\bigg]^{\frac{1}{2}}\\
&=\bigg[{\int}_{-{\infty}}^{\infty}u_{n,k}^*(r)u_{n,k}(r)dr\bigg]^{\frac{1}{2}}
\end{split}
\label{eq16}
\end{equation}
(with $p=2$) is required to be finite. In physics, Eq.~\ref{eq16} is comprehended as the total probability (or the total number of particles) in the space should be finite. In the above, we have used Bloch's Theorem 
\begin{equation}
\begin{split}
\psi_{n,k}(r)=e^{ikr}u_{n,k}(r)
\end{split}
\label{eq17}
\end{equation}
| $u_{n,k}(r)$ is a periodic function of the lattice constant $a$.
Evidently, the norm for $\mathcal{H}_B$ diverges, i.e., the integration Eq.~\ref{eq16} diverges. (It suffices to consider the special case where $u_{n,k}(r)$ is a constant function). 

The norm can be induced from the inner product: the self-product of a vector yields its norm. Thus, defining the norm boils down to defining inner products with continuous indices. Definitions like Eq.~\ref{eq16} arise from the analog
\begin{equation}
\begin{split}
\text{Inner Prod.}{\sum}_j^N{\psi}_j^*{\psi}_j{\rightarrow}{\int}_{-\infty}^{+\infty}{\psi}^*(r){\psi}(r)dr\,,
\end{split}
\label{eq18}
\end{equation}
where continuous $r$ assumes the role of the discrete values $j$, the sum over $j$ becoming an integration over infinity. Eq.~\ref{eq18} is an obvious transition from the discrete to the continuous case, but it is not the only one, and may not be the proper one. It will be modified later (Sec. 4), as a key step to casting $\hat{r}$ onto discrete bases. 
\begin{equation}
\begin{split}
{\langle}{\psi}_{m,k'}|{\psi}_{n,k}{\rangle}={\delta}_{m,n}{\delta}_{k',k}\xrightarrow{\overset{N{\rightarrow}{\infty}}{}}{\delta}_{m,n}{\delta}(k'-k).
\end{split}
\label{eq19}
\end{equation}
As a convention, discrete variables are denoted in subscripts; if $m=n$, ${\delta}_{m,n}=1$. For continuous variables, we denote them in brackets; if $k=k'$, ${\delta}(k-k'){\rightarrow}{\infty}$. With Eq.~\ref{eq19}, the self-product $m=n$, $k=k'$ produces an infinitely “long” vector, meaning an infinite probability. Besides, it also makes the derivative $|{\partial}_k{\psi}_{n,k}{\rangle}$ diverge, a similar problem to that inherent in Eq.~\ref{eq9}. 

As the second example of bad behavior, we observe that $\mathcal{H}_B$ is incomplete for $\hat{r}$, which is a serious concern since transport arises from position change. The matrix of the position operator is expressed as
\begin{equation}
\begin{split}
{\int}_{-\infty}^{+\infty}{\psi}_{m,k'}^*(r)r{\psi}_{n,k}(r)dr,
\end{split}
\label{eq20}
\end{equation}
which unfortunately diverges. This is easily seen from translating the diagonal terms
\begin{equation}
\begin{split}
\hat{T}_a{\langle}\hat{r}{\rangle}={\langle}\hat{r}{\rangle}-a,
\end{split}
\label{eq21}
\end{equation}
where $\hat{T}_a$ is the translation operator by $a$ (here with $a{\neq}0$). Plugging Eq.~\ref{eq20} into ${\langle}\hat{r}{\rangle}$, we obtain the contradiction
\begin{equation}
\begin{split}
\hat{T}_a\bigg[{\int}{\psi}_{n,k}^*(r)r{\psi}_{n,k}(r)dr\bigg]&={\int}{\psi}_{n,k}^*(r)e^{-ika}re^{ika}{\psi}_{n,k}(r)dr \\
&={\int}{\psi}_{n,k}^*(r)r{\psi}_{n,k}(r)dr={\langle}\hat{r}{\rangle}.
\end{split}
\label{eq22}
\end{equation}
Here, we have applied $\hat{T}_a{\psi}_{n,k}(r)=$
\begin{equation}
\begin{split}
{\psi}_{n,k}(r+a)=e^{ik(r+a)}u_{n,k}(r+a)=e^{ika}{\psi}_{n,k}(r).
\end{split}
\label{eq23}
\end{equation}
The periodic function is a function of crystal momentum $k$ and band $n$, transcribed from $\psi_{n,k}(r)$.  Now $u_{n,k}(r+a)$ is obtained by translating $u_{n,k} (r)$ by $a$ in the negative direction, thus ${\langle}\hat{r}{\rangle}$ in Eq.~\ref{eq21} is shifted by $-a$.

The inconsistency between Eq.~\ref{eq21} and Eq.~\ref{eq22} indicates that the integration in Eq.~\ref{eq20} cannot converge, for otherwise the contradiction ${\langle}\hat{r}{\rangle}-a={\langle}\hat{r}{\rangle}$ would be obtained. This divergence is genuine and inevitable. It arises from the fact that it is impossible to pin down the “center” of an infinitely extensive wave function. One may attribute this divergence to the infinite dimension of $\mathcal{H}_B$ (the infinite dimension arising from the infinite number of possible values for ($n$, $k$), where $k$ might be either discretely infinite or continuous), since a sum over a finite numbers of terms should never diverge. Note that each distinct ($n$, $k$) corresponds to a linearly independent basis. Note the ambiguous meanings of “bases”: the bases span $\mathcal{H}_B$, rather than the vector label $k$. It is very possible that $k{\cdot}k'{\neq}0$, but ${\langle}{\psi}_{n,k}|{\psi}_{n,k'}{\rangle}=0$.

To resolve these problems, we next construct a well-behaved product space related to $\mathcal{H}_B$ with isomorphism; CRM will be established on the well-behaved space. We begin by introducing a well-defined inner product, on the basis of which procedures associated with vectors, operators, etc. may be defined \cite{38}. We adopt the following procedure to force convergence of Eq.~\ref{eq20}:
\begin{equation}
\begin{split}
&{\int}{\psi}_{m,k_p}^*(r){\psi}_{n,k_q}(r)dr=\\
{\cdots}&+{\int}_{-a}^0+{\int}_0^a{\psi}_{m,k_p}^*(r){\psi}_{n,k_q}(r)dr+{\int}_a^{2a}+{\cdots} \\
&={\sum}_i^Ne^{-i(k_p-k_q)R_i}{\int}_0^a{\psi}_{m,k_p}^*(r){\psi}_{n,k_q}(r)dr.
\end{split}
\label{eq24}
\end{equation}
For simplicity, consider a 1D atomic chain of $N$ sites with lattice constant $a$. Set $R_i=(i-1){\cdot}a$ and $k_p=(p-1){\cdot}\frac{2\pi}{Na}$ ($i,p,q=1,\ldots,N$). The sum over $R_i$ is independent of $r$, and thus can be factored out: the infinite integration is reduced to a finite integration. We have
\begin{equation}
\begin{split}
{\sum}_i^Ne^{-i(k_p-k_q)R_i}=N{\cdot}{\sum}_l^N{\delta}_{k_p,k_q-G_l}\xrightarrow{\overset{1^{st}B.Z.}{}}N{\cdot}{\delta}_{k_p,k_q}.
\end{split}
\label{eq25}
\end{equation}
If $k$ and $k'$ are restricted to the first B.Z. by convention, the above sum can be reduced to a single $\delta$-function. For the other term, in-cell integration, we require
\begin{equation}
\begin{split}
&{\int}_0^a{\psi}_{m,k_p}^*(r){\psi}_{n,k_p}(r)dr\\
&={\int}_0^ae^{-ik_{p}r}u_{m,k_p}^*(r)e^{ik_{q}r}u_{n,k_p}(r)dr={\delta}_{m,n}
\end{split}
\label{eq26}
\end{equation}
for the case where $k_p=k_q$ . On the other hand, for $k_p{\neq}k_q$, even if $m{\neq}n$, Eq.~\ref{eq26} may not necessarily vanish, because it is Eq.~\ref{eq25} which governs the orthogonality for distinct $k$-values. In that case, Eq.~\ref{eq26} plays a role of a normalization factor, as discussed in connection with Remark~3.3 below. 

We have expressed the inner product Eq.~\ref{eq24} as a product of two $\delta$-functions. This leads one to think that $\mathcal{H}_B$ can be isomorphic to a tensor product space. We invoke ${\delta}_{k_p,k_q}$ from Eq.~\ref{eq25} to generate one quotient space $E$ of dimension $N$ (the number of possible values of $k$), and invoke ${\delta}_{m,n}$ to generate a second quotient space $\mathcal{V}$ of dimension $\mathcal{N}$ (the number of bands), yielding the isomorphism
\begin{equation}
\begin{split}
{\Pi}\colon \mathcal{H}_B\to\mathcal{V}{\otimes}E;
|{\psi}_{n,k}{\rangle}{\mapsto}|A_{n,k}{\rangle}{\otimes}|E_k{\rangle}
\end{split}
\label{eq27}
\end{equation}
with $|A_{n,k}{\rangle}{\in}\mathcal{V}$ and $|E_{k}{\rangle}{\in}E$.
Here, both ``$\rightarrow$" and ``${\mapsto}$" indicate a map. The difference is ``$\rightarrow$" connects two sets: from map's domain to co-domain; ${\mapsto}$ connects elements belonging to the two sets. Thus, ``$\rightarrow$" and ``${\mapsto}$" in Eq.~\ref{eq27} stand for two conventions for denoting a map. These map denotations will be frequently used in this paper, especially in Sec. 5 and 6.

Isomorphism means a linear map which preserves the inner product (inner product value is invariant with map)
\begin{equation}
\begin{split}
{\langle}{\psi}_{m,k'}|{\psi}_{n,k}{\rangle}={\langle}A_{m,k'}|A_{n,k}{\rangle}\cdot{\langle}E_{k'}|E_{k}{\rangle}
\end{split}
\label{eq28}
\end{equation}
for $|{\psi}_{m,k'}{\rangle},|{\psi}_{n,k}{\rangle}{\in}\mathcal{H}_B$. Intuitively speaking, we seek a replacement vector $|A_{n,k}{\rangle}{\otimes}|E_k{\rangle}$ for the original $|{\psi}_{n,k}{\rangle}$, such that after the replacement the inner product value remains unchanged as indicated by Eq.~\ref{eq28}.

To facilitate analysis, we introduce maps
\begin{equation}
\begin{split}
&{\Pi}_1:I_B{\rightarrow}\mathcal{V} \\
&{\Pi}_2:I_B{\rightarrow}E \\
&{\Pi}={\Pi}_1\otimes{\Pi}_2:I_B{\rightarrow}\mathcal{V}{\otimes}E
\end{split}
\label{eq29}
\end{equation}
where $I_B$ is the basis 
$$
I_B={\lbrace}{\psi}_{n,k}|n=1,2,{\ldots},\mathcal{N};k=p{\cdot}\tfrac{2{\pi}}{Na},p=0,1,{\ldots},N{\rbrace}
$$
of $\mathcal{H}_B$ which contains $\mathcal{N}{\times}N$ elements.

Since each basic Bloch vector is characterized by band $n$ and crystal momentum $k$, one can write these maps in terms of
\begin{equation}
\begin{split}
{\Pi}_i({\psi}_{n,k})~\text{or}~{\Pi}_i(n,k).
\end{split}
\label{eq30}
\end{equation}
In particular, $\Pi$ can be expressed as
\begin{equation}
\begin{split}
{\Pi}({\psi}_{n,k})={\Pi}_1({\psi}_{n,k}){\otimes}{\Pi}_2({\psi}_{n,k})\,.
\end{split}
\label{eq31}
\end{equation}
The existence of these maps ${\Pi}_i$ is constrained by 
\begin{equation}
\begin{split}
{\int}_0^a{\psi}_{m,k_p}^*(r){\psi}_{n,k_q}(r)dr={\sum}_{i=1}^{\mathcal{N}}a_i^{(m)*}(k_p){\cdot}a_i^{(n)}(k_q)
\end{split}
\label{eq32}
\end{equation}
as discussed in Appx. E. 
The maps ${\Pi}_1$, ${\Pi}_2$ are then defined as
\begin{equation}
\begin{split}
&{\Pi}_1\colon |{\psi}_{n,k}{\rangle}{\mapsto}|A_{n,k}{\rangle}=\begin{pmatrix} a_1^{(n)}(k) \\ a_2^{(n)}(k) \\ {\vdots} \\ a_{\mathcal{N}}^{(n)}(k) \end{pmatrix}\ \mbox{ and} \\
{\Pi}_2&\colon |{\psi}_{n,k}{\rangle}{\mapsto}|E_{k}{\rangle}=\frac{1}{\sqrt{N}}\begin{pmatrix} e^{-ikR_0} \\ e^{-ikR_1} \\ {\vdots} \\ e^{-ikR_{N-1}} \end{pmatrix}\,,
\end{split}
\label{eq33}
\end{equation}
with
\begin{align}\label{InnProdA}
&{\langle}A_{m,k_p}|A_{n,k_q}{\rangle}
={\sum}_{i=1}^{\mathcal{N}}a_i^{(m)}(k_p)^*a_i^{(n)}(k_q)\ \mbox{ and} \\
&{\langle}E_{k_p}|E_{k_q}{\rangle}=
\frac{1}{N}{\sum}_{j=0}^{N-1}e^{-i(k_p-k_q)R_j}
\label{InnProdE}
\end{align}
as the corresponding inner product rules.
Since
\begin{equation}
\begin{split}
&{\langle}A_{m,k_p}|A_{n,k_q}{\rangle}{\otimes}{\langle}E_{k_p}|E_{k_q}{\rangle}\\
&=\frac{1}{N}{\sum}_{j=0}^{N-1}e^{-i(k_p-k_q)R_j}{\cdot}{\int}_0^a{\psi}_{m,k_p}^*(r){\psi}_{n,k_q}(r)dr \\
&={\langle}{\psi}_{m,k_p}|{\psi}_{n,k_q}{\rangle}\,,
\end{split}
\label{eq36}
\end{equation}
the conjectured map $\Pi$ preserve inner products.

\textbf{Remark 3.1.} Recall the main goal of this section: building a well-behaved isomorphic space to substitute for $\mathcal{H}_B$. The goal is realized by ${\Pi}$ map constructed with Eqs.~\ref{eq27}--\ref{eq36}. The map ${\Pi}$ reads
\begin{equation}
\begin{split}
&|{\psi}_{n,k}{\rangle}{\mapsto}|A_{n,k}{\rangle}{\otimes}|E_{k}{\rangle} =\frac{1}{\sqrt{N}}\begin{pmatrix} a_1^{(n)}(k) \\ a_2^{(n)}(k) \\ {\vdots} \\ a_{\mathcal{N}}^{(n)}(k) \end{pmatrix}{\otimes}\begin{pmatrix} e^{-ikR_0} \\ e^{-ikR_1} \\ {\vdots} \\ e^{-ikR_{N-1}} \end{pmatrix}
\end{split}
\label{eq34}
\end{equation}
to circumvent integration over $r$, facilitating the evaluation of the $r$-matrix and other expressions (Sec. 4). 

\textbf{Remark 3.2.} Roughly speaking, $\Pi$ creates a new basis equivalent to $I_B$ 
by switching the representation from $\mathcal{H}_B$ to $\mathcal{V}{\otimes}E$. But in fact, $\Pi$ does not map vectors to vectors. In the ad hoc terminology introduced in Sec.~5, it maps vectors to \emph{ribbon bands} or \emph{ribbons}. Within the framework of bundle theory \cite{37,38}, a ribbon is a feature of the bundle space $K{\times}\mathcal{H}_B$, where $K$ is the Brillouin zone and Bloch space $\mathcal{H}_B$ is the fiber space (see Sec. 5). Precisely speaking, $\Pi$ is to map $N$ mutually orthogonal vectors ($I_B$) to a ribbon (Sec. 5), on which the vectors at ${\lbrace}k_i{\rbrace}_N$ will comprise a set of basis vectors isomorphic with the original set $I_B$. In plain but less accurate words, this means that the components of the vectors (Eq.~\ref{eq34}) should be functions of $k$ (e.g., $a_i^{(n)}(k), e^{-ikR_0}$) instead of constant values. For example, the following map also defines an isomorphism on $I_B$ but does not work:
\begin{equation}
\begin{split}
|{\psi}_{n,k_p}{\rangle}{\mapsto}{\delta}_{j,n}{\otimes}{\delta}_{l,p}=\begin{pmatrix} 0 \\ {\vdots} \\ 1^{n^{th} row} \\ {\vdots} \end{pmatrix}{\otimes}\begin{pmatrix} 0 \\ {\vdots} \\ 1^{p^{th} row} \\ {\vdots} \end{pmatrix}.
\end{split}
\label{eq37}
\end{equation}
The crucial difference is that in Eq.~\ref{eq34} we have $\lim_{k_p{\to}k_q}{\Pi}[|{\psi}_{n,k_p}{\rangle}]={\Pi}[|{\psi}_{n,k_q}{\rangle}]$ when the $k_p$'s are viewed as variables. For maps like Eq.~\ref{eq37}, however, we have $\lim_{k_p{\to}k_q}{\Pi}[|{\psi}_{n,k_p}{\rangle}]={\Pi}[|{\psi}_{n,k_p}{\rangle}]$, since the components in Eq.~\ref{eq37} are constants. If the $k$-independent Eq.~\ref{eq37} interacts with partial derivatives, the $r$-matrix will vanish. Thus, $\Pi$ cannot be chosen with an arbitrary isomorphism defined from $I_B$. 

\textbf{Remark 3.3.} How should we visuaize the map $\Pi$? The wave functions ${\psi}_{n,k}$ are the eigenstates of the translations $\hat{T}_a$. Thus, we seek representations of the translation group on $\mathcal{H}_B$ and its quotients. It is conceivable that one part of space affords a trivial repesentation, while the rest affords a non-trivial one characterized by $k$. Spaces can be described by orthogonal bases: the $k$-independent ${\delta}_{m,n}$ in Eq.~\ref{eq26} seem appropriate for the trivial representation, with the ${\delta}_{k,k'}$ for the non-trivial part. Thus, we introduce ${\Pi}_1$ to project to vectors whose inner product is meant to reproduce ${\delta}_{m,n}$, while ${\Pi}_2$ projects to the other quotient space. Constructing a function space requires more than just expressing the functions: the inner product must also be specified. Thus, we simultaneously move from Eq.~\ref{eq24} to Eq.~\ref{eq36}.

We can rewrite the image of $|{\psi}_{n,k_p}{\rangle}$ under $\Pi$ from  Eq.~\ref{eq34} in a single Kronecker product column as 
\begin{equation}
\begin{split}
\begin{pmatrix} a_1^{(n)}(k) \\ {\vdots} \\ a_{\mathcal{N}}^{(n)}(k)\end{pmatrix}_{N}{\otimes}\begin{pmatrix} e^{-ikR_1} \\ {\vdots} \\ e^{-ikR_N} \end{pmatrix}_{\mathcal{N}}=\begin{pmatrix} a_1^{(n)}(k){\cdot}e^{-ikR_1} \\ {\vdots} \\ a_1^{(n)}(k){\cdot}e^{-ikR_N} \\ a_2^{(n)}(k){\cdot}e^{-ikR_1} \\ {\vdots} \\ a_2^{(n)}(k){\cdot}e^{-ikR_N} \\ {\vdots} \end{pmatrix}_{N\mathcal{N}}
\end{split}
\label{eq39}
\end{equation}
so that the inner product Eq.~\ref{eq36} returns to its familiar row-times-column form with rows and columns of length $N\mathcal{N}$. In the conjugate transpose row of the column from Eq.~\ref{eq39}, element $a_i^{(n)}(k){\cdot}e^{ikR_j} $ is the component for the $n$-th Bloch wave at the point $k$ projecting to band $i$ and local site $j$. There are two pairs of conjugated variables: $n{\sim}i$, and $k{\sim}R_j$. Note that $k$ is not conjugated with $r$, since $r$ merely provides the normalization factor of Eq.~\ref{eq26}. The translation group acts trivially on the quotient space $\mathcal{V}$ | translation does not change vectors $v{\in}\mathcal{V}$. This is why $R_j$ is only involved in the quotient space $E$.

Bearing in mind that components in the column vector of Eq.~\ref{eq39} must correlate to certain inner products ${\langle}X|{\psi}_{n,k}{\rangle}$, then what basis vectors $|X_{i,R_j}{\rangle}$ are chosen for $|{\psi}_{n,k}{\rangle}$ to be projected onto? We consider the map $\Pi$ as exhibited in terms of basis vectors by Eq.~\ref{eq37} of Remark~3.1.  The basis elements can be regarded as generalized Wannier functions
\begin{equation}
\begin{split}
|{\psi}_{n,k_p}{\rangle}={\sum}_{i=1}^{\mathcal{N}}{\sum}_{j=1}^Na_i^{(n)}(k_p){\cdot}e^{-ik_pR_j}|X_{i,R_j}{\rangle}\,.
\end{split}
\label{eq38}
\end{equation}
If $a_i^{(n)}(k_p)={\delta}_{n,i}$, we recover the normal definition of Wannier functions: $|X_{i,R_j}{\rangle}{\rightarrow}|w_{i,R_j}{\rangle}$. While Eq.~\ref{eq38} is not the standard Fourier transformation, it is still invertible. The space spanned by the generalized Wannier functions $|X_{i,R_j}{\rangle}$ is denoted by $\mathcal{H}_W$. Evidently, $\mathcal{H}_B$ and $\mathcal{H}_W$ are isomorphic. 

The isomorphism (Eq.~\ref{eq34}) is a major result of this work, which allows us to switch from $\mathcal{H}_B$ to $\mathcal{V}{\otimes}E$ and avoid the continuous coordinate $r$ that appears in ${\psi}_{n,k}(r)$ and $u_{n,k}(r)$. Inner products like Eq.~\ref{InnProdA}--\ref{InnProdE} sum over discrete indices, without referring to the integral ${\int}{\cdots}dr$ within a unit cell volume. The pre-factor $\frac{(2{\pi})^d}{V_{\text{cell}}}$ arising from the integration is also avoided. 

We should be cautious about designations such as
\begin{equation}
\begin{split}
&{\psi}_{n,k}(r)={\langle}r|{\psi}_{n,k}{\rangle} \\
&u_{n,k}(r)={\langle}r|u_{n,k}{\rangle}.
\end{split}
\label{eq40}
\end{equation}
These denotations are often seen in literatures but not perfectly accurate. Since $|r{\rangle}{\in}\mathcal{H}$, $|{\psi}_{n,k}{\rangle}{\in}\mathcal{H}_B$, and $\mathcal{H}{\ncong}{\mathcal{H}_B}$ (later we will show $|u_{n,k}{\rangle}{\in}\mathcal{V}$). Rigorously speaking, inner products are illegal to be defined between  vectors in different spaces, such as ${\langle}r|{\psi}_{n,k}{\rangle}$ and ${\langle}r|u_{n,k}{\rangle}$. Given Eq.~\ref{eq40} is accepted, one obtains
\begin{equation}
\begin{split}
{\langle}r|{\psi}_{n,k}{\rangle}&={\psi}_{n,k}(r)=e^{ikr}u_{n,k}(r) \\
&=e^{ikr}{\langle}r|u_{n,k}{\rangle}={\langle}r|e^{ik\hat{r}}|u_{n,k}{\rangle}. 
\end{split}
\label{eq41}
\end{equation}
If ${\langle}r|$ is taken away from the left and only ket is kept, we have following expressions.
\begin{equation}
\begin{split}
&|{\psi}_{n,k}{\rangle}=e^{ik\hat{r}}|u_{n,k}{\rangle} \\
&|u_{n,k}{\rangle}=e^{-ik\hat{r}}|{\psi}_{n,k}{\rangle}.
\end{split}
\label{eq42}
\end{equation}
Eq.~\ref{eq42} is about using a unitary operator $e^{ik\hat{r}}$ to link two vectors $|u_{n,k}{\rangle}$ and $|{\psi}_{n,k}{\rangle}$. In general, a unitary operator is invertible and may only connect spaces of the same dimension. Remember $|{\psi}_{n,k}{\rangle}$ and $|u_{n,k}{\rangle}$ are belonging to spaces $\mathcal{H}_B$ and $\mathcal{V}$, which are of different dimensions. The concern is that $e^{ik\hat{r}}$ promotes a lower-dimension vector $|u_{n,k}{\rangle}$ to a higher-dimension one $|{\psi}_{n,k}{\rangle}$; its inverse $e^{-ik\hat{r}}$ degrades  $|{\psi}_{n,k}{\rangle}$ to $|u_{n,k}{\rangle}$. Having the same dimension is the sufficient and necessary condition for two vector spaces being isomorphic. $|{\psi}_{n,k}{\rangle}=e^{ik\hat{r}}|u_{n,k}{\rangle}$ (or $|u_{n,k}{\rangle}=e^{-ik\hat{r}}|{\psi}_{n,k}{\rangle}$) would suggest $\mathcal{H}_B{\cong}\mathcal{V}$. However, $\mathcal{V}$ is merely a quotient space of $\mathcal{H}_B$. Similarly, expression like $H(k)=e^{-ik\hat{r}}He^{ik\hat{r}}$ seen in literature (e.g., Eq.2 of \cite{41}) deserves special attention. The dimensions of operators ($e^{ik\hat{r}}$, $H(k)$, etc.) are summarized in Table~\ref{tab5} in Sec. 7.

Another point, as we should be aware of, is that the below orthogonality is false (whether $k$ can be either discrete or continuous)
\begin{equation}
\begin{split}
{\sum}_n^{\mathcal{N}}{\int}{\psi}_{n,k}^*(r){\psi}_{n,k}(r'){\cdot}dk{\neq}{\delta}(r-r').
\end{split}
\label{eq43}
\end{equation}
Otherwise, if equality in Eq.~\ref{eq43} holds, $\mathcal{H}_B$ is complete for $\hat{r}$. The set of functions ${\psi}_{n,k}(r)$ can expand arbitrary functions. However, this is false. To demonstrate the incompleteness, we construct counter examples in Appx. B, i.e., functions unachievable by superposition of ${\psi}_{n,k}(r)$. 

In band contexts, $\mathcal{H}_B$ is the Hilbert space, thus $\mathcal{H}_B$ should be complete --- this idea has been taken for granted. However, although $\mathcal{H}_B$ is complete for operators defined within $\mathcal{H}_B$, $\mathcal{H}_B$ is incomplete for operators defined beyond $\mathcal{H}_B$, such as $\hat{r}$. Evidence includes:
\vskip 0.5mm
(1) Space $\mathcal{H}_B$ spanned by Bloch waves ${\lbrace}|{\psi}_{n,k}{\rangle}{\rbrace}_{N{\times}{\mathcal{N}}}$ has a lower dimension than space $\mathcal{H}$ spanned by eigenstates ${\lbrace}|r{\rangle}{\rbrace}_{\mathbb{R}}$ of position operator; that is, the population (cardinality) of elements in the two basis sets ${\lbrace}|r{\rangle}{\rbrace}_{\mathbb{R}}$ and ${\lbrace}|{\psi}_{n,k}{\rangle}{\rbrace}_{N{\times}{\mathcal{N}}}$ are unequal.
\vskip 0.5mm
(2) Matrix of position operator (shortly seen in Sec. 4) is diverging. 
\vskip 0.5mm
(3) ${\delta}(r-r')$ is false (Eq.~\ref{eq43}), and functions that cannot be achieved by superposition of Bloch waves ${\psi}_{n,k}(r)$ are constructed (Appx. B).
\vskip 0.5mm

A conviction is that different quantum bases are equivalent. Thus, one vaguely believes $|r{\rangle}$ and $|{\psi}_{n,k}{\rangle}$ are equivalent; continuous functions (e.g., ${\psi}_{n,k}(r)$) and discrete bases are equivalent, fancying that these bases could be linked by unitary transformations. However, on a second thought, how can a continuously infinite bases possibly be linked to discrete bases? In fact, these bases are not equivalent. 


We conclude this section by emphasizing some important points:
\vskip 0.5mm
(1) The space $\mathcal{H}_B$ is complete for operators defined within $\mathcal{H}_B$, while $\mathcal{H}_B$ is incomplete for the $\hat{r}$ operator, as $\hat{r}$ is defined in $\mathcal{H}$.
\vskip 0.5mm
\vskip 0.5mm
(2) The sets ${\lbrace}|r{\rangle}{\rbrace}_{\mathbb{R}}$ and ${\lbrace}|{\psi}_{n,k}{\rangle}{\rbrace}_{N{\times}\mathcal{N}}$ are bases of different dimensions (different cardinality); they cannot be linked by a unitary transformation, and one cannot be obtained from the other by a change-of-basis transformation.
\vskip 0.5mm
(3) The respective spaces spanned by ${\lbrace}|r{\rangle}{\rbrace}_{\mathbb{R}}$ and ${\lbrace}|{\psi}_{n,k}{\rangle}{\rbrace}_{N{\times}\mathcal{N}}$ are not isomorphic.
\vskip 0.5mm

Nevertheless, the map $\Pi$ (Eq.~\ref{eq27}) is a precise tool for introducing the $N$-th Weyl algebra (Sec. 4) and obtaining convergent matrices for the position operator. Moreover, geometrical quantities, such as Berry connections and curvatures, are defined unambiguously as operators on the quotient space $\mathcal{V}$ of $\mathcal{H}_B$.

\section{4. Matrices of the position operator.}

In this section, we will identify converging $r$-matrices (CRM). We start by examining $r$-matrices in $\mathcal{H}_B$ as they have appeared in previous work \cite{3}. These matrices inevitably contain divergent terms. The matrix elements are defined by integration of the Bloch function ${\psi}_{n,k}(r)$ over an infinite range. For reasons discussed in Sec. 5, we avoid basis-free designations for the position operator, such as ${\langle}{\psi}_{m,k'}|\hat{r}|{\psi}_{n,k}{\rangle}$. For the moment, we consider the matrix element 
\begin{equation}
\begin{split}
{\int}{\psi}_{m,k'}^*(r)r{\psi}_{n,k}(r){\cdot}dr
\end{split}
\label{eq44}
\end{equation}
given by the original integration. Making the substitution $r{\mapsto}i\frac{\partial}{{\partial}k}$ in Bloch's Theorem (Eq.~\ref{eq17}) , we obtain
\begin{equation}
\begin{split}
r{\psi}_{n,k}(r)&=re^{ikr}u_{n,k}(r) \\ 
&=-i\tfrac{\partial}{{\partial}k}\big[e^{ikr}u_{n,k}(r)\big]+ie^{ikr}\tfrac{\partial}{{\partial}k}u_{n,k}(r)\,.
\end{split}
\label{eq45}
\end{equation}
The matrix elements then take the form
\begin{equation}
\begin{split}
{\int}{\psi}_{m,k'}^*(r){\lbrace}&-i\tfrac{\partial}{{\partial}k}[e^{ikr}u_{n,k}(r)]+ie^{ikr}\tfrac{\partial}{{\partial}k}u_{n,k}(r){\rbrace}{\cdot}dr= \\
&-i\tfrac{\partial}{{\partial}k}{\int}_r{\psi}_{m,k'}^*(r){\psi}_{n,k}(r)\\
&
\rule{10mm}{0mm}
+i{\int}_re^{i(k-k')r}u_{m,k'}^*(r)\tfrac{\partial}{{\partial}k}u_{n,k}(r)
\end{split}
\label{eq46}
\end{equation}
on plugging Eq.~\ref{eq45} back into Eq.~\ref{eq44}. Note that the first term in Eq.~\ref{eq46} becomes $-i{\partial}_k{\delta}_{m,n}{\delta}(k-k')$. Recall that for the continuous $\delta$-function, arguments appear in brackets | e.g., $\delta(k)$, and $\delta(0){\rightarrow}{\infty}$. On the other hand, for the discrete $\delta$-function, the arguments are in the subscripts | e.g., ${\delta}_{m,n}$, and ${\delta}_{m,m}=1$. The second term in Eq.~\ref{eq46} is
\begin{equation}
\begin{split}
&i{\int}e^{i(k-k')r}u_{m,k'}^*(r)\tfrac{\partial}{{\partial}k}u_{n,k}(r){\cdot}dr \\
&={\sum}_i^Ne^{i(k-k')R_i}{\int}_{V_{\text{cell}}}e^{i(k-k')r}iu_{m,k'}^*(r)\tfrac{\partial}{{\partial}k}u_{n,k}(r){\cdot}dr \\
&=\frac{(2\pi)^d}{V_{\text{cell}}}{\delta}(k-k'){\int}_{V_{\text{cell}}}e^{i(k-k')r}iu_{m,k'}^*(r)\tfrac{\partial}{{\partial}k}u_{n,k}(r){\cdot}dr.
\end{split}
\label{eq47}
\end{equation}
For a general function $F(k)$, we have
\begin{equation}
\begin{split}
F(k){\delta}(k)=F(0){\delta}(k)\,.
\end{split}
\label{eq48}
\end{equation}
Thus Eq.~\ref{eq47} becomes
\begin{equation}
\begin{split}
\frac{(2\pi)^d}{V_{\text{cell}}}{\delta}(k-k'){\cdot}{\int}_{V_{\text{cell}}}iu_{m,k}^*(r)\tfrac{\partial}{{\partial}k}u_{n,k}(r){\cdot}dr\,.
\end{split}
\label{eq49}
\end{equation}
We define
\begin{equation}
\begin{split}
A_{m,n}(k):=i{\int}_{V_{\text{cell}}}u_{m,k}^*(r)\tfrac{\partial}{{\partial}k}u_{n,k}(r){\cdot}dr\,,
\end{split}
\label{eq50}
\end{equation}
yielding
\begin{equation}
\begin{split}
&{\int}{\psi}_{m,k'}^*(r)r{\psi}_{n,k}(r){\cdot}dr \\
&=-i{\delta}_{m,n}\frac{\partial}{{\partial}k}{\delta}(k-k')+\frac{(2\pi)^d}{V_{\text{cell}}}{\delta}(k-k')A_{m,n}(k).
\end{split}
\label{eq51}
\end{equation}
Except for a sign difference in the first term, our derivation is consistent with previous work \cite{3}, the second term differing by a normalization convention. Due to the appearence of ${\partial}_k{\delta}(k-k')$, the diagonal terms evidently diverge. Thus, the matrix elements defined by Eq.~\ref{eq51} are merely formal.

Now, instead of working directly with $\mathcal{H}_B$, we construct a Weyl algebra on its isomorphic copy $\mathcal{V}{\otimes}E$, obtaining a non-singular matrix which converges at both the diagonal and off-diagonal entries. The basic rule for the differential operator is the coproduct 
\begin{equation}
\begin{split}
i{\partial}_k=i{\partial}_k{\otimes}1+1{\otimes}i{\partial}_k
\end{split}
\label{eq52}
\end{equation}
in the tensor algebra \cite{38,40}. The partial derivative acts on each of the two tensor factors $\mathcal{V}$ and $E$. Thus,
\begin{equation}
\begin{split}
&({\langle}E_{k'}|{\otimes}{\langle}A_{m,k'}|)(i{\partial}_k{\otimes}1+1{\otimes}i{\partial}_k)(|A_{n,k}{\rangle}{\otimes}|E_k{\rangle})= \\
&{\langle}A_{m,k'}|i{\partial}_kA_{n,k}{\rangle}\cdot{\langle}E_{k'}|E_k{\rangle}
+{\langle}A_{m,k'}|A_{n,k}{\rangle}\cdot{\langle}E_{k'}|i{\partial}_kE_k{\rangle}.
\end{split}
\label{eq53}
\end{equation}
For momentum and position, we have
\begin{equation}
\begin{split}
{\lambda}{\cdot}p=h\
\mbox{ and }\
k=\frac{2\pi}{\lambda}=\frac{2\pi{\cdot}p}{h}=\frac{p}{\hbar}.
\end{split}
\label{eq54}
\end{equation}
The commutator becomes
\begin{equation}
\begin{split}
[\hat{r},k]=i.
\end{split}
\label{eq55}
\end{equation}
To satisfy Eq.~\ref{eq55}, we may choose the position operator
\begin{equation}
\begin{split}
\hat{r}=i{\partial}_k\,.
\end{split}
\label{eq56}
\end{equation}
The operator of Eq.~\ref{eq56} is the generator of the 1-st Weyl algebra $A_1$, as there is a single variable $k$.   

\subsection{4A. The $N$-th Weyl algebra $A_N$.}

The choice $\hat{r}=i{\partial}_k$ made in Eq.~\ref{eq56} is not the only solution of the commutator equation~(\ref{eq55}). Instead, we may consider $N$ variables ${\lbrace}k_i{\rbrace}_N$, thereby obtaining the \emph{$N$-th Weyl algebra} $A_N$ \cite{10}. In this algebra, take the new position operator
\begin{equation}
\begin{split}
\hat{r}=\frac{1}{N}{\sum}_{m=1}^Ni\frac{\partial}{{\partial}k_m}
\end{split}
\label{eq56+}
\end{equation}
in place of Eq.~\ref{eq56}. The commutator equation~(\ref{eq55}) is then solved as
\begin{equation}
\begin{split}
&\bigg[\frac{1}{N}{\sum}_{m=1}^Ni\tfrac{\partial}{{\partial}k_m},{\sum}_{n=1}^N{k_n}\bigg] \\
&
=\frac{1}{N}\bigg({\sum}_{1\le m,n\le N}i\tfrac{\partial}{{\partial}k_m}k_n-{\sum}_{1\le m,n\le N}ik_n\tfrac{\partial}{{\partial}k_m}\bigg)= \\
&
\frac{1}{N}\bigg({\sum}_{1\le m,n\le N}i(k_n\tfrac{\partial}{{\partial}k_m}+{\delta}_{m,n})-{\sum}_{1\le m,n\le N}ik_n\tfrac{\partial}{{\partial}k_m}\bigg) \\
&
=\frac{1}{N}{\sum}_{1\le m,n\le N}i{\cdot}{\delta}_{m,n}=\frac{i}{N}{\sum}_{m=1}^N1=i.
\end{split}
\label{eq57}
\end{equation}
We set the number of variables involved in the Weyl algebra equal to the dimension $N$ of the quotient space $E$. It is absolutely essential for $N$ to be finite, to ensure that the partial derivatives are well-defined without recourse to any infinite limits. Note that Eq.~\ref{eq57} is a concrete realization of Eq.~\ref{eq15} in Sec. 2. The state vector will be parameterized by $k_1,{\ldots},k_N$, assuming the role of $f_{m_1,{\ldots},m_N}(p_1,{\ldots},p_N)$ in Eq.~\ref{eq15}. (Appx. F)

\textbf{Remark 4.1} At this stage of the construction, we will not endow $k_m$ or ${\partial}_{k_m}$ with any physical significance, like taking the $k_m$ as quantum numbers of “one-particle” or “many particle’” states. Thus, $k_m$ is not yet interpreted as the crystal momentum. Currently, we are working at the purely algebraic level, just making the substitutions $\hat{r}{\rightarrow}\frac{1}{N}{\sum}_m^Ni\frac{\partial}{{\partial}k_m}$ and $k{\rightarrow}{\sum}_n^Nk_n$.

\textbf{Remark 4.2} Eq.~\ref{eq57} is at the same fundamental level as  Eq.~\ref{eq56}: both model the Weyl algebra relation of Eq.~\ref{eq55}. 

\textbf{Remark 4.3} It is tempting to think that the variable $k$ appearing in ${\partial}_k$ must be continuous, because otherwise the derivative would not be defined. This misconception is based on the narrow calculus definition of a derivative as ${\partial}_kf(k)=\lim_{{\Delta}k{\to}0}(f(k+{\Delta}k)-f(k))/{\Delta}k$. This definition requires extraneous apparatus, such as a division operation, a limit process, and so on. In fact, it suffices to work with the formal definition by ${\partial}_kk^n=nk^{n-1}$, which only requires multiplication and addition (mathematically, a ring structure \cite{40}). Using power series, one can extend the action of ${\partial}_k$'s to generic analytic functions $f(k)$. 
As an example, taken from \cite{42}, we have
\begin{equation}
\begin{split}
cf(c^{\dagger})|{\Psi}_0{\rangle}=(\frac{{\partial}f(c^{\dagger})}{{\partial}c^{\dagger}})|{\Psi}_0{\rangle},
\end{split}
\label{eq58}
\end{equation}
where $c$ and $c^{\dagger}$ are fermion operators with ${\lbrace}c,c^{\dagger}{\rbrace}=1$, the function $f$ is analytic, and $|{\Psi}_0{\rangle}$ is the vacuum state, i.e., $c|{\Psi}_0{\rangle}=0$. Eq.~\ref{eq58} is an example of a derivative appearing in an operator acting on a term $c^{\dagger}$ which is not required to be a continuous numerical function.


We now interpret the action of the $N$-th Weyl algebra $A_N$ on the space $\mathcal{V}{\otimes}E$ as it appears in basis form in Eq.~\ref{eq34}, obtaining the matrix of the position operator $\mathfrak{r}_{m,n}(k_p,k_q)$ with respect to the Bloch basis. We have
\begin{equation}
\begin{split}
\mathfrak{r}_{m,n}(k_p,k_q)={\delta}_{k_p,k_q}{\cdot}{\sum}_{l=1}^{\mathcal{N}}(a_l^{(m)}(k_p))^*i{\partial}_{k_q}a_l^{(n)}(k_q) \\
+\bigg[{\sum}_{l=1}^{\mathcal{N}}(a_l^{(m)}(k_p))^*a_l^{(n)}(k_q)\bigg]{\cdot}\bigg[\frac{1}{N}{\sum}_{j=1}^Ne^{i(k_p-k_q)R_j}\bigg]\,.
\end{split}
\label{eq59}
\end{equation}
The second term can be expressed as
\begin{equation}
\begin{split}
&\bigg[{\sum}_l^{\mathcal{N}}(a_l^{(m)}(k_p))^*a_l^{(n)}(k_q)\bigg]{\cdot}\bigg[\frac{1}{N}{\sum}_j^Ne^{i(k_p-k_q)R_j}\bigg] \\
&={\delta}_{k_p,k_q}{\delta}_{m,n}\bar{R}\\
&
\rule{10mm}{0mm}
+(1-{\delta}_{k_p,k_q})K_{m,n}(k_p,k_q)\frac{1}{N}{\sum}_j^Ne^{i(k_p-k_q)R_j}R_j
\end{split}
\label{eq60}
\end{equation}
with $\bar{R}$ as the average position $
\dfrac{1}{N}{\sum}_{j=1}^NR_j$ of the $N$-site chain. This constant, independent of $k$, is the mass center of the crystal. The term $K_{m,n}(k_p,k_q)$ depends on the particular forms of $a_l^{(n)}(k_q)$, and we shall shortly evaluate $K_{m,n}(k_p,k_q)$ in a concrete two-band model. In general, $K_{m,n}(k_p,k_q)$ cannot be reduced to a $\delta$-function in terms of  either $k_p$ and $k_q$, or $m$ and $n$. Recall that $N$ must be finite to have the $N$-th Weyl algebra defined. Thus, CRMs are \textit{always} based on finite $N$. On the other hand, if we consider $N{\rightarrow}{\infty}$, will the CRMs approach the DRM? Or, if we let $N$ be finite, will the DRM become a CRM? The answer is no! The CRMs are fundamentally distinct from the DRM, and one cannot relate them. A more detailed comparison appears later.

Evidently, the matrix $\mathfrak{r}_{m,n}(k_p,k_q)$ of Eq.~\ref{eq59} is well-defined, with both diagonal and off-diagonal terms converging. Note ${\delta}_{k_p,k_q}=1$ if $k_p=k_q$. We call $\mathfrak{r}_{m,n}(k_p,k_q)$ the \emph{convergent $r$-matrix} (CRM). It is an $({\mathcal{N}{\cdot}N})$-dimensional square matrix. To emphasize this point, we can write the matrix as $\mathfrak{r}_{{\lbrace}m,k_p{\rbrace}, {\lbrace}n,k_q{\rbrace}}$. In this context, we call the matrix of Eq.~\ref{eq51} a \emph{divergent $r$-matrix} (DRM). Although DRMs frequently appear in the literature, their dimensions have not explicitly been stated \cite{3,25}. 

\subsection{4B. Geometry defined on the quotient space $\mathcal{V}$.}

The first term in Eq.~\ref{eq59} is the Berry connection, which naturally emerges when $k_p=k_q$ and $m=n$. Comparing with Eq.~\ref{eq51}, we obtain the correspondence
\begin{equation}
\begin{split}
{\delta}_{k_p,k_q}{\cdot}&{\sum}_{l=1}^{\mathcal{N}}(a_l^{(m)}(k_p))^*i{\partial}_{k_q}a_l^{(n)}(k_q) \\&{\mapsto}
\frac{(2\pi)^d}{V_{\text{cell}}}{\delta}(k-k'){\int}_{V_{\text{cell}}}u_{m,k}^*(r)i{\partial}_ku_{n,k}(r){\cdot}dr\,.
\end{split}
\label{eq61}
\end{equation}
This is a map
\begin{equation}
\begin{split}
|u_{n,k}{\rangle}{\mapsto}|A_{n,k}{\rangle}=\begin{pmatrix} a_1^{(n)}(k) \\ {\vdots} \\ a_{\mathcal{N}}^{(n)}(k) \end{pmatrix}_{\mathcal{N}}{\in}\mathcal{V}
\end{split}
\label{eq62}
\end{equation}
Eq.~\ref{eq62} indicates that the $|u_{n,k}{\rangle}$ map injectively to vectors $|A_{n,k}{\rangle}$ in the quotient space $\mathcal{V}$. In precise language, the inner product space spanned by the $|u_{n,k}{\rangle}$ is isomorphic to $\mathcal{V}$. In other words, there exists a map between the two spaces which preserves the inner product. Note that $|u_{n,k}{\rangle}$ is an $\mathcal{N}$-dimensional vector, and $a_j^{(n)}(k)$ is its $j$-th component.

\textbf{Remark 4.4} Eq.~\ref{eq62} builds a vector space associated with the functions $u_{n,k}(r)$. Rigorously speaking, the $u_{n,k}(r)$ do not yet provide a function space. For that purpose, we should have $\delta$-functions as extra structure (just like a norm or Lie brackets on vector spaces). The answer to the question “what is the dimension of the space containing the $u_{n,k}(r)$” is indeterminate \cite{Note2}. Thus, we \textit{cannot} set $u_{n,k}(r)=|u_{n,k}{\rangle}$. 
Consider the false implication 
\begin{equation}
\begin{split}
{\psi}_{n,k}(r)=e^{ikr}u_{n,k}(r)\ {\Rightarrow}\|{\psi}_{n,k}{\rangle}=e^{ikr}|u_{n,k}{\rangle}
\end{split}
\label{eq63}
\end{equation}
While the hypothesis of ${\Rightarrow}$ is correct, the conclusion is not. Unfortunately, $u_{n,k}(r)$ and $|u_{n,k}{\rangle}$ are often used indisciminately \cite{4,5,25}. It is tempting to interpret ${\psi}_{n,k}(r)$ and $u_{n,k}(r)$ as basis elements, and then convert these basis elements from functions into bra/ket forms. But that is incorrect. 

Equating $u_{n,k}(r)$ and $|u_{n,k}{\rangle}$ may lead to vagueness and misconceptions. For example, if we (mistakenly) infer $|{\psi}_{n,k}{\rangle}=e^{ikr}|u_{n,k}{\rangle}$ from the Bloch Theorem, we might be led to thinking that there would be a linear relation between $|{\psi}_{n,k}{\rangle}$ and $|u_{n,k}{\rangle}$, and further that every Bloch basis element $|{\psi}_{n,k}{\rangle}$ corresponds linearly to a basis element $|u_{n,k}{\rangle}$, such that their two spans have the same dimension. Besides, we would have difficulty with the boundary conditions. Bloch waves should be continuous over the B.Z., i.e., $|{\psi}_{n,k=0}{\rangle}=|{\psi}_{n,k=2{\pi}}{\rangle}$. Given the (mistaken) assumption that $|{\psi}_{n,k}{\rangle}=e^{ikr}|u_{n,k}{\rangle}$, there is no way to make $|u_{n,k}{\rangle}$ continuous, since $|u_{n,k=0}{\rangle}{\neq}|u_{n,k=2\pi}{\rangle}$ for $k{\neq}0$.

In fact, there is no phase correlation between $|{\psi}_{n,k}{\rangle}$ and $|u_{n,k}{\rangle}$, because when $|u_{n,k}{\rangle}$ is first introduced, the definition (Eq.~\ref{eq62}) merely considers inner products, allowing the freedom to adjust phases. Perhaps, it might be more accurate to adopt a different vector notation (e.g., $|A_{n,k}{\rangle}$) for $|u_{n,k}{\rangle}$, to avoid confusion between $u_{n,k}(r)$ and $|u_{n,k}{\rangle}$. (Note that $u_{n,k}(r)$ is a function of $r$, while $|u_{n,k}{\rangle}$ is not labeled with $r$). However, given the wide use of $|u_{n,k}{\rangle}$ in literature, we have adopted this notation. 

It is often stated that the Berry connection is defined on the periodic part of the wave function $u_{n,k}(r)$, instead of on the Bloch functions ${\psi}_{n,k}(r)$ \cite{25}. 
Now we have the accurate statement that the Berry connection is defined on the $|u_{n,k}{\rangle}$, 
and the map Eq.~\ref{eq62} establishes the identity of the $|u_{n,k}{\rangle}$-space as the quotient space $\mathcal{V}$ of $\mathcal{H}_B$.

We may ask why the Berry connection is defined in terms of the $|u_{n,k}{\rangle}$, rather than in terms of the $|{\psi}_{n,k}{\rangle}$. This is commonly explained by showing that $|{\psi}_{n,k}{\rangle}$ does not work. Consider a discrete formalism for the Berry phase $\vartheta$ | the system goes through a series of discrete states $|k_1{\rangle}{\rightarrow}|k_2{\rangle}{\rightarrow}{\ldots}$ \cite{25}. Then
\begin{equation}
\begin{split}
{\vartheta}=-{\Im}\log({\langle}k_1|k_2{\rangle}{\langle}k_2|k_3{\rangle}{\cdots}{\langle}k_N|k_1{\rangle}).
\end{split}
\label{eq64}
\end{equation}
If we plug in Bloch waves $|k{\rangle}=|{\psi}_{n,k}{\rangle}$, orthogonality will force ${\langle}{k_j}|{k_{j+1}}{\rangle}{=}0$. Thus, $|k{\rangle}=|{\psi}_{n,k}{\rangle}$ make $\vartheta$ trivially zero for arbitrary band structures. But this only precludes $|{\psi}_{n,k}{\rangle}$ from appearing in $\vartheta$, and does not show that $|k{\rangle}=|u_{n,k}{\rangle}$ must be the case.

Consideration of the DRM does at least show that the Berry connection $A_{m,n}(k)$ is contained in the matrix \cite{3,4}. However, there are at least three shortcomings. First, the DRM itself is ill-defined. In particular, the divergence on the diagonals directly influences displacement and transport. Secondly, the substitution $\hat{r}{\rightarrow}i{\partial}_k$ is not well justified. Naively, one may argue that $i{\partial}_k$ is the fundamental form of $\hat{r}$, as quantum mechanics suggests. However, the quantum oracle merely suggests the commutation $[\hat{r},k]=i$, and $\hat{r}=i{\partial}_k$ is not the unique solution. The operator $\hat{r}$ might also take the form of Eq.~\ref{eq57}, for instance. In fact, both Eq.~\ref{eq56} and Eq.~\ref{eq57} can serve as appropriate forms for the operator $\hat{r}$. Thirdly, the Berry connection matrix $A_{m,n}(k)$ belongs to the space spanned by $u_{n,k}(r)$, but the dimension of this space is left uncertain. One is led to (mistakenly) consider the continuous parameter $r$ as the index for the basis elements, under the vague impression that “the space is infinite-dimensional,” which hinders a comparison with $\mathcal{H}_B$. 

With the CRM and $N$-th Weyl algebra, we obtain all Berry connection signatures contained in $r$-matrix as indicated by DRM and 1-st Weyl algebra; moreover, the divergence disappears, and the matrix is well-defined. The constant term $\bar{R}$ is precisely cancelled by subtraction, establishing a rigorous link between diagonal terms and transport. In the absence of a well-defined renormalization protocol, one cannot just drop or cancel two diverging terms in the DRM. We stress that the DRM cannot connect to the CRMs by a limiting process. Secondly, compared with the DRM, construction of the CRM has taken two steps: 
\begin{itemize}
\item[(1)] Express $\hat{r}$ with the $N$-th Weyl algebra $A_N$ in $\mathcal{H}_B$; 
\item[(2)] Reduce it to the 1-st Weyl algebra $A_1$ established on the lower-dimensional quotient space $\mathcal{V}$,
\end{itemize}
as summarized in Table~\ref{tab2}. It is risky to directly replace $\hat{r}$ with $i{\partial}_k$, without referring to the hosting space. Thirdly, the dimension of the space spanned by the $|u_{n,k}{\rangle}$ is specified, identifying it as the quotient space $\mathcal{V}$ of $\mathcal{H}_B$. However, this issue has been concealed by the vague idea that both spaces are infinite-dimensional, which also hides the method to make the Berry phase $\vartheta$ non-zero. Now, we understand the procedure of obtaining $\vartheta{\neq}0$ by “folding” $\mathcal{H}_B$ into a product space, and defining the Berry connection (and other geometric objects) on the quotient space $\mathcal{V}$, instead of on $\mathcal{H}_B$. The dimension of $\mathcal{V}$ is finite (and the norm can be defined). Usually, it is equal to the number $\mathcal{N}$ of bands, which might either be given when $\mathcal{H}_B$ is first introduced, or obtained by a truncation.
\begin{table}
\caption{\label{tab:table2} It is correct to assert that the CRM could eventually be brought down to the 1-st Weyl algebra $A_1$, but it is mistaken to have $A_1$  act directly on $\mathcal{H}_B$, as this will lead to divergence. Instead, it is $A_N$ which acts on $\mathcal{H}_B$ or its isomorphic copy $\mathcal{V}{\otimes}E$. One may subsequently reduce this action to one of $A_1$ on $\mathcal{V}$. \label{tab2}}
\begin{ruledtabular}
\begin{tabular}{c c c c}
\multicolumn{1}{c}{} & \multicolumn{1}{c}{DRM} & \multicolumn{2}{c}{CRM} \\
\hline           
Vector space & $\mathcal{H}_B$ & Step 1:~$\mathcal{V}{\otimes}E$ & Step 2:~$\mathcal{V}$ \\ 
Weyl algebra & $A_1$ & $A_N$ & $A_1$
\end{tabular}
\end{ruledtabular}
\end{table}

\subsection{4C. How has the convergence been achieved?} 

Let us revisit the divergent expressions
\begin{equation}
\begin{split}
{\int}_{-\infty}^{+\infty}{\psi}_{n,k}^*(r)r{\psi}_{n,k}(r){\cdot}dr~\text{or}~{\int}_{-\infty}^{+\infty}{\psi}_{n,k}^*(p)i{\partial}_p{\psi}_{n,k}(p){\cdot}dp.
\end{split}
\label{eq65}
\end{equation}
Here, the divergence arises from the use of either of the unbounded coordinates $r$ or $p$. Since they are conjugate variables, use of one is no better than the other. Recall that $\hat{r}{\rightarrow}i{\partial}_p$ is different from $\hat{r}{\rightarrow}i{\partial}_k$, because $k$ is the quantum number for Bloch vectors, while $p$ is the real momentum. Contrasting the conjugate pairs, $\hat{r}$ is conjugate with $p$, while it is $R_j$ which is conjugate with $k$ according to Eq.~\ref{eq25}.

Convergence has now been achieved thanks to two modifications:
\begin{itemize}
\item[(i)] The choice of the operator $\hat{r}=\frac{1}{N}{\sum}_{m=1}^Ni\frac{\partial}{{\partial}k_m}$ in the Weyl algebra $A_N$;
\item[(ii)] Application of the isomorphism 
$$
\Pi\colon\mathcal{H}_B\to\mathcal{V}{\otimes}E;{\psi}_{n,k}(r){\rightarrow}a_i^{(n)}(k){\cdot}e^{ikR_j}\,,
$$ 
i.e., $\mathcal{H}_B$ is replaced by tensor product space $\mathcal{V}{\otimes}E$. 
\end{itemize}
Rethinking both the operator $\hat{r}$ and the wave functions ${\psi}_{n,k}(r)$ reflects that the Weyl algebra $A_N$, as a ring, involves not only the differential operators, but also the functions $f_m$ on which they act, as seen in the definition of Eq.~\ref{eq11}. Physically, the functions correspond to the wave functions. Intuitively speaking, after making the replacement (i), one still needs the replacement (ii) to specify the arguments on which these differential operators act, in order to complete the representation of the Weyl algebra.

From a different perspective, we are specifying a new pair $r{\leftrightarrow}k$ of conjugate variables. The position variable $r$ is paired with the crystal momentum $k$, rather than with real momentum $p$. Although (i) and (ii) do not have the explicit form of a declaration of conjugate variables, the declaration is implicit within them. It would not be enough merely to state that $i{\partial}_k$ (or $\frac{1}{N}{\sum}_m^Ni\frac{\partial}{{\partial}k_m}$) is the expression of $\hat{r}$, since that would only invoke modification (i), not (ii).

It is incomplete and misleading to assert that $\hat{r}$ is ``equal to" $i{\partial}_k$. For example, it is easily seen that
\begin{equation}
\begin{split}
{\int}{\psi}_{m,k'}^*(r)r{\psi}_{m,k}(r){\cdot}dr{\neq}{\int}{\psi}_{m,k'}^*(r)i{\partial}_k{\psi}_{m,k}(r){\cdot}dr\,.
\end{split}
\label{eq66}
\end{equation}
One may encounter attempts in the literature to use  $\hat{r}=i{\partial}_k$ to suggest a form like ${\langle}u_{m,k}|i{\partial}_ku_{n,k}{\rangle}$ for the $r$-matrix. However, in other situations, if we replace $i{\partial}_k$ by $r$, we may obtain the contradiction ${\langle}u_{m,k}|i{\partial}_ku_{n,k}{\rangle}={\langle}u_{m,k}|r|u_{n,k}{\rangle}={\delta}_{m,n}r$ suggesting that $|u_{n,k}{\rangle}$ would be an eigenstate of $\hat{r}$.

\subsection{4D. Can the DRM be a limit of CRMs?}
We now show that letting $N{\rightarrow}{\infty}$ does not produce the DRM as a limit of the CRMs. Recall that $k_p$ and $R_j$ are conjugate variables. Thus, increasing the population $N$ of $R_j$-values corresponds to making the $k_p$-values denser, bringing us to the limit where $k$ is continuous. The first term of Eq.~\ref{eq59} approaches $${\delta}(k-k'){\cdot}{\sum}_{l=1}^{\mathcal{N}}(a_l^{(m)}(k))^*i{\partial}_ka_l^{(n)}(k)\,,$$ coinciding with the Berry connection ${\delta}(k-k'){\cdot}A_{m,n}(k)$ in Eq.~\ref{eq51}. However, the second terms cannot match. The CRM does not invoke ${\delta}_{m,n}$ when $k_p{\neq}k_q$, in contradiction to the separation of the factor ${\delta}_{m,n}$ in Eq.~\ref{eq51}. 

This can be seen on a concrete example with $\mathcal{N}=2$ | a two-band model. In this case, the quotient space $\mathcal{V}$ is 2-dimensional, spanned by two basis elements $|u_{1,k_p}{\rangle}$ and $|u_{2,k_p}{\rangle}$ (following the notation of Eq.~\ref{eq33}). We take
\begin{equation}
\begin{split}
&\begin{pmatrix} a_1^{(1)}(k_p) \\ a_2^{(1)}(k_p) \end{pmatrix}=\begin{pmatrix} \cos(\frac{\theta(k_p)}{2}) \\ \sin(\frac{\theta(k_p)}{2}){\cdot}e^{i{\phi}(k_p)} \end{pmatrix}\ \mbox{ and} \\
&\begin{pmatrix} a_1^{(2)}(k_p) \\ a_2^{(2)}(k_p) \end{pmatrix}=\begin{pmatrix} -\sin(\frac{\theta(k_p)}{2}){\cdot}e^{-i{\phi}(k_p)} \\ \sin(\frac{\theta(k_p)}{2}) \end{pmatrix}\,,
\end{split}
\label{eq67}
\end{equation}
where $\theta$ and $\phi$ are functions of $k$. Given Eq.~\ref{eq67}, we find
\begin{equation}
\begin{split}
K_{1,1}=K_{2,2}^*&=\cos(\frac{{\theta}(k_p)}{2})\cos(\frac{{\theta}(k_q)}{2}) \\
&+\sin(\frac{{\theta}(k_p)}{2})\sin(\frac{{\theta}(k_q)}{2})e^{-i({\phi}(k_p)-{\phi}(k_q))}\,. \\
\end{split}
\label{eq68}
\end{equation}
The off-diagonals are 
\begin{equation}
\begin{split}
K_{1,2}=-K_{2,1}^*&=-\cos\Big(\frac{{\theta}(k_p)}{2}\Big)\sin\Big(\frac{{\theta}(k_q)}{2}\Big)e^{-i{\phi}(k_q)} \\
&+\sin\Big(\frac{{\theta}(k_p)}{2}\Big)\cos\Big(\frac{{\theta}(k_q)}{2}\Big)e^{-i{\phi}(k_p)}.
\end{split}
\label{eq69}
\end{equation}
In general, $K_{m,n}$ is non-vanishing. Eq.~\ref{eq69} shows that the $r$-matrix does not vanish when $k_p{\neq}k_q$. The entries non-diagonal with $k$ identify a clear distinction from the DRM. 

In practice, the Hamiltonian $H(k)$ mainly focuses on the diagonal terms $k_p=k_q=k$. In terms of observables, we are interested in a subset of the elements of the $r$-matrix, but this does not mean that the $r$-matrix is block diagonal in $k$. 
We have $r_{1\le m,n\le 2}
(k)=$
\begin{equation}
\begin{split}
&
{\small
\begin{pmatrix} -\sin^2(\theta){\partial}_k\phi+\bar{R} & -\frac{i}{2}e^{-i{\phi}}{\partial}_k\theta-\frac{1}{2}\sin(\theta)e^{-i\phi}{\partial}_k\phi \\ \frac{i}{2}e^{i{\phi}}{\partial}_k\theta-\frac{1}{2}\sin(\theta)e^{i\phi}{\partial}_k\phi & \sin^2(\theta){\partial}_k\phi+\bar{R} \end{pmatrix}\,.
}
\end{split}
\label{eq70}
\end{equation}
Recall that  $\theta$ and $\phi$ are functions of $k$. In a particular model of graphene, for instance, they take the specific forms
\begin{equation}
\begin{split}
\theta(\textbf{k})&=\frac{\pi}{2}, \\
\phi(\textbf{k})&=-\arg(e^{i\textbf{k}{\cdot}{\delta}_1}+e^{i\textbf{k}{\cdot}{\delta}_2}+e^{i\textbf{k}{\cdot}{\delta}_3}) \\
&=-\arg(e^{ik_xa}+e^{i(-\frac{1}{2}k_x+\frac{\sqrt{3}}{2}k_y)a}+e^{i(-\frac{1}{2}k_x-\frac{\sqrt{3}}{2}k_y)a})
\end{split}
\label{eq71}
\end{equation}
where the ${\delta}_i$ are the position vectors of the three nearest neighbor (NN) carbon atoms, $\arg$ denotes argument of a complex number, and $a$ is the carbon bond length \cite{44}. The two components physically represent the two bands due to the mutual independence of the $A/B$ atoms in a primitive cell of graphene. How can we interpret that $\theta(\textbf{k}){\equiv}\pi/2$ is independent of $\textbf{k}$? The conduction band and valence band are formed with $\pi$ bonding with equal weights from the $A/B$ orbitals, which requires $\theta(\textbf{k}){\equiv}\pi/2$ to make the magnitudes of the two components equal.
\vskip 0.5mm

We may summarize the logic of our process as follows: 
\begin{itemize}
\item[(1)] Employ the Weyl algebra to define $\hat{r}$ and $p$; 
\item[(2)] Based on these physical operators, extract bases to build a vector space to serve as Hilbert space; 
\item[(3)] The first (unsuccessful) attempt (left arrow below) built the space with $\hat{r}$- (or $p$-) eigenstates as bases; however, the eigenvalues of $\hat{r}$ (or $p$) cover all of $\mathbb{R}$, whose cardinality is uncountably infinite, leading to a diverging norm, unsuitable for a Hilbert space; 
\item[(4)] The second (successful) attempt (right arrow below) recognizes the bases differently.
\end{itemize}
\begin{equation}
\begin{split}
{\psi}_i\xleftarrow{1st: i\leftrightarrow r}{\psi}(r)\xrightarrow{2nd:~\text{label}~m}{\psi}_m(r).
\end{split}
\label{eq72}
\end{equation}
We add index $m$ to label the bases, and $r$ serves as a “parameter”; different from the first attempt (the left arrow in Eq.~\ref{eq72}), for which $r$ was taken as the label of bases and was discretized into finite intervals. In the second attempt with CRM, the dimension of vector space depends on label $m$ instead of $r$. Therefore, CRM represents a modified means (compared with DRM) of assigning a vector space to Weyl algebra, such that the constructed vector space is equipped with a converged norm. This addresses the question raised by Eq.~\ref{eq18} in Sec. 3, representing a different route for discrete crossover to continuous situations. DRM also arises from Weyl algebra; however, the resultant space has diverging norm, unsuitable for Hilbert space.

From a math point of view, Weyl algebra is defined as a ring (an algebra equipped with addition and multiplication). A vector space, in terms of ring’s definition, is not an intrinsic notion; thus, it is an art to associate a vector space to the ring. If this is improperly done, one ends up with a space of diverging norm (e.g., a space of infinite dimensions), which hinders evaluating ${\langle}\hat{r}{\rangle}$, a fatal issue for transport. Our scheme is that the dimension of vector space should not be characterized by $r$ (eigenvalues of $\hat{r}$) nor its conjugated variable $p$, but by $\mathcal{N}{\cdot}N$-dimensional vector space, on which $N$-th Weyl algebra acts on, resulting $\mathcal{N}$-dimensional $r$-matrix. (Mind $N{\neq}\mathcal{N}$.) Since $\mathcal{N}$ is arbitrary, this approach represents a generic approach of projecting $\hat{r}$ to arbitrary finite dimensions. In previous deriving of $r$-matrix, the implicit belief that the notion continuity of $k$ is indispensable for partial derivative $\partial_k$ has prevented the extension to $N$-th Weyl algebra $A_N$.

\textbf{$r$-matrix $\mathfrak{r}_{m,n}(k,k')$, reduced $r$-matrix $r_{m,n}(k)$ and Berry connection matrix $A_{m,n}(k)$}. CRM represents a way of mapping $\hat{r}$ to a finite-dimensional Hermitian matrix (but the matrix does not form a representation of Weyl algebra). Next, we sharpen terminology “matrix”.

The $r$-matrix is originally introduced on Bloch bases; thus, its dimension is equal to the dimension of Bloch waves: ${\mathcal{N}{\times}N}$.
\begin{equation}
\begin{split}
\mathfrak{r}_{m,n}(k,k'):\mathcal{H}_B{\rightarrow}\mathcal{H}_B.
\end{split}
\label{eq73}
\end{equation}
We further introduce “reduced $r$-matrix” $r_{m,n}^{(\mathcal{N})}(k)$ (or $r_{m,n}(k)$ for short), i.e., project ${\mathcal{N}{\times}N}$ dimensional $\mathfrak{r}_{m,n}(k,k')$ to quotient space $\mathcal{V}$ of dimension $\mathcal{N}$ by setting $k=k'$:
\begin{equation}
\begin{split}
r_{m,n}^{(\mathcal{N})}(k)=A_{m,n}(k)+{\delta}_{m,n}\bar{R},
\end{split}
\label{eq74}
\end{equation}
where $A_{m,n}(k)$ is Berry connection matrix of $\mathcal{N}$ dimension. Noteworthy, $\bar{R}$ in the diagonal term of $\mathfrak{r}_{m,n}(k,k')$ (also the reduced $r_{m,n}^{(\mathcal{N})}(k)$ has a clear physical meaning: the mass center of the crystal, which is a $k$-independent constant. That means term $\bar{R}$ for different bands are exactly the same, which will be cancelled (in evaluating displacement, one will take the difference of two diagonal terms and $\bar{R}$ will be exactly cancelled). This converts a problem defined in $\mathcal{H}_B$ to its quotient space $\mathcal{V}$. 

Accurately speaking, $r_{m,n}^{(\mathcal{N})}(k)$ (also $A_{m,n}(k)$) is not a single matrix, but “a continuous series of matrices” for variable $k$. Formally, it is a map, 
\begin{equation}
\begin{split}
r_{m,n}^{(\mathcal{N})}(k):K{\rightarrow}\mathcal{A},
\end{split}
\label{eq75}
\end{equation}
where $\mathcal{A}$ represents Hermitian matrices on quotient space $\mathcal{V}$ (not $\mathcal{H}_B$). Berry connection matrix $A_{m,n}(k)$ is also such a map. Eq.~ \ref{eq70} gives an example of $r_{m,n}^{(\mathcal{N})}(k)$ with $\mathcal{N}=2$. It is not that CRM reduces an operator of infinite dimension to one of 2-dimension, which immediately raises the concern how the information can be encoded into such a small matrix? Instead, it is a single matrix of higher dimensions to be mapped to a series of lower-dimensional matrices. Formally speaking, the higher-dimensional matrix is mapped to a map whose codomain elements are 2-dimensional matrices. 

In terms of bundle theory \cite{37,38}, one may interpret that the reduced $r$-matrix $r_{m,n}^{(\mathcal{N})}(k)$ transforms the problem originally defined in high-dimension vector space $\mathcal{H}_B$ to a bundle whose fiber space is a lower dimensional space $\mathcal{V}$. It is incorrect to regard $r_{m,n}^{(\mathcal{N})}(k)$ as an $\mathcal{N}$-dimensional matrix, neither as the lower dimensional counterpart of DRM.

CRM does not form representation of Weyl algebra (i.e., $[\hat{r},k]{\neq}i$). We shall point out CRM does not satisfy commutation, whether $N$ is finite or $N{\rightarrow}{\infty}$. (In fact, DRM does not satisfy neither, which is concealed by its divergence.) The matrix for $k$ can be found in a similar fashion as Eq.~\ref{eq70}.
\begin{equation}
\begin{split}
{\langle}{\psi}_{m,k_p}|{\sum}_n^Nk_n|{\psi}_{n,k_q}{\rangle}={\delta}_{m,n}{\delta}_{k_p,k_q}k_p.
\end{split}
\label{eq76}
\end{equation}
Project it to quotient space, we have
\begin{equation}
\begin{split}
k_{m,n}^{(\mathcal{N}=2)}(k)=\begin{pmatrix} k & 0 \\ 0 & k \end{pmatrix}
\end{split}
\label{eq77}
\end{equation}
The commutation yields 
\begin{equation}
\begin{split}
r_{m,j}^{(\mathcal{N}=2)}k_{j,n}^{(\mathcal{N}=2)}-k_{m,j}^{(\mathcal{N}=2)}r_{j,n}^{(\mathcal{N}=2)}=0.
\end{split}
\label{eq78}
\end{equation}
The commutation does not yield the expected $i$. Thus, $r$-matrix together with $k$ matrix does not form the generator of Weyl algebra. This is different from spin’s matrices, which are meant to preserve the Lie-algebra (Lie brackets). Therefore, it involves new principles to define $r$-matrix, for which we give more discussions in Sec. 7. The way of defining matrices is hinged to the way of extracting observables. In addition, when we work with Berry connection in the quotient space, one vector is not one-to-one corresponding to a physical state if a vector in $\mathcal{H}_B$ represents a physical state. 

\textbf{Remarks 4.5} We shall stress a few points about CRM and DRM:
\vskip 0.5mm
(1) 1D CRM is still convergent, different from DRM; thus, DRM is \textit{not} the 1D special case of CRM.
\vskip 0.5mm
(2) The continuous limit of CRM will \textit{not} approach to DRM.
\vskip 0.5mm
(3) CRM is not an inferior or approximate form of $\hat{r}$, it is not achieved by representing $\hat{r}$ in a subspace of $\mathcal{H}$.
\vskip 0.5mm
(4) The matrix will \textit{not} reproduce the commutator; $[\hat{r},k]=i$ cannot serve as the principle in defining the form of $r$-matrix.
\vskip 0.5mm

\section{5. Properties of the $\hat{r}$ operator and $r$-matrix}

In Sec. 5 and 6, the language of maps has to be used to illustrate concepts, especially the $\Pi$ map built earlier, and the paper is organized with a few progressive definitions. This may cause some discomfort, but we try to keep it at the minimum level. Besides, in view of the mistakes made by authors themselves, it seems necessary to underscore certain algebraic rules. Although these parts of discussions might appear not quite ``physical", they are needed to make the raised concepts and algebraic derivation unambiguous. 

One is familiar with the bases to represent a spin (Lie algebra) and such notions as basis transformations. It is natural to wonder about the counterparts for $\hat{r}$ (Weyl algebra). Because DRM contains divergence, these issues were left open, since one cannot perform any calculation in the presence of ``$\infty$". With CRM, we find the notion “ribbon band”, on which $r$-matrix is defined, is in analog with “bases” on which spin is defined. Physically, the ribbon band is related to the description of electronic states in crystals. 

On top of ribbons, $\hat{r}$ will be handled like a matrix when it interplays with ribbons or other matrix operators without reference to its origin. Procedures associated with ${\partial}_k$, such as effectual range, one-sided acting on the right, are incarnated in matrix multiplication. Intuitively speaking, we disguise a differential operator like a matrix as much as possible; however, a differential operator may never really become a matrix due to the distinctions in their bottom algebras. Therefore, the cost one must pay is the differential operator’s matrix follows distinctive rules for transformation, exactly where the gauge transformation makes entrance as articulated in the next section. In this context, terms “differential operator” and “the matrix of the differential operator” should be discriminated.

Next, we shall follow the logic line of introducing bases to introduce ribbon bands (Def. 1) and associated concepts like inner product of ribbons (Def. 2), orthogonal ribbons (Def. 3), components of ribbons (Def. 4), ribbon transformation (Def. 5), etc.
\vskip 0.5mm

\textbf{Definition 1}: A ribbon band (or “ribbon” for short) over smooth manifold $K$ to vector space $V$ is a map 
\begin{equation}
\begin{split}
\mathscr{R}:K{\rightarrow}V~\text{or}~\mathscr{R}:k{\mapsto}v,~k{\in}K,v{\in}V.
\end{split}
\label{eq79}
\end{equation}
In band context, $K$ is the B.Z., topologically, an $n$-dimensional torus $T^n=S^1{\times}{\ldots}S^1$; $V$ is a vector space, e.g., quotient space $\mathcal{V}$ of $\mathcal{H}_B$. The rank of $\mathscr{r}$ is defined as the dimension of $V$. If continuity is globally satisfied for map $\mathscr{R}$, it is called continuous ribbon band; otherwise, we say the ribbon band is discontinuous at point $k_0$. For example, if degeneracy exists, eigenstate $|u_{n,k}{\rangle}$ of $H(k)$ specifies a ribbon, which is discontinuous at the degenerate $k$.

A ribbon band could be induced by $\Pi$ maps (Eq.~\ref{eq34}). For example
\begin{equation}
\begin{split}
\mathscr{R}:k{\mapsto}{\Pi}^{(n)}(k):={\Pi}({\psi}_{n,k}),
\end{split}
\label{eq80}
\end{equation}
and
\begin{equation}
\begin{split}
\mathscr{R}:k{\mapsto}{\Pi}_1^{(n)}(k):={\Pi}_1({\psi}_{n,k}),
\end{split}
\label{eq81}
\end{equation}
Eq.~\ref{eq80} is a ribbon $K{\rightarrow}\mathcal{V}{\times}E$ of rank $\mathcal{N}{\times}N$; Eq.~\ref{eq81} is another ribbon $K{\rightarrow}\mathcal{V}$ of rank $\mathcal{N}$. These ribbons defined for either the product space or the quotient space. From Eq.~\ref{eq80},\ref{eq81}, we notice that given $\Pi$ or ${\Pi}_1$ maps, $\mathcal{N}$ branches of ribbon bands will be induced (for there are $\mathcal{N}$-fold eigenstates). 
\vskip 0.5mm

\textbf{Definition 2}: Inner product between ribbons is defined as a linear binary map
\begin{equation}
\begin{split}
{\langle}{\cdot},{\cdot}{\rangle}:\mathscr{R}{\times}\mathscr{R}~{\mapsto}f,
\end{split}
\label{eq82}
\end{equation}
where $f$ is a map
\begin{equation}
\begin{split}
f:K{\mapsto}\mathbb{C}.
\end{split}
\label{eq83}
\end{equation}
For two arbitrary ribbons $|{\varphi}(k){\rangle}$ and $|{\psi}(k){\rangle}$, commonly defined over $K$ to vector space $V$, inner product of two ribbons can be induced by inner product for vectors in $V$
\begin{equation}
\begin{split}
f(k)={\langle}{\phi}(k)|{\psi}(k){\rangle}
\end{split}
\label{eq84}
\end{equation}
\vskip 0.5mm

\textbf{Definition 3}: Orthogonal ribbons are two ribbons whose inner products (Def. 2) are \textit{constantly} zero. Consider a set of ribbons $\mathcal{I}={\lbrace}|m(k){\rbrace}_\mathcal{N}$ with the number of elements equal to the ranks of these ribbons. If ribbons in $\mathcal{I}$ are mutually orthogonal, $\mathcal{I}$ is a set of ribbon bases.
\begin{equation}
\begin{split}
{\langle}m(k)|n(k){\rangle}={\delta}_{m,n}~{\forall}k{\in}K,~|m(k){\rangle},|n(k){\rangle}{\in}\mathcal{I}.
\end{split}
\label{eq85}
\end{equation}
Just like a vector is characterized by the dimension and can be represented by a set of bases of the same dimension; a ribbon can be characterized by its rank and represented by a set of orthogonal ribbons. We may denote a general ribbon over $K$ as, for example, $|{\varphi}(k){\rangle}$ with an extra parameter $k{\in}K$, in analog of a general vector $|{\varphi}{\rangle}$.

We shall use $|m(k){\rangle}$, $|n(k){\rangle}$, etc. to denote different elements in the same set of ribbon bases $\mathcal{I}={\lbrace}|m(k){\rangle}{\rbrace}_{\mathcal{N}}$, i.e., the $m^{th}$, $n^{th}$ elements in $\mathcal{I}$. When different ribbon bases are involved, we will add primes $\mathcal{I}'={\lbrace}|m'(k){\rangle}{\rbrace}_{\mathcal{N}}$. 
\vskip 0.5mm

\textbf{Definition 4}. In analog to arbitrary vectors being expressed in components on a set of orthogonal bases, we define the \textit{components of a ribbon} projected to ribbon bases $\mathcal{I}={\lbrace}|m(k){\rangle}{\rbrace}_{\mathcal{N}}$ as
\begin{equation}
\begin{split}
{\varphi}_m^*(k):={\langle}{\varphi}(k)|m(k){\rangle};~{\varphi}_m(k):={\langle}m(k)|{\varphi}(k){\rangle}.
\end{split}
\label{eq86}
\end{equation}
${\varphi}_m^*(k)$ and ${\varphi}_m(k)$ are with respect to a set of ribbons, instead of a set of bases of $V$. Thus,
\begin{equation}
\begin{split}
{\varphi}_m^*(k)\neq{\langle}{\varphi}(k)|m{\rangle};~{\varphi}_m(k)\neq{\langle}m|{\varphi}(k){\rangle},
\end{split}
\label{eq87}
\end{equation}
where ${\lbrace}|m{\rangle}{\rbrace}_{\mathcal{N}}$ is a set of bases for $V$, which could also be viewed as $k$-independent ribbon bases. ${\lbrace}|m{\rangle}{\rbrace}_{\mathcal{N}}$ is likely to be different from ${\lbrace}|m(k){\rangle}{\rbrace}_{\mathcal{N}}$. Thus the $k$ label should \textit{not} be discarded.
\vskip 0.5mm

\textbf{Definition 5}. In an analog of basis transformation, one may introduce ribbon transformation
\begin{equation}
\begin{split}
T_R:V_{\mathscr{R}}{\rightarrow}V_{\mathscr{R}}~\text{or}~T_R:\mathscr{R}\mapsto\mathscr{R}',
\end{split}
\label{eq88}
\end{equation}
where $V_{\mathscr{R}}$ stands for a ribbon space that consists of ribbon bands $\mathscr{R}$. $T_R$ transforms a ribbon space just like basis transformation transforms a vector space. $T_R$ turns a ribbon band into another $\mathscr{R}\mapsto\mathscr{R}'$, which can be realized by a rotation of vector space $V$ at a local $k$. 
\begin{equation}
\begin{split}
{\varphi}_m(k)\xrightarrow{\overset{T_R}{}}U_{m,n}(k){\varphi}_n(k),
\end{split}
\label{eq89}
\end{equation}
where ${\varphi}_m(k)$ is the component of a ribbon. Thus, ribbon transformation $T_R$ can be written in an equivalent form
\begin{equation}
\begin{split}
T_R:K{\rightarrow}\text{Aut}(V),
\end{split}
\label{eq90}
\end{equation}
where $\text{Aut}(V)$ stands for automphism group. Automorphisms refer to inversible self-maps ($V{\rightarrow}V$) that will preserve the inner product ${\langle}{\varphi}|{\psi}{\rangle}={\langle}{\varphi}|U^{\dagger}U|{\psi}{\rangle}$. In other contexts, $\text{Aut}(V)$ may preserve other structures equipped on $V$ than inner products. This requires $U$ to be a unitary transformation. Then the information of $T_R$ is fully encoded in a unitary matrix $U_{m,n}(k)$ indexed by $k{\in}K$.
\vskip 0.5mm

\textbf{Definition 6}: The matrix of a matrix operator (e.g., spin) defined on ribbon space $\mathscr{R}:K{\rightarrow}V$, spanned by orthogonal ribbons $\mathcal{I}={\lbrace}|m(k){\rangle}{\rbrace}_{\mathcal{N}}$, is a matrix function of $k$ that commits to the following ribbon transformations:
\begin{equation}
\begin{split}
O_{m,n}(k)\xrightarrow{\overset{T_R}{}}O_{m,n}'(k)=U_{m,i}(k)O_{i,j}(k)U_{j,n}^{\dagger}(k).
\end{split}
\label{eq91}
\end{equation}
\vskip 0.5mm

\textbf{Definition 7}: The matrix of a differential operator on ribbon space $\mathscr{R}:K{\rightarrow}V$ spanned by ribbon bases $\mathcal{I}$ is defined as a matrix function of $k$ subject to the following ribbon transformation.
\begin{equation}
\begin{split}
M_{m,n}(k)\xrightarrow{\overset{T_R}{}}M_{m,n}'(k)=U_{m,i}(k)M_{i,j}(k)U_{j,n}^{\dagger}(k) \\
+U_{m,j}(k)i{\partial}_k(U_{j,n}^{\dagger}(k)).
\end{split}
\label{eq92}
\end{equation}
Note that in the last term the effectual range of $\partial_k$ is limited to $U_{j,n}^{\dagger}(k)$, and will not act all the way to the right. For the rules about $\partial_k$, refer to Appx. D. In this work, we adopt a convention: the matrix of matrix operator is denoted with $O_{m,n}$ or $O_{m,n}(k)$; the matrix of differential operator is $M_{m,n}(k)$.

\textbf{Remarks 5.1} Matrix is the denotation of an operator on a specific space. Eq.~\ref{eq91} generalizes such a denotation for a matrix operator: from vector space $V$ to ribbon space $V_{\mathscr{R}}$. Such generalization is equivalent to introducing independent replicas (labelled by $k$) of the operator. Transformation at different $k$ is separate, in principle, not requiring $U_{m,i}(k)$ to be continuous or smooth with $k$. On the other hand, Eq.~\ref{eq92} defines a matrix denotation for differential operator. Transformation of $M_{m,n}(k)$ at different $k$ is not irrelevant but requires neighborhood knowledge of $k$ (due to the term $U_{m,j}(k)i{\partial}_k(U_{j,n}^{\dagger}(k))$), such that the global topology begins to enter.

\textbf{Remarks 5.2} Differential operator is the motivation to introduce the ribbon band $\mathscr{R}$ (such generalization is trivial for matrix operators). Nonetheless, ribbon $\mathscr{R}$ allows the two types of operators to be examined on a common ground. It is not that the ribbon band is solely associated to differential operators, nor “bases” are solely associated with matrix operators. 

\textbf{Remarks 5.3} Regarding linear maps. $\hat{O}(k)$ is a linear map $V{\rightarrow}V$ at a $k$ point because $\hat{O}(k)(|m(k){\rangle}+|n(k){\rangle})=\hat{O}|m(k){\rangle}+\hat{O}(k)|n(k){\rangle}$ and $\hat{O}(k)(c|m(k){\rangle})=c\hat{O}(k)|m(k){\rangle}$. Although $\partial_k$ is often referred to as a linear operation because of $\partial_k(F_1(k)+F_2(k))={\partial}_kF_1(k)+{\partial}_kF_2(k)$ and $\partial_kF(ck)=c\partial_kF(k)$, it is \textit{not} a linear map $V{\rightarrow}V$ at a local $k$. In other words, it is a linear map: $\mathcal{H}{\rightarrow}\mathcal{H}$ ($\mathcal{H}$ is the first space introduced in Sec. 2), but not for vector space $V$. This could be seen from $\partial_k$ acting on vector $v{\in}V$.
\begin{equation}
\begin{split}
{\partial}_k(c(k){\cdot}v)=c(k){\cdot}{\partial}_k(v)+v{\cdot}{\partial}_k(c(k)),
\end{split}
\label{eq93}
\end{equation}
producing an extra term $v{\partial}_k(c)$, where $c$ is a function of $k$. In addition, we have seen the matrix of $\partial_k$ is subject to a different transformation rule Eq.~\ref{eq92},\ref{eq93} \cite{Note3}

Next, we underscore a fortuitous finding during our clarifying the fundamentals about CRM and the ribbon space. The fact that position $\hat{r}$ is Hermitian, in matrix context, indicates $r_{m,n}=r_{n,m}^*$; in the operator context, this is denoted as $\hat{r}=\hat{r}^{\dagger}$ --- the two are usually considered identical. However, $\hat{r}=\hat{r}^{\dagger}$ has two implicit connotations which turn out stronger arguments: (1) $\hat{r}$ is associative, i.e., two-sided action; (2) $\hat{r}$ being free of index indicates its basis-invariance.

Our argument is that position is a Hermitian operator and $r$-matrix is a Hermitian matrix, while this fact cannot be expressed with associative operator in basis-independent forms, because (1) $i{\partial}_k$ is not associative, (2) basis-free denotation should not be taken for granted due to the distinct transformation properties of $r$-matrix. This idea could be be compactly expressed as
\begin{equation}
\begin{split}
r_{m,n}=r_{n,m}^*~{\nLeftrightarrow}~\hat{r}=\hat{r}^{\dagger}.
\end{split}
\label{eq94}
\end{equation}
Noteworthy, $O_{m,n}=O_{n,m}^*$ is a property about a matrix, which involves a particular set of bases, while $\hat{O}=\hat{O}^{\dagger}$ is basis-free designation, which is usually applied to bra/ket. 

Basis-free designation means “it works for arbitrary bases”, therefore, it is implicitly conditioned by invariance under basis (or ribbon) transformation. For example, a ket state $|{\psi}{\rangle}=|m{\rangle}{\langle}m|{\psi}{\rangle}=|m'{\rangle}{\langle}m'|{\psi}{\rangle}=\cdots$ works for arbitrary bases ${\lbrace}|m{\rangle}{\rbrace}$, ${\lbrace}|m'{\rangle}{\rbrace}$, etc. Thus, we erase the subscripts and denote it as $|{\psi}{\rangle}$. The same idea for $\hat{r}$, which has no subscripts associated with particular bases. Note that basis-free designation is not always justified. It is true for matrix operators, while might lead to mistakes for differential operators. 

Matrix operator $\hat{O}(k)$ is an example of ribbon-invariant map. Accordingly, one develops the notion that elements in the domain or co-domain sets are objects whose identities are independent of bases, endowed by the following invariance under ribbon transformation. 
\begin{equation}
\begin{split}
&{\varphi}_m(k)O_{m,n}(k){\psi}_n(k)\xrightarrow{\overset{T_R}{}}{\varphi}_m'(k)O_{m,n}'(k){\psi}_n'(k) \\ 
&=({\varphi}_m(k)U_{m,l}^{\dagger}(k))(U_{l,i}(k)O_{i,j}(k)U_{j,g}^{\dagger}(k))(U_{g,n}(k){\psi}_n(k))
\end{split}
\label{eq95}
\end{equation}
Unitary matrix leads to
\begin{equation}
\begin{split}
U_{m,j}^{\dagger}(k)U_{j,n}(k)={\delta}_{m,n}.
\end{split}
\label{eq96}
\end{equation}
Thus, it is invariant with $T_R$
\begin{equation}
\begin{split}
{\varphi}_m(k)O_{m,n}(k){\psi}_n(k)={\varphi}_m'(k)O_{m,n}'(k){\psi}_n'(k)
\end{split}
\label{eq97}
\end{equation}
In Eq.~\ref{eq95}, $T_R$ is to replace the vector and the operator with their counterparts under the updated ribbon bases, which are given by Def. 5,6. The idea is the components of vector and operators are alterable, but the inner product Eq.~\ref{eq97} is invariant. For this, one can introduce a denotation as below, ignoring the indices associated with specific ribbons
\begin{equation}
\begin{split}
&{\langle}{\varphi}(k)|\hat{O}(k)|{\psi}(k){\rangle}\xrightarrow{\overset{T_R}{}}{\langle}{\varphi}'(k)|\hat{O}'(k)|{\psi}'(k){\rangle} \\
&=({\langle}{\varphi}(k)|U^{\dagger}(k))(U(k)\hat{O}(k)U^{\dagger}(k))(U(k)|{\psi}(k){\rangle}).
\end{split}
\label{eq98}
\end{equation}
In above, the basis-free designation such as $|{\psi}(k){\rangle}$, $\hat{O}(k)$, etc. (belonging to Dirac’s ket/bra symbolism) is not subject to specific ribbons nor attached with subindices $m$, $n$, etc. One can interpret $|{\psi}(k){\rangle}$ is not representing a single vector but a class of equivalent vectors that are linked by the ribbon transformation Eq.~\ref{eq98}. Since the equivalence class cover all possible choices of orthogonal ribbons, the class becomes “ribbon independent”. Then, one may designate it without explicitly referring to the choice of ribbon bases. This gimmick is commonly used in defining coordinate-independent fiber bundle, origin-free space (affine space), etc. \cite{37}

On the other hand, if the invariance fails, as in the case of differential operator below, basis-free designations should \textit{not} be taken for granted. 
\begin{equation}
\begin{split}
&{\varphi}_m(k){\cdot}r_{m,n}^{(\mathcal{N})}(k){\cdot}{\psi}_n(k)\xrightarrow{\overset{T_R}{}}{\varphi}_m'(k){\cdot}r_{m,n}^{(\mathcal{N})'}(k){\cdot}{\psi}_n'(k) \\
&=({\varphi}_m(k)U_{m,l}^{\dagger}(k)){\lbrace}U_{l,i}(k)r_{i,j}^{(\mathcal{N})}(k)U_{j,g}^{\dagger}(k) \\
&+U_{l,j}(k)i{\partial}_k(U_{j,g}^{\dagger}(k)){\rbrace}(U_{g,n}(k){\psi}_n(k)) \\
&={\varphi}_m(k){\cdot}r_{m,n}^{(\mathcal{N})}(k){\cdot}{\psi}_n(k)+{\varphi}_m(k)i{\partial}_k(U_{j,g}^{\dagger}(k))U_{g,n}(k){\psi}_n(k).
\end{split}
\label{eq99}
\end{equation}
In doing ribbon transformation Eq.~\ref{eq99}, we have substituted $O_{m,n}(k)$ with $r_{m,n}(k)$ in Eq.~\ref{eq97} and applied the transformation rule of differential operator Eq.~\ref{eq92}. The basis-free designations could be problematic. Obviously, there is an extra term, and invariance is lost. That is why in Eq.~\ref{eq92} we define the operator with a form of pure matrix components, without referring to basis-free designations, such as bra or ket.

We give an example of bra/ket notations causing problems in handling complex conjugation. For matrix operator $O_{m,n}^{\dagger}:=[O^{\dagger}]_{m,n}$, i.e., $O^{\dagger}$ stands for a matrix as a whole, just like $O$, and $\dagger$ is not acting on a specific matrix elements such like $O_{m,n}$),
\begin{equation}
\begin{split}
O_{m,n}^{\dagger}=O_{n,m}^*
\end{split}
\label{eq100}
\end{equation}
This is true for generic matrix operator, without requiring $\hat{O}$ to be Hermitian. If $\hat{O}$ is a Hermitian operator, we further have
\begin{equation}
\begin{split}
O_{m,n}=O_{n,m}^*
\end{split}
\label{eq101}
\end{equation}
Since a matrix operator is invariant under ribbon/basis transformation, one may employ basis-independent notation
\begin{equation}
\begin{split}
\hat{O}^{\dagger}=\hat{O}
\end{split}
\label{eq102}
\end{equation}
Then, evaluate the expectation value of Hermitian $\hat{O}$
\begin{equation}
\begin{split}
&{\partial}_{\lambda}{\langle}\hat{O}{\rangle}={\partial}_{\lambda}{\langle}{\varphi}|\hat{O}|{\varphi}{\rangle}={\langle}{\partial}_{\lambda}{\varphi}|\hat{O}|{\varphi}{\rangle}+{\langle}{\varphi}|\hat{O}|{\partial}_{\lambda}{\varphi}{\rangle} \\
&={\langle}{\partial}_{\lambda}{\varphi}|\hat{O}|{\varphi}{\rangle}+({\langle}{\partial}_{\lambda}{\varphi}|\hat{O}^{\dagger}|{\varphi}{\rangle})^*=2{\Re}[{\langle}{\partial}_{\lambda}{\varphi}|\hat{O}|{\varphi}{\rangle}].
\end{split}
\label{eq103}
\end{equation}
When taking the complex conjugation, we have employed formulas in Appx. D.

Eq.~\ref{eq103} is true for matrix operator $\hat{O}$, but not for differential operator $\hat{r}$. If we plug in $\hat{O}=\hat{r}$, and replace $\hat{r}{\rightarrow}i{\partial}_k$, we achieve a celebrated result (ch. 4 of \cite{25})
\begin{equation}
\begin{split}
{\partial}_{\lambda}{\langle}\hat{r}{\rangle}=2\Re[{\langle}{\partial}_{\lambda}{\varphi}(k)|i{\partial}_k{\varphi}(k){\rangle}].
\end{split}
\label{eq104}
\end{equation}
The derivative of displacement is linked to the Berry curvature defined in the $({\lambda},k)$ space, which is the kernel for developing Berry phase formalism of electric polarization. Eq.~\ref{eq104} is also essential for path-independent formulation of polarization field $P$.

Noteworthy, the validity of elegant Eq.~\ref{eq104} \cite{25} relies on implicit preconditions. Compare it with a second way of handling it: make the replacement $\hat{r}{\rightarrow}i{\partial}_k$ in the first place. 
\begin{equation}
\begin{split}
{\partial}_{\lambda}{\langle}\hat{r}{\rangle}&={\partial}_{\lambda}{\langle}{\varphi}(k)|i{\partial}_k{\varphi}(k){\rangle} \\
&={\langle}{\partial}_{\lambda}{\varphi}(k)|i{\partial}_k{\varphi}(k){\rangle}+{\langle}{\varphi}(k)|i{\partial}_{\lambda}{\partial}_k{\varphi}(k){\rangle}.
\end{split}
\label{eq105}
\end{equation}
In order to yield a consistent result as Eq.~\ref{eq104}, the following equality must be true.
\begin{equation}
\begin{split}
&{\langle}{\varphi}(k)|i{\partial}_{\lambda}{\partial}_k{\varphi}(k){\rangle}
\xrightarrow{\overset{?}{}}-{\langle}{\partial}_k{\varphi}(k)|i{\partial}_{\lambda}{\varphi}(k){\rangle} \\
&={\langle}i{\partial}_k{\varphi}(k)|{\partial}_{\lambda}{\varphi}(k){\rangle} =({\langle}{\partial}_{\lambda}{\varphi}(k)|i{\partial}_k{\varphi}(k){\rangle})^*.
\end{split}
\label{eq106}
\end{equation}
Plug in Eq.~\ref{eq106} to Eq.~\ref{eq105}, we will get the Berry curvature results Eq.~\ref{eq104}. However, the deriving is based on “$\xrightarrow{\overset{?}{}}$” in Eq~\ref{eq106} is an equality, which is only conditionally true if 
\begin{equation}
\begin{split}
{\partial}_k({\langle}{\varphi}(k)|{\partial}_{\lambda}{\varphi}(k){\rangle}){\equiv}0.
\end{split}
\label{eq107}
\end{equation}
It is straightforward to show Eq.~\ref{eq107} does not hold \textit{locally} in general. 
Thus, we ought to discriminate $r_{m,n}=r_{n,m}^*~{\nLeftrightarrow}~\hat{r}=\hat{r}^{\dagger}$. Such equivalence is true for matrix operators. 

As a consequence, the equality in Eq.~\ref{eq104} does not hold for local $k$. Only if one integrates the left and right sides of Eq.~\ref{eq104} on a closed manifold, such as B.Z., the total integral will be equal although each local $k$ might make different contributions. As such, the celebrated Berry curvature formula for adiabatic currents relies on a closed topology. 

It appears the adiabatic current is infinitely fragile to missing (or adding) even a single particle that will break a closed topology. Since thermal excitation is existing even at low temperature, it seems necessary to extensively examine the stability of Eq.~\ref{eq104} with the presence of excitation, although the purpose of Eq.~\ref{eq104} is for adiabatic limit. This will be given in a separate work. Nonetheless, the finding of this work shows the substitution as Eq.~\ref{eq104} is false in a local sense.

In short, position operator is a Hermitian (differential) operator (all its eigenvalues are $\mathbb{R}$) and $r$-matrix is a Hermitian matrix (the left side of Eq.~\ref{eq94}); but the fact of position being Hermitian operator does not guarantee position operator should exhibit behaviors like a two-sided associative operator as the basis-free designation $\hat{r}=\hat{r}^{\dagger}$ allude to. Given the formal invariance Eq.~\ref{eq97} is absent, Eq.~\ref{eq94} is an example of mistake caused by basis-free notations. 

To elude the problem, one may either replace the ket/bra designation system with matrix component formalism, as Weinberg does \cite{46}; or keep using it but with special attention paid when $\hat{r}$ is involved.
\begin{equation}
\begin{split}
\text{Recommend}&:{\langle}{\varphi}(k)|i{\partial}_k{\psi}(k){\rangle}. \\
\text{Not Recommend}&: {\langle}{\varphi}(k)|\hat{r}|{\psi}(k){\rangle}. \\
\end{split}
\label{eq108}
\end{equation}
The difference is clear: $i{\partial}_k$ has effectual range (in this case, confined to ${\psi}(k)$) and is acting on one side (right); $\hat{r}$ is interpreted as associative (acting on both sides) without an effectual range. It is never an equivalent replacement $\hat{r}{\leftrightarrow}i{\partial}_k$. Thus, we shall not directly inherit the designation designed for matrix operator and apply it to $\hat{r}$. Recall that in expressing $r$-matrix element in Sec. 4, we adopt the original integration Eq.~\ref{eq44} instead of using ${\langle}{\psi}_{m,k'}|\hat{r}|{\psi}_{n,k}{\rangle}$. Eq.~\ref{eq108} is exactly the reason. 

Next, we summarize the algebraic rules (a)-(e) for matrix of differential operator in comparison with the matrix of matrix operator.
\vskip 0.5mm

\textbf{(a) Matrix elements}. $\hat{r}$ operator is directly defined by matrix elements.
\begin{equation}
\begin{split}
r_{m,n}(k)={\langle}m(k)|i{\partial}_kn(k){\rangle}.
\end{split}
\label{eq109}
\end{equation}
In contrast, matrix operator has
\begin{equation}
\begin{split}
O_{m,n}(k)={\langle}m(k)|\hat{O}(k)|n(k){\rangle}.
\end{split}
\label{eq110}
\end{equation}
Matrix operator may use basis-free designation $\hat{O}$ in the midst of ${\langle}{\ldots}{\rangle}$; while for the reason listed above, we should avoid using $r_{m,n}(k)={\langle}m(k)|\hat{r}|n(k){\rangle}$. Additionally, $\hat{O}(k)$ may intrinsically depend on $k$, not just due to ribbons being $k$-dependent; thus the $k$ label in $\hat{O}(k)$ should \textit{not} be ignored.
\vskip 0.5mm

\textbf{(b) Complex conjugation}.
\begin{equation}
\begin{split}
r_{m,n}^*(k)=({\langle}m(k)|i{\partial}_kn(k){\rangle})^*={\langle}i{\partial}_kn(k)|m(k){\rangle}
\end{split}
\label{eq111}
\end{equation}
where
\begin{equation}
\begin{split}
|i{\partial}_kn(k){\rangle}&=i|{\partial}_kn(k){\rangle}, \\
{\langle}i{\partial}_kn(k)|&=(-i){\langle}{\partial}_kn(k)|.
\end{split}
\label{eq112}
\end{equation}
For matrix operator,
\begin{equation}
\begin{split}
({\langle}m(k)|\hat{O}(k)|n(k){\rangle})^*={\langle}n(k)|\hat{O}^{\dagger}(k)|m(k){\rangle}.
\end{split}
\label{eq113}
\end{equation}
Note that $i{\partial}_k$ cannot take the position of $\hat{O}$, and “$\dagger$” should not be attached to $i{\partial}_k$ since $(i{\partial}_k)^{\dagger}$ is ill-defined. Given $\hat{O}$ co-exists with $i{\partial}_k$, the algebra rule is, for instance,
\begin{equation}
\begin{split}
({\langle}m(k)|\hat{O}(k)|i{\partial}_kn(k){\rangle})^*={\langle}i{\partial}_kn(k)|\hat{O}^{\dagger}(k)|m(k){\rangle}.
\end{split}
\label{eq114}
\end{equation}
The following could be used to express the fact that $\hat{r}$ is a Hermitian operator (i.e., $r$-matrix is a Hermitian matrix)
\begin{equation}
\begin{split}
{\langle}m(k)|i{\partial}_kn(k){\rangle}=({\langle}n(k)|i{\partial}_km(k){\rangle})^*{\Leftrightarrow}r_{m,n}=r_{n,m}^*.
\end{split}
\label{eq115}
\end{equation}
However,
\begin{equation}
\begin{split}
r_{m,n}=r_{n,m}^*{\nRightarrow}\hat{r}=\hat{r}^{\dagger}
\end{split}
\label{eq116}
\end{equation}
For matrix operators,
\begin{equation}
\begin{split}
{\langle}m(k)|\hat{O}(k)|n(k){\rangle}&={\langle}n(k)|\hat{O}(k)|m(k){\rangle}^*{\Leftrightarrow}O_{m,n}=O_{n,m}^* \\
O_{m,n}&=O_{n,m}^*~{\Leftrightarrow}~\hat{O}=\hat{O}^{\dagger}
\end{split}
\label{eq117}
\end{equation}
The difference between Eq.~\ref{eq116} and Eq.~\ref{eq117} is due to differential operators lacking the basis-free designation. For $\partial_k$ acting on generic vectors,
\begin{equation}
\begin{split}
({\langle}{\varphi}(k)|i{\partial}_k{\psi}(k){\rangle})^*={\langle}i{\partial}_k{\psi}(k)|{\varphi}(k){\rangle}.
\end{split}
\label{eq118}
\end{equation}
For matrix operators,
\begin{equation}
\begin{split}
({\langle}{\varphi}(k)|\hat{O}(k)|{\psi}(k){\rangle})^*={\langle}{\psi}(k)|\hat{O}^{\dagger}(k)|{\varphi}(k){\rangle}.
\end{split}
\label{eq119}
\end{equation}
A mistaken expression is
\begin{equation}
\begin{split}
\text{Mistaken}:(i{\partial}_k)^{\dagger}=i{\partial}_k~\text{or}~\hat{r}=\hat{r}^{\dagger},
\end{split}
\label{eq120}
\end{equation}
which leads to mistakes
\begin{equation}
\begin{split}
&\text{Incorrect}: \\
&[{\langle}{\varphi}(k)|i{\partial}_k|{\psi}(k){\rangle}]^*={\langle}{\psi}(k)|(i{\partial}_k)^{\dagger}|{\varphi}(k){\rangle}={\langle}{\psi}(k)|i{\partial}_k|{\varphi}(k){\rangle}.
\end{split}
\label{eq121}
\end{equation}
Obviously, the Eq.~\ref{eq121} is against the correct result Eq.~\ref{eq111}.
\vskip 0.5mm

\textbf{(c) Dimensions and Effectual range.} Differential operator ${\partial}_k$ is not with a fixed dimension of matrix. This can be seen that ${\partial}_k$ might act on both $|{\psi}_{n,k}{\rangle}{\in}\mathcal{H}_B$ and on $|u_{n,k}{\rangle}{\in}\mathcal{V}$. On the other hand, a matrix operator $\hat{O}(k)$ is associated with a determined dimension when first introduced.

Another feature of ${\partial}_k$ is effectual range. ${\partial}_k$ should always be specified with its effectual range, which is denoted by ${\partial}_k(\ldots)$. For example
\begin{equation}
\begin{split}
{\partial}_k|n(k){\rangle}=|{\partial}_kn(k){\rangle}+|n(k){\rangle}{\partial}_k.
\end{split}
\label{eq122}
\end{equation}
Thus, we shall distinguish ${\partial}_k|n(k){\rangle}$ from $|{\partial}_kn(k){\rangle}$ because $|{\partial}_kn(k){\rangle}$ has ${\partial}_k$'s effect restricted to $|n(k){\rangle}$; ${\partial}_k|n(k){\rangle}{\ldots}$, will affect every term all the way to the right. (Appx. D)
\vskip 0.5mm

\textbf{(d) Inner product}. For differential operator (Einstein convention)
\begin{equation}
\begin{split}
{\langle}{\varphi}(k)|{\partial}_k{\psi}(k){\rangle}&={\langle}{\varphi}(k)|m(k){\rangle}{\langle}m(k)|{\partial}_k(|n(k){\rangle}{\langle}n(k)|{\psi}(k){\rangle}) \\
&={\langle}{\varphi}(k)|m(k){\rangle}{\langle}m(k)|{\partial}_kn(k){\rangle}{\langle}n(k)|{\psi}(k){\rangle} \\
&+{\langle}{\varphi}(k)|m(k){\rangle}{\partial}_k({\langle}m(k)|{\psi}(k){\rangle})
\end{split}
\label{eq123}
\end{equation}
That is
\begin{equation}
\begin{split}
{\varphi}_m^*(k)r_{m,n}(k){\psi}_n(k)+{\varphi}_m^*(k){\partial}_k({\psi}_m(k))
\end{split}
\label{eq124}
\end{equation}
In contrast, matrix operator has
\begin{equation}
\begin{split}
&{\langle}{\varphi}(k)|\hat{O}(k)|{\psi}(k){\rangle}= \\
&{\langle}{\varphi}(k)|m(k){\rangle}{\langle}m(k)|\hat{O}(k)|n(k){\rangle}{\langle}n(k)|{\psi}(k){\rangle}.
\end{split}
\label{eq125}
\end{equation}
That is,
\begin{equation}
\begin{split}
{\varphi}_m^*(k)O_{m,n}(k){\psi}_n(k)
\end{split}
\label{eq126}
\end{equation}
Compared with Eq.~\ref{eq126}, the inner product of differential operator features an inhomogeneous term. Only with the $k$-independent ribbons, Eq.~\ref{eq124} will reduce to the same form as the matrix operator. 
\vskip 0.5mm

\textbf{(e) Ribbon band transformations}. The transformation rules for matrix and differential operators are specified by Def. 6, 7. The unitary matrix is stipulated as
\begin{equation}
\begin{split}
U_{n,j}(k)={\langle}n'(k)|j(k){\rangle}={\langle}n(k)|U(k)|j(k){\rangle},
\end{split}
\label{eq127}
\end{equation}
where $|n'(k){\rangle}=U^{\dagger}(k)|n(k){\rangle}$. $\mathcal{I}$ and $\mathcal{I}'$ are two sets of orthogonal ribbon bases, and $|n(k){\rangle}{\in}\mathcal{I}$ and $|n'(k){\rangle}{\in}\mathcal{I}'$. ($|n(k){\rangle}$ and $|n'(k){\rangle}$ mean the $n^{th}$ elements in sets $\mathcal{I}$ and $\mathcal{I}'$, respectively.)

Note that the rule for differential operators (Def. 7) is specified in matrix forms; unfortunately, the rule is virtually incompatible with the basis-independent designation. Neither of the followings is proper! 
\begin{equation}
\begin{split}
&{\partial}_k\xrightarrow{\overset{T_R}{}}U(k){\partial}_kU^{\dagger}(k) \\
&{\partial}_k\xrightarrow{\overset{T_R}{}}U(k){\partial}_k(U^{\dagger}(k))
\end{split}
\label{eq128}
\end{equation}
These two expressions are motivated by an analog with $\hat{O}\xrightarrow{\overset{T_R}{}}U\hat{O}U^{\dagger}$, i.e., ${\partial}_k$ takes the position of $\hat{O}$ in the middle and transformation takes a similarity form. The difference is merely about the effectual range. In the first line of Eq.~\ref{eq128}, ${\partial}_k$ will act all the way to the right. Consider an inner product with two arbitrary ribbons under ribbon transformations
\begin{equation}
\begin{split}
&{\langle}{\varphi}(k)|i{\partial}_k{\psi}(k){\rangle}\xrightarrow{\overset{T_R}{}}{\langle}{\varphi}(k)|U^{\dagger}(k)(U(k)i{\partial}_kU^{\dagger}(k))U(k){\psi}(k){\rangle} \\
&={\langle}{\varphi}(k)|i{\partial}_k{\psi}(k){\rangle}
\end{split}
\label{eq129}
\end{equation}
That is, ${\langle}{\varphi}(k)|i{\partial}_k{\psi}(k){\rangle}$ is invariant under the ribbon transformation (due to $U^{\dagger}(k)$ canceling with $U(k)$). Such invariance is against the ribbon transformation defined with matrix forms (Def. 7, Eq.~\ref{eq92}), which gives an extra term ${\varphi}_m(k)i{\partial}_k(U_{j,g}^{\dagger}(k))U_{g,n}(k){\psi}_n(k)$. The invariance of Eq.~\ref{eq129} is also against the common knowledge that Berry-connection-like quantity should be variant under transformation. 

On the other hand, such a designation is not a total failure, as ${\partial}_k{\rightarrow}U(k){\partial}_kU^{\dagger}(k)$ may correctly deduce matrix forms of ribbon transformation when $\partial_k$ is “isolated”, i.e., it does not act upon other ribbons (detailed in Sec. 7). That is why denotation like $U(k){\partial}_kU^{\dagger}(k)$ has been adopted in some literatures. However, it will encounter difficulty when the two parts “work some interplay” demonstrated by Eq.~\ref{eq129}. Such a designation cannot constantly stay harmonic with itself, nor yield consistent results with matrix forms (unfortunately, these issues often elude people’s notices).

How about using $U(k){\partial}_k(U^{\dagger}(k))$, restricting the effectual range to $U^{\dagger}(k)$? In that case, $U(k){\partial}_k(U^{\dagger}(k))$ will become a pure matrix operator (one may just regard ${\partial}_k(U^{\dagger}(k))$ as a matrix, and $\partial_k$ becomes a product of two matrices, which yield another matrix). Then, a differential operator has decayed into a matrix operator, which is obviously incorrect. 

Ribbon transformation is defined on the matrix of differential operator rather than differential operator. The fundamental mistake for Eq.~\ref{eq128} is that we try to find a denotation that directly expressed with the differential operator; instead, we shall first introduce the matrix of $\partial_k$, and define transformation on the matrix elements. In other words, it implicates that the analog between $\hat{O}$ and $\partial_k$ is improper, although people tend to call both of them operators. Thus, we see a second example for the deep incompatibility between differential operator and basis-free designations, adding to the earlier issue on complex conjugation (Eq.~\ref{eq116}). 

We compare the two designations in Table~\ref{tab3}, showing that basis-free designation will encounter problems occasionally; seemingly matrix designation is advantageous. That is why Def. 6,7 and those afterward are given in matrix forms, rather than basis-free forms. Algebraic rules about incorporating differential operators with ket/bra designations are summarized in Appx. D. 
\begin{table*}
\caption{\label{tab:table3} Comparison of basis-free (bra/ket) and matrix designations in terms of complex conjugation, inter product (${\langle}{\varphi}|\hat{O}|{\psi}{\rangle}$ or ${\langle}{\varphi}|i{\partial}_k{\psi}{\rangle}$), and ribbon transformation $T_R$. "N/A" means no appropriate or self-consistent designation is found, as in the case of differential operators in inter-product and ribbon transformation. Hence, Basis-free designation is considered unsuitable for differential operators. ($k$ labels are ignored for matrix operator) \label{tab3}}
\begin{ruledtabular}
\begin{tabular}{c c c c c}
\multicolumn{1}{c}{} & \multicolumn{2}{c}{Matrix Oper.} & \multicolumn{2}{c}{Differential Oper.} \\
\hline
  & Basis-free & Matrix form & Basis-free & Matrix form \\
\hline
Conj. & ${\langle}{\varphi}|\hat{O}|{\psi}{\rangle}^*$ & $({\varphi}_m^*O_{m,n}{\psi}_n)^*={\varphi}_mO_{m,n}^*{\psi}_n^*$ & ${\langle}{\varphi}(k)|i{\partial}_k{\psi}(k){\rangle}^*$ &	$(\varphi_m^*(k)r_{m,n}(k){\psi}_n(k))^*=\varphi_m(k)r_{m,n}^*(k){\psi}_n^*(k)$ \\
 & $={\langle}{\psi}|\hat{O}^{\dagger}|{\varphi}{\rangle}$ & & $={\langle}i{\partial}_k{\psi}(k)|{\varphi}(k){\rangle}$ & $(\varphi_m^*(k)i{\partial}_k{\psi}_n(k))^*=\varphi_m(k)(-i){\partial}_k{\psi}_n^*(k)$ \\
Prod. & ${\langle}{\varphi}|\hat{O}|{\psi}{\rangle}$ & ${\varphi}_m^*O_{m,n}{\psi}_n$ & N/A \cite{Note4} & ${\varphi}_m(k)r_{m,n}(k){\psi}_n(k)+{\varphi}_m(k)i{\partial}_k({\psi}_m(k))$ \\
$T_R$ & $\hat{O}\xrightarrow{\overset{T_R}{}}U\hat{O}U^{\dagger}$ & $O_{m,n}\xrightarrow{\overset{T_R}{}}U_{m,i}O_{i,j}U_{j,n}^{\dagger}$ & N/A \cite{Note5} & $M_{m,n}(k)\xrightarrow{\overset{T_R}{}}M_{m,n}'(k)$ \\
 & $|{\psi}{\rangle}\xrightarrow{\overset{T_R}{}}U|{\psi}{\rangle}$ & ${\varphi}_m\xrightarrow{\overset{T_R}{}}U_{m,n}{\varphi}_n$ & & $=U_{m,i}M_{i,j}(k)U_{j,n}^{\dagger}(k)+U_{m,j}(k)i{\partial}_k(U_{j,n}^{\dagger}(k))$
\end{tabular}
\end{ruledtabular}
\end{table*}

\section{6. Gauge transformation, ribbon transformation, and basis transformation.}
Although gauge invariance is common in constructing transport theory \cite{8,18,25}, there is still vagueness in concepts and the relationship between different proposals. Here we particularly focus on gauge transformation's relation with CRM and differential operators. We address in which cases one should concern gauge issues. We define gauge transformation unambiguously (Def. 8) as the frame to characterize various gauge transformations. We shall emphasize the gauge invariant could mean quite differently in different contexts. \cite{8,18,25} This will shed light on understanding different transport theories. \cite{8,18,20,25,26}

In the classical, gauge transformation arises from modifying vector potential $\textbf{A}(\textbf{r})$  but preserving magnetic field \textbf{B} (\textbf{B} is considered as physical reality). The generic expression is to add a curl-less field ${\nabla}{\Lambda}$ \cite{49}
\begin{equation}
\begin{split}
\textbf{A}(\textbf{r}){\rightarrow}\textbf{A}'(\textbf{r})=\textbf{A}(\textbf{r})+{\nabla}{\Lambda}(\textbf{r}).
\end{split}
\label{eq130}
\end{equation}
The only notion involved is the (differential) vector field. In quantum (especially in the context of geometric phases), however, (Abelian) gauge transformation refers to “phase shift” of eigenstate $|u_m(k){\rangle}$, \cite{15,22}
\begin{equation}
\begin{split}
|u_m(k){\rangle}{\rightarrow}e^{i\xi_m(k)}|u_m(k){\rangle}.
\end{split}
\label{eq131}
\end{equation}
It is more like a convention change, not involving preservation of physical quantities. Moreover, Eq.~\ref{eq131} relies on notions absent in the classical Eq.~\ref{eq130}, such as eigenstates and complex phase ${\xi}_m(k)$, which require a vector space established on $\mathbb{C}$; in contrast, the classical $\textbf{A}$ and $\Lambda$ are built on $\mathbb{R}$. 

Why Eq.~\ref{eq130} and Eq.~\ref{eq131} are both referred to as gauge transformations despite these distinctions? In this section, we shall define gauge transformation and clarify its relations with ribbon and basis transformations, and its physical implications. 

\textbf{Definition 8}. Gauge transformation $T_G$ associated with manifold $K$ is defined as a matrix map of a particular form
\begin{equation}
\begin{split}
T_G:M_{m,n}^{(\upsilon)}(k){\mapsto}&M_{m,n}^{(\upsilon)'}(k)=U_{m,i}(k)M_{i,j}^{(\upsilon)}(k)U_{j,n}^{\dagger}(k) \\ 
&+U_{m,j}(k)i{\partial}_{k_{\upsilon}}(U_{j,n}^{\dagger}(k)),
\end{split}
\label{eq132}
\end{equation}
where $M_{m,n}^{(\upsilon)}(k)$ is a matrix and its elements are indexed by $m$, $n$. $k{\in}K$ and ${\upsilon}=1,2{\ldots},d$, and $d=\text{dim}(K)$. $U_{m,i}(k)$ is unitary matrix with $k$ being a shorthand for coordinates ${\lbrace}k_1,{\ldots} k_{\upsilon},\ldots{\rbrace}_d$. $T_G$ can be denoted with a generic form 
\begin{equation}
\begin{split}
T_G:K{\rightarrow}f,
\end{split}
\label{eq133}
\end{equation}
where $f:{\oplus}_{\upsilon}^dV_O^{(\upsilon)}\rightarrow{\oplus}_{\upsilon}^dV_O^{(\upsilon)}$, i.e., direct sum a series of operator space $V_O^{(\upsilon)}$, and the number of sums depends on $d=\text{dim}{\lbrace}K{\rbrace}$. $M_{m,n}^{(\upsilon)}{\in}V_O^{(\upsilon)}$, i.e., elements in $V_O^{(\upsilon)}$ are matrices. 

\textbf{Remarks 6.1} Consider a concrete case $\text{dim}(K)=3$ (3D B.Z.) and $\text{dim}(V_O^{(\upsilon)})=1$ and $V_O^{(\upsilon)}$ is defined on $\mathbb{R}$. In that case, $V_O^{(\upsilon)}{\cong}\mathbb{R}$ and ${\oplus}_{\upsilon}^3V_O^{(\upsilon)}\cong\mathbb{R}^3$. Thus, $f:\mathbb{R}^3\rightarrow\mathbb{R}^3$. Thus, at a local $k$, $T_G$ is equivalent to transforming a 3D vector transformation. Matrix $M_{m,n}^{(\upsilon)}{\in}V_O^{(\upsilon)}$ reduces to a real number and is commutative, and unitary matrices $U(k)$ have become complex phases $e^{i{\xi}_{\upsilon}(k)}$ and will bypass $M^{(\upsilon)}$ and cancel its conjugation. Manifold $K$ refers to the space that hosts $\textbf{r}$, thus $K=\mathbb{R}^3$, 3D real space. This is exactly the situation of Eq.~\ref{eq130} the classical gauge transformation corresponds to Abelian $T_G$. Note that $\nabla\Lambda(\textbf{r})$ is a merely a means to specify map $f$, not indispensable for gauge definition. 

\textbf{Remarks 6.2} Consider non-Abelian case $\text{dim}(K)=1$ (1D B.Z.). In band model, $\text{dim}(V_O^{(\upsilon)})=\text{dim}(\mathcal{V})=\mathcal{N}$, i.e., the dimension of matrices in $V_O^{(\upsilon)}$ is equal to the dimension of quotient space $\mathcal{V}$ of Bloch space $\mathcal{H}_B$ (the band number). In fact, gauge transformation is exactly the transformation rule of reduced $r$-matrices under ribbon transformation. 
\begin{equation}
\begin{split}
r_{m,n}^{(\mathcal{N})}(k)\xrightarrow{\overset{T_R}{}}r_{m,n}^{(\mathcal{N})'}(k)=&U_{m,i}(k)r_{i,j}^{(\mathcal{N})}(k)U_{j,n}^{\dagger}(k) \\
&+U_{m,j}(k)i{\partial}_k(U_{j,n}^{\dagger}(k)).
\end{split}
\label{eq134}
\end{equation}
Thus, the Abelian Eq.~\ref{eq131} is a special case of $T_G$. 

\textbf{Remarks 6.3} $T_G$ can be viewed as a modified form of matrix rotation, which should have obeyed a similarity form
\begin{equation}
\begin{split}
O_{m,n}(k){\rightarrow}O_{m,n}'(k)=U_{m,i}(k)O_{i,j}(k)U_{j,n}^{\dagger}(k).
\end{split}
\label{eq135}
\end{equation}
Evidently, $T_G$ has an extra term $\lambda{\cdot}U_{m,j}(k)i{\partial}_k\left(U_{j,n}^\dag\left(k\right)\right)$ with $\lambda=1$; while $\lambda{\rightarrow}0$, $T_G$ will reduce to the transformation behavior of matrix operator. 

\textbf{Remarks 6.4} $T_G$ is not an arbitrary transformation from $K$ to $f$, but one that conforms to a particular form specified by Eq.~\ref{eq132}. Why is this form chosen? Because it is the transformation of $r$-matrix, or more generally, it is the transformation response of matrices of differential operators. That means, gauge transformation is closely relevant to differential operator, i.e., the transformation form is determined from differential operator’s matrix behaviors. On the other hand, if differential operator is not involved, for example, for spin operators (or other matrix operators), gauge transformation is trivial, as it is identical to many independent versions of basis transformation.

\textbf{Remarks 6.5} $T_G$ is not for a single point but defined for ${\forall}k{\in}K$ (thus it is associated with $K$). In band scenarios, $K$ is B.Z. In terms of topological space, B.Z. is a torus $T^d$. 

\textbf{Remarks 6.6} The map Eq.~\ref{eq132} is directly defined with matrices without involving the matrices’ identities or natures. An $n{\times}n$ matrix can denote a linear operator (or map) in an $n$-dimensional vector space $V$, but only until its response under the basis rotation is specified. Intuitively speaking, an operator is a matrix that is hinged with basis transformation: for an operator, its matrix $\sigma_1$ might have to change to a different “version”, say $\sigma_2$, under new bases; while a pure matrix is just “free” when it is not associated with vector spaces or bases, but just a set of numbers arranged in a rectangular box. In the level of defining $T_G$, the target is just a pure matrix, without worrying if the entries in the domain form linear operators or not. (In fact, these matrices will not be linear operators, as the non-linear terms in $T_G$ violates the linearity.)

\textbf{Remarks 6.7} Note that $T_G$ is not a linear transformation, as $T_G(M_1+M_2){\neq}T_G(M_1)+T_G(M_2)$, $T_G(cM_1){\neq}cT_G(M_1)$. On the other hand, it has property, $T_G(M_1)-T_G(M_2)=M_1-M_2$. 

\textbf{Remarks 6.8} There are multiple spaces involved in $T_G$’s definitions, thus there are multiple dimensions associated with $T_G$. The first one is the dimension of the manifold $K$ that corresponds to B.Z. in this context. Second, the dimension of matrix $M$ which, in band models, is set equal to $\text{dim}(\mathcal{V})$, i.e., equal to the band number. Thirdly, the $\upsilon^{th}$ component of $M^{(\upsilon)}$, whose dimension is equal to $\text{dim}(K)$. 

\textbf{Remarks 6.9} $T_G$ is defined as an abstract map, not associated with classical or quantum physics, neither with Hilbert space. 
\vskip 0.5mm

\textbf{Relation between gauge transformation $T_G$ and ribbon transformation $T_R$}. Ribbon transformation $T_R:K{\rightarrow}\text{Aut}(V)$ (Eq.~\ref{eq90}) is associated with vector space $V$ and manifold $K$; gauge transformation $T_G:K{\rightarrow}f$ (Eq.~\ref{eq133}) is only defined with $K$, not involving any vector space. $T_R$ is to transform a ribbon (which is a map); while $T_G$ is to transform a matrix --- distinct transformation targets. In other words, the domains (also the co-domains) of the two transformations are different: the domain of $T_R$ is ribbon space $V_\mathscr{R}$, while $T_R$’s domain is a matrix set. Additionally, $T_R:\mathscr{R}{\mapsto}\mathscr{R}'$ is a linear map, while $T_G$ is not linear.

Despite these conceptual distinctions, $T_G$ and $T_R$ are closely related: they are both established on manifold $K$; moreover, the entire information about $T_G$ is encoded in unitary matrix $U_{m,i}(k)$. This is easily seen from Eq.~\ref{eq132} given the unitary matrix $U_{m,i}(k)$, output $M_{m,n}^{(\upsilon)'}(k)$ is determined, which is exactly the matrix designation for a ribbon transformation. Therefore, $T_R$ can induce $T_G$. In other words, a correspondence exists between $T_G{\sim}T_R$ via $U_{m,i}(k)$ matrix. 

Additionally, both $T_G$ and $T_R$ can be classified by unitary group U($N$), where $N$ refers to the highest dimension of the group’s irreducible representation (IR) on $V$. Thus, we may utilize a superscript ``$N$" in $T_G^{(N)}$(or $T_R^{(N)}$) to denote U($N$) gauge (ribbon) transformations. The dimension of U($N$) could be different from the dimension of matrices. (The dimension of a group is a property about a set of matrices \cite{38}, while the dimension of a matrix is about a single matrix.) For example, consider $T_G^{(1)}$ on two bands, i.e., $e^{i{\xi}_m(k)}$ ($m=0,1$). Aut($V$) takes a form of $2{\times}2$ diagonal matrix as below, 
\begin{equation}
\begin{split}
\begin{pmatrix} |u_0'(k){\rangle} \\ |u_1'(k){\rangle} \end{pmatrix}=\begin{pmatrix}  e^{i{\xi}_0(k)} & 0 \\ 0 & e^{i{\xi}_1(k)} \end{pmatrix} \begin{pmatrix} |u_0(k){\rangle} \\ |u_1(k){\rangle} \end{pmatrix}.
\end{split}
\label{eq136}
\end{equation}
The matrix above is not representing a single one, but a set of matrices parameterized by $k$ that forms 2D reducible representation of  U(1)  group, in which the highest IR is 1D, i.e., $U(1){\oplus}U(1)$. Thus, $T_G$ is 1D, while the matrix (or the vector space) is 2D.

Back to the question raised earlier: why Eq.~\ref{eq130} and Eq.~\ref{eq131} are both regarded as gauge transformation? Accurately speaking, based on Def. 8 (Eq.~\ref{eq133}) only Eq.~\ref{eq130} is $T_G^{(1)}$, while Eq.~\ref{eq131} is transforming a ribbon with a unitary matrix $U(k)$. Eq.~\ref{eq131} being considered a gauge transformation requires an additional step: the correspondence $T_R{\sim}T_G$, i.e., $e^{i{\xi}(k)}$ in Eq.~\ref{eq131} contains all the information to deduce the gauge transformation. Forgetting that might lead to conceptual vagueness and confusion. For example, one may mistakenly believe $T_G$ is established on the notion of eigenstate and complex phase shift. In fact, $T_G$ can be defined as an abstract map, without referring to eigenvectors. 

Next, we shall extend the familiar notion “basis-invariance” to another concept ``gauge invariance". Basis invariance is associated with a specific function of matrix operator. For example,
\begin{equation}
\begin{split}
F(O_{m,n})={\varphi}_m^*O_{m,n}{\psi}_n
\end{split}
\label{eq137}
\end{equation}
The basis invariance refers to
\begin{equation}
\begin{split}
T_B[F(O_{m,n})]=F(O_{m,n})
\end{split}
\label{eq138}
\end{equation}
The basis transformation $T_B$ means that we need to find the updated components for the vector ${\varphi}_m^*$, ${\psi}_n$
\begin{equation}
\begin{split}
{\varphi}_m^*\xrightarrow{\overset{T_R}{}}{\varphi}_j^*U_{j,m}^{\dagger},~{\psi}_n\xrightarrow{\overset{T_R}{}}U_{n,l}{\psi}_l,
\end{split}
\label{eq139}
\end{equation}
and for matrix operator $O_{m,n}$
\begin{equation}
\begin{split}
O_{m,n}\xrightarrow{\overset{T_R}{}}U_{m,g}O_{g,h}U_{h,n}^{\dagger}.
\end{split}
\label{eq140}
\end{equation}
Then,
\begin{equation}
\begin{split}
T_B[F(O_{m,n})]&={\varphi}_j^*U_{j,m}^{\dagger}U_{m,g}O_{g,h}U_{h,n}^{\dagger}U_{n,l}{\psi}_l \\
&={\varphi}_m^*O_{m,n}{\psi}_n=F(O_{m,n})
\end{split}
\label{eq141}
\end{equation}
Then function $F$ is said to be invariant under basis transformation $T_B$. This invariance is associated with vector space $V$ and function $F$. 

In the same vein, we try to define invariance for ribbon transformation. In this case, function $F$ needs to be replaced by a functional $\mathfrak{F}$ about the matrix $M_{m,n}(k)$. Functional can be viewed as a generalized function whose variable is a map. $M_{m,n}(k)$ is the map $K{\rightarrow}V_O$ (from manifold $K$ to matrix (operator) space $V_O$) that serves as the functional’s variable, thus functional is denoted as $\mathfrak{F}(M_{m,n}(k))$. Additionally, we need to replace $T_B$ by $T_R$.
\begin{equation}
\begin{split}
T_R[\mathfrak{F}(M_{m,n}(k))]:=\mathfrak{F}(T_R[M_{m,n}(k)]),
\end{split}
\label{eq142}
\end{equation}
for which $T_R[M_{m,n}(k)]$ is given by Def. 6, 7. That means $T_R[M_{m,n}(k)]$ relies on whether matrix $M_{m,n}(k)$ belongs to matrix or differential operators: the two are subject to distinct behaviors under $T_R$. The ribbon band is a common platform for both the two types of operators, but only when it is defined for matrix operator, $T_R$ may effectively reduce to the $T_B$ since transformations at different $k$ are independent. 

On the other hand, $T_R$ applies non-trivially to the matrix of differential operators. Consider an example of $M_{m,n}(k)$ belonging to a differential operator.
\begin{equation}
\begin{split}
\mathfrak{F}(M_{m,n}(k))=\text{Tr}[{\oint}M_{m,n}(k){\cdot}dk]
\end{split}
\label{eq143}
\end{equation}
Plug in Eq.~\ref{eq143} with Eq.~\ref{eq92} 
\begin{equation}
\begin{split}
&T_R[\mathfrak{F}(M_{m,n}(k))]= \\ 
&\text{Tr}[{\oint}(U_{m,i}(k)M_{i,j}(k)U_{j,n}^{\dagger}(k)){\cdot}dk] \\
+&\text{Tr}[{\oint}(U_{m,i}(k)i{\partial}_k(U_{j,n}^{\dagger}(k))){\cdot}dk]
\end{split}
\label{eq144}
\end{equation}
Using that Tr and $\oint$ may exchange the sequence and similarity transformation will preserve trace. Eq.~\ref{eq144} becomes 
\begin{equation}
\begin{split}
&{\oint}\text{Tr}(U_{m,i}(k)M_{i,j}(k)U_{j,n}^{\dagger}(k)){\cdot}dk \\
&+\text{Tr}{\oint}(U_{m,i}(k)i{\partial}_k(U_{j,n}^{\dagger}(k))){\cdot}dk \\
&={\oint}\text{Tr}(M_{i,j}(k)){\cdot}dk+\text{Tr}{\oint}U_{m,j}(k)id(U_{j,n}^{\dagger}(k)).
\end{split}
\label{eq145}
\end{equation}
A generic fact for unitary matrices
\begin{equation}
\begin{split}
U=e^{iH}
\end{split}
\label{eq146}
\end{equation}
where $H$ is a Hermitian matrix.
\begin{equation}
\begin{split}
Ud(U^{\dagger})=e^{iH}d(e^{-iH})=e^{iH}e^{-iH}d(H)=dH.
\end{split}
\label{eq147}
\end{equation}
Combined with Eq.~\ref{eq145}, we have 
\begin{equation}
\begin{split}
&\text{Tr}[{\oint}M_{i,j}(k)dk]+\text{Tr}[{\oint}id(H_{m,n}(k))] \\
&=\text{Tr}{\oint}M_{i,j}(k)dk+\text{Tr}[H_{m,n}(k)|_0^{2\pi}].
\end{split}
\label{eq148}
\end{equation}
The last term in Eq.~\ref{eq148} will be vanishing for continuity of $H_{m,n}(k)$ on torus. Thus $\mathfrak{F}$ is invariant under ribbon transformation, i.e., $T_R[\mathfrak{F}(M_{m,n}(k))]=\mathfrak{F}(M_{m,n}(k))$. On the other hand, if the integration is not for closed manifold or the trace Tr is absent, ribbon invariance fails.  

Since gauge transformation $T_G$ is induced by ribbon transformation $T_R$. We may introduce the notion of gauge invariance and gauge symmetry in a similar line as ribbon invariance. 
\vskip 0.5mm

\textbf{Definition 9}. Gauge invariance is the following property associated with functional $\mathfrak{F}:M_{m,n}(k){\mapsto}\mathbb{R}$ or $\mathbb{C}$.
\begin{equation}
\begin{split}
T_G[\mathfrak{F}(M_{m,n}(k))]:=\mathfrak{F}(T_G[M_{m,n}])=\mathfrak{F}(M_{m,n}(k)),
\end{split}
\label{eq149}
\end{equation}
which holds ${\forall}T_R{\in}\text{U}(N)$. $T_G[M_{m,n}(k)]$ is given by Def. 8. If Eq.~\ref{eq149} is fulfilled, functional $\mathfrak{F}$ about matrix $M_{m,n}(k)$ is said to be invariant under U($N$) gauge transformation, or $\mathfrak{F}$ has U($N$) gauge symmetry. Obviously, if U($N$) is the gauge symmetry of $\mathfrak{F}$, the subgroups of U($N$) is also gauge symmetry of $\mathfrak{F}$.

\textbf{Remarks 6.10} $T_G$ for functional $\mathfrak{F}(M_{m,n}(k))$ generalizes $T_G$ for matrix $M_{m,n}(k)$ (Def. 8). Then, gauge invariance is a notion established on $T_G$ for $\mathfrak{F}$. Since transformation of $M_{m,n}(k)$ is “fixed” (by Def. 8), whether it is gauge invariant entirely depends on the form of $\mathfrak{F}$. 

\textbf{Remarks 6.11} Since $T_G$ (Def. 8) is defined for $M_{m,n}(k)$ that follows the transformation of matrix of differential operator (Def. 7), thus gauge invariance is more pertinent to the matrix of differential operators (e.g., $\hat{r}$). Ribbon invariance is a more generalized notion in this context, that works both differential and matrix operators.

\textbf{Remarks 6.12} Gauge invariance is a notation subject to functional $\mathfrak{F}$ and matrix $M_{m,n}(k)$. It is also implicitly subject to the manifold $K$ and vector space $V$, which are ingredients for the definition of matrix $M_{m,n}(k)$, since $M_{m,n}(k)$ is a map:$K{\rightarrow}f$, where it is a linear self-map $f:V{\rightarrow}V$.

\textbf{Remarks 6.13} For the correspondence $T_R{\sim}T_G$, gauge invariance and gauge symmetry can be characterized by U($N$) group. 

\textbf{Remarks 6.14} “Gauge invariance” might vary slightly in its meanings and emphasis in different contexts, thus showing different facets. These distinctions can all be attributed to specific constructions of functional $\mathfrak{F}$.

Consider an $\mathfrak{F}$ without gauge invariance.
\begin{equation}
\begin{split}
&A_{n,n}(k)=\mathfrak{F}(A_{m,n}(k')) \\ 
&=\lim_{k{\to}k'}\frac{1}{k-k'}{\int}_{k'}^k{\delta}_{m,n}A_{m,n}(k')dk'
\end{split}
\label{eq150}
\end{equation}
and
\begin{equation}
\begin{split}
&T_R[\mathfrak{F}(A_{m,n}(k'))]=U_{n,i}(k)A_{i,j}(k)U_{j,n}^{\dagger}(k) \\
&+U_{n,i}(k)i{\partial}_k(U_{j,n}^{\dagger}(k)){\neq}\mathfrak{F}(A_{m,n}(k'))
\end{split}
\label{eq151}
\end{equation}
In fact, $\mathfrak{F}(A_{m,n}(k'))$ is no more than a functional construction for Berry connection, is not gauge invariant as well known. Note that in above all the invariances are defined with matrices and matrix transformation behavior, without reference to whether these matrices arise from differential operator or bra/ket. In other words, one does not need to refer to $A_{m,n}(k')={\langle}u_{m,k'}|i{\partial}_{k'}u_{n,k'}{\rangle}$, but only the matrix transformation. 

Consider an $\mathfrak{F}$ respecting gauge invariance.
\begin{equation}
\begin{split}
{\vartheta}=\mathfrak{F}(A_{m,n}(k))={\oint}{\delta}_{m,n}A_{m,n}(k){\cdot}dk
\end{split}
\label{eq152}
\end{equation}
It can be shown that
\begin{equation}
\begin{split}
T_R[\mathfrak{F}(A_{m,n}(k))]={\oint}U_{n,i}(k)A_{i,j}(k)U_{j,n}^{\dagger}(k){\cdot}dk.
\end{split}
\label{eq153}
\end{equation}
The inhomogeneous term ${\oint}U_{n,j}(k)id(U_{j,n}^{\dagger}(k))$ is vanishing. Eq.~\ref{eq153} is invariant when $U_{n,i}(k)$ commutes with $A_{i,j}(k)$; however, in general, Eq.~\ref{eq153} is not invariant. Thus, such a construction of $\mathfrak{F}$ has U(1) gauge symmetry but not U($N$).
\begin{figure}
\includegraphics[scale=0.38]{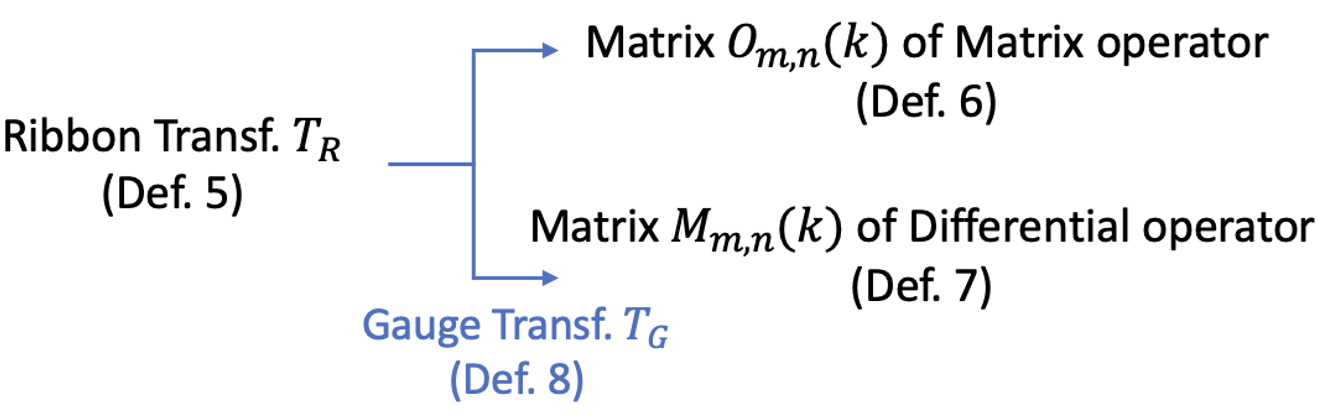}
\caption{\label{fig:epsart}(color online): Ribbon transformations apply to matrices of both matrix and differential operators. When $T_R$ is applied to matrix of differential operator, it induces transformation $T_G$. Both $T_G$ and $T_R$ are directly defined with matrices, independent of the ket/bra such basis-free designations. \label{f1}}
\end{figure}

In short, gauge transformation $T_G$ reflects the response of the matrix of a differential operator under ribbon transformation $T_R$. Thus, $T_G$ can be induced by $T_R$. Both of them can be characterized by the unitary group U($N$) involved in the ribbon transformation. In certain cases, the two may be exchangeable. But their definitions in terms of map are not the same. 
\vskip 0.5mm

\textbf{Relation between gauge transformation $T_G$ and basis transformation $T_B$}. Basis is a notion associated with a vector space $V$. Basis transformation $T_B$ will affect vectors and operators defined in space $V$. It is expressed as a map:
\begin{equation}
\begin{split}
T_B:V{\rightarrow}V.
\end{split}
\label{eq154}
\end{equation}
Conceptually, one can interpret $T_B$ changes components ${\varphi}_m$ of a vector under the updated bases, but does not alter the vector; alternatively, one may interpret $T_B$ does alter the vector, sending it to another vector. The two interpretations only differ by the “reference”, actually equivalent.

$T_G$ and $T_R$ are notions associated with product space $K{\otimes}V$ (technically, $K{\otimes}V$ can be called a bundle space with $K$ being base space and $V$ being fiber space; $K{\otimes}V$ is locally like a tensor product of spaces, but not necessary globally \cite{37}). Roughly speaking, $T_G$ and $T_R$ are defined for a “bigger” space. Then, could we view $T_G$ and $T_R$ as “basis transformation” for the bigger space? This view could be mistaken, since $K{\otimes}V$ may not form a vector space (e.g., when $K$ is B.Z.), but only a topological space. $T_G$ and $T_R$ are different maps from $T_B$ as summarized in Table~\ref{tab4}.

$T_G$ and $T_R$ will affect ribbons and operators defined in $K{\otimes}V$, just like $T_B$ will modify vectors and operators in $V$. Quantum operators are originally defined in Hilbert space, which is a vector space; in treating transport, as mentioned in Sec. 2, we encounter diverging norms, and transcribe operators into non-vector spaces, wherein $T_G$ and $T_R$ are defined. In the band context, $V$ is quotient space $\mathcal{V}$ of $\mathcal{H}_B$, and $T_B$ is a map $\mathcal{V}{\rightarrow}\mathcal{V}$. Thus, one can practically understand $T_B$ is a “single” $\mathcal{N}$-dimensional matrix at $k$; while $T_G$ (also $T_R$) involves ${\forall}k{\in}K$, thus is a matrix field.

It is inaccurate to regard $T_G^{(1)}$ just as “phase shifts”, as $T_B$ can also produce that. For example, a 2D vector space spanned by $|u_j{\rangle}$,($j=0,1$). Consider a $T_B$ of a phase shift, $|u_j{\rangle}{\rightarrow}e^{i{\xi}_j}|u_j{\rangle}$. Is this phase shifting a $T_G^{(1)}$? No. Because it does not involve manifold $K$, i.e., ${\xi}_j$ is not a field about $k$. If $T_G^{(1)}$ is a “special case” of $T_B$? Since a general $T_B$ allows mixing of bases, e.g., $|u_0'{\rangle}=c_0|u_0{\rangle}+c_1|u_1{\rangle}$, while $T_G^{(1)}$ constrains $c_0=e^{i{\xi}_0}$  and $c_1=0$. This view is inaccurate for the same reason: ${\xi}_j$ is not a field about $k$. On the other hand, at a local $k$, $T_G^{(1)}$ can be viewed as a constrained $T_B$; if one relaxes the constraint and mixing of orthogonal bases is allowed, $T_G^{(N)}$ is achieved.  Thus, one may use the dimension of local $T_B$ to classify $T_G$ and $T_R$.

We have seen $T_G^{(1)}$ invariance is an important criterion for constructing observables. One may wonder whether invariance of $T_B$ should be tested too? Fortunately, associative operators (i.e., matrix operators) defined in vector space are automatically endowed with U(1) symmetry. Thus, gauge invariance is only pertinent to differential operators, such as position operators, Berry connection, etc.

In other words, if an observable involves ${\partial}_k$, etc., one shall test if this quantity is invariant under $T_G^{(1)}$; if gauge invariance is true, in principle, the quantity could be detectable. On the other hand, for matrix operator serving as observables (e.g., spin), there is not such an issue of gauge invariance. For operators other than observables, e.g., propagation operator, shall we worry about basis or gauge transformation? Fortunately, the answer is no, again. But for a different reason. Consider an evolution operator ($\mathfrak{T}$ is time-ordered)
\begin{equation}
\begin{split}
|u_{n,k}(t){\rangle}=\mathfrak{T}\text{exp}[-i{\int}_0^tH(k,{\tau})d{\tau}]|u_{n,k}(0){\rangle}.
\end{split}
\label{eq155}
\end{equation}
We notice that the evolution operator is composed by product of matrix operators and their sums (expansion of exponential functions under time ordering), which again form a matrix operator. Therefore, when we are evaluating a spin operator, and its evolution, we are not facing a gauge issue.
\begin{table}
\caption{\label{tab:table4} Comparison of ribbon transformation, gauge transformation, and basis transformation in terms of designation symbols, identities as maps, and application scopes, either on matrix or differential operators, or both. \label{tab4}}
\begin{ruledtabular}
\begin{tabular}{c c c c}
 Transf. & Symbol & Map & Oper. \\
\hline           
Ribbon & $T_R$ & $K{\rightarrow}\text{Aut}(V)$ & Matrix \& Diff. \\
Gauge  & $T_G$ & $K{\rightarrow}f$ & Diff. \\
Basis  & $T_B$ & $V{\rightarrow}V$ & Matrix \\
\end{tabular}
\end{ruledtabular}
\end{table}

At last, we shall remark $T_R$ or $T_G$ are not only “transforming” the ribbon, but also affect the matrix defined in the ribbon space; just like basis transformation will also affect operators defined in the vector space. Thus, ribbon transformation is defined as map of ribbons, while as shown by Eq.~\ref{eq90}, the operators will also be affected. 
\vskip 0.5mm

\textbf{Extracting observables}. From Fig.~\ref{f1}, we realize $T_R$ is a general transformation applicable to both matrix and differential operators. When it is applied to differential operator, it induces gauge transformation $T_G$; gauge invariance is a special case of $T_R$ invariance, as $T_R$ is applied to differential operators. 

By reviewing the observable of a matrix operator, we find it features U(1) symmetry, i.e., invariant under $T_R^{(1)}$. It corresponds to a function $F$, namely \textit{observable function}, that links an observable with the diagonal terms of matrix $O_{m,n}$.
\begin{equation}
\begin{split}
F(O_{m,n})=O_{n,n}
\end{split}
\label{eq156}
\end{equation}
Function $F$ is invariant under U(1) ribbon transformation
\begin{equation}
\begin{split}
T_R^{(1)}[F(O_{m,n})]&=F(T_R^{(1)}[O_{m,n}])=F(e^{i{\xi}_m}O_{m,n}e^{-i{\xi}_n}) \\
&=e^{i{\xi}_n}O_{n,n}e^{-i{\xi}_n}=O_{n,n}=F(O_{m,n}),
\end{split}
\label{eq157}
\end{equation}
a property not shared by off-diagonal terms (thus, observable is linked to diagonals rather than off-diagonals). In the case of $N>1$, we generally have $T_R^{(N)}[F(O_{m,n})]{\neq}F(O_{m,n})$, i.e., $F$ does not enjoy U($N$) symmetry. U(1) symmetry is believed indispensable for observables. In band scenarios, the ground state has a fixed occupancy (all states below Fermi level), but phase is flexible due to dynamic evolution $e^{-iE(k)t/{\hbar}}$. That means, even without disturbance, each quasi-particle keeps evolving, and the ground state is composed by a collection of quasi-particles with random phases. Thus, robustness to phase fluctuations, i.e., U(1) symmetry, ensures a quantity to be stable over time (given occupancy is unchanged) and thus detectable during a measurement. Since our “vision” depends on measurement conditions, time/spatial scales, etc., the meaning and criteria for observables might vary with cases. In non-Abelian gauge theory, observables might have higher symmetries. Nonetheless, U(1) symmetry should be of outstanding importance.

We try to extend gauge symmetry principle for observables to differential operators. Evidently, for the matrix of differential operators, $F(M_{m,n})$ is not invariant under $T_G^{(1)}$ (i.e., $T_R^{(1)}$).
\begin{equation}
\begin{split}
&T_G^{(1)}[F(M_{m,n}(k))]=F(T_G^{(1)}[M_{m,n}(k)]) \\
&=F(e^{i{\xi}_m(k)}M_{m,n}(k)e^{-i{\xi}_n(k)}+e^{i{\xi}_m(k)}i{\partial}_ke^{-i{\xi}_n(k)}) \\
&=e^{i{\xi}_n(k)}M_{n,n}(k)e^{-i{\xi}_n(k)}+e^{i{\xi}_n(k)}i{\partial}_ke^{-i{\xi}_n(k)} \\
&=M_{n,n}(k)+{\partial}_k{\xi}_n(k){\neq}F(M_{m,n}(k))
\end{split}
\label{eq158}
\end{equation}
That means function $F$ is unsuitable for observables associated with differential operators. Thus, we shall reconstruct a form that ensures U(1) symmetry. Consider the following
\begin{equation}
\begin{split}
\mathfrak{F}(M_{m,n}(k)):={\oint}M_{n,n}(k){\cdot}dk,
\end{split}
\label{eq159}
\end{equation}
which is a satisfaction that gives $F$ minimum modification (thus maximum elegance). $\mathfrak{F}$ still involves the diagonal terms $M_{n,n}(k)$, but adds an integrand over $k$. This form has emerged in different fields of physics \cite{15,16,22,25} and is privileged with U(1) symmetry.
\begin{equation}
\begin{split}
&T_G^{(1)}[\mathfrak{F}(M_{m,n}(k))]=\mathfrak{F}(T_G^{(1)}[M_{m,n}(k)]) \\
&=\mathfrak{F}(e^{i{\xi}_m(k)}M_{m,n}(k)e^{-i{\xi}_n(k)}+e^{i{\xi}_m(k)}i{\partial}_ke^{-i{\xi}_n(k)}) \\
&={\oint}(e^{i{\xi}_n(k)}M_{n,n}(k)e^{-i{\xi}_n(k)}+e^{i{\xi}_n(k)}i{\partial}_ke^{-i{\xi}_n(k)})dk \\
&={\oint}M_{n,n}(k){\cdot}dk+{\oint}{\partial}_k{\xi}_n(k){\cdot}dk \\
&={\oint}M_{n,n}(k){\cdot}dk=\mathfrak{F}(M_{m,n}(k)).
\end{split}
\label{eq160}
\end{equation}
$F(O_{m,n})$ and $\mathfrak{F}(M_{m,n}(k))$ represent two distinct ways of obtaining observables. The dichotomy seems weird that one has to follow separate principles. Now we argue the two are linked by a common principle: U(1) symmetry, subject to a common physical origin of “stability to dynamical phases”. If U(1) symmetry is the principle for observables (at least in condensed matter scenarios), subject to which the distinct ways of yielding observables, $F(O_{m,n})$ and $\mathfrak{F}(M_{m,n}(k))$, can be unified.

Conventionally, observables are evaluated by inner products with observable’s operator, i.e., diagonal terms of $O_{m,n}$ (or their supposition), expressed by $F(O_{m,n})$, which is a necessary result if the followings are true: (i) every observable has a corresponding (Hermitian) operator; (ii) the corresponding operator is an associative operator (i.e., a matrix operator), which will ensure U(1) symmetry. However, counterexamples are now known for both (i)(ii). In non-relativistic quantum mechanics, time does not have such a corresponding operator; in relativistic scope, boson lacks its position operator \cite{46}. Thus, observable-operator correspondence is not guaranteed. Moreover, when the corresponding operator exists (e.g., $\hat{r}$ as discussed throughout Sec. 3), it might not be associative; that is why we have seen the divergence in DRM and incompleteness of vector space for $\hat{r}$, which has motivated our seeking CRM with $N$-th Weyl algebra $A_N$, leading to distinct transformation behaviors for $r$-matrix. In history, based on mistaken presumptions (i)(ii), Von Neumann “proved” hidden local variables in quantum mechanics. \cite{2}

The traditional view is that $F$ is the generic form of generating observables, subject to which $\mathfrak{F}$ should belong to the frame of $F$. However, the difficulty is that there is not a counterpart for integration over $k$ in $F$. In other words, $F$ is for local $k$, while $\mathfrak{F}$ is for global $K$: employing $\mathfrak{F}$ to determine the value of observable, one must come into knowledge about $M_{m,n}(k)$ all over B.Z., while $F$ is only about a vector at a single $k$. This issue was noticed in Vanderbilt’s book (ch. 4). \cite{25} The essential argument is that electric polarization cannot be expressed as the expectation value of a quantum operators as the case of most quantum operators; instead, it is related to Berry phases which are defined by global means. 

For the new principle, we firstly establish matrix and differential operators on a common ground: ribbon space; on top of it, U($N$) symmetry is to classify both of them. Then we argue $F(O_{m,n})$ is no more fundamental than $\mathfrak{F}(M_{m,n}(k))$. Instead, we take the U(1) symmetry as the major principle and seek the robust forms for each type of operators.
\begin{figure}
\includegraphics[scale=0.38]{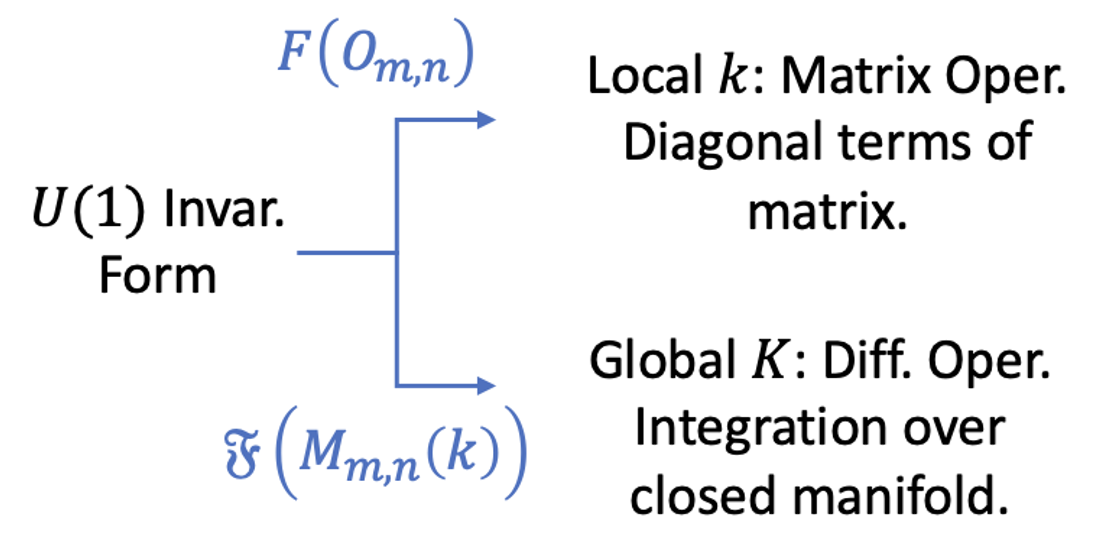}
\caption{\label{fig:epsart}(color online): Gauge symmetry becomes the fundamental principle of deriving observables from matrix and differential operators. To ensure gauge symmetry, function $F$ (functional $\mathfrak{F}$) is applied to matrices of matrix (differential) operators. \label{f2}}
\end{figure}

Recall that an important reason for $F(O_{m,n})$ to be an observable is its U(1) symmetry; off-diagonal terms are unqualified for observables in view of the absence of U(1). In a similar fashion, if we find another U(1) invariance form $\mathfrak{F}(M_{m,n}(k))$, which works for differential operators, $\mathfrak{F}(M_{m,n}(k))$ should be treated with equity as $F(O_{m,n})$. That is, $\mathfrak{F}$ is not subject to $F$, nor deduced by $F$. U(1) could be extended to U($N$) invariance form, such as Eq.~\ref{eq143}, by including trace.


We shall call for attention on the following hinged aspects:
\vskip 0.5mm
(1) Differential operators are not associative, fundamentally distinct from matrix operators
\vskip 0.5mm
(2) Distinct transformation rules for matrices of the two types of operators
\vskip 0.5mm
(3) Different meanings of U(1) symmetry
\vskip 0.5mm
(4) Means of extracting observables
\vskip 0.5mm

Differential operator is not associative but “one-sided”, never equivalent to matrix operators. Consequently, the matrix of differential operators features a distinct transition rule with an extra term (Eq.~\ref{eq92}). The inhomogeneous transformation renders a different meaning of U(1) invariance since it is a notion (Def. 9) associated with specific functionals. Although both matrix and differential operators have U(1) symmetry, their functionals are different, the meanings of U(1) is not the same. Observable is about constructing U(1) gauge since the functional are different, the two operators have different ways of extracting observables. 

Thus, $\mathfrak{F}$ is implicitly linked to behaviors of $M_{m,n}(k)$ and its transformation symmetries, and the way of extracting observable Eq.~\ref{eq159} is a generic consequence for differential operators, not limited to $\hat{r}$.

\section{7. Discussion \& Outlook.}
\textbf{The spaces related to $r$-matrix}. In this work, we transcribe $\hat{r}$, originally defined as an linear infinite-dimensional operator in space $\mathcal{H}$, to a ribbon space $K{\otimes}\mathcal{V}$, in which $\hat{r}$ exhibits as a finite-dimensional matrix, namely CRM. In the course, several spaces are involved as summarized in Table~\ref{tab5}.

CRM arises from $\hat{r}$, while it is not a representation of Weyl algebra (Sec. 2), thus not a representation of generator $\hat{r}$. CRM should be viewed as a matrix incarnation of $\hat{r}$ that encodes the information to evaluate ${\langle}\hat{r}{\rangle}$. CRM is hosted by a “ribbon space” spanned by ribbon bases $\mathcal{I}$ (Def. 3) just as a vector is hosted by a vector space spanned by vector bases. The ribbon basis is a product space $K{\otimes}\mathcal{V}$ (which may or may not form a vector space relying on whether $K$ is a vector space). In $K{\otimes}\mathcal{V}$, notions such as ribbon $T_R$ and gauge transformations $T_G$ emerge.

\begin{table}
\caption{\label{tab:table5} Summary of spaces relevant to CRM, in terms of dimensions, identity of matrix as map, and transformation defined in the space. The dimension of $\mathcal{H}$ is uncountable infinite, while matrix requires its labels to be either finite or countably infinite; thus matrix is ill-defined in space $\mathcal{H}$. Matrices are well defined in other spaces, while their meanings (in terms of maps) vary with spaces. The ribbon space provides a platform that allows to express both differential operators and matrix operators. \label{tab5}}
\begin{ruledtabular}
\begin{tabular}{c c c c c}
\multicolumn{1}{c}{} & \multicolumn{3}{c}{Vector space} & \multicolumn{1}{c}{Ribbon space} \\
\hline
  & $\mathcal{H}$ & $\mathcal{H}_B$ & $\mathcal{V}$ & $K{\otimes}\mathcal{V}$ \\
\hline
Dim. & $\infty$ & $N{\times}\mathcal{N}$ & $\mathcal{N}$ & dim($K$)$+\mathcal{N}$ \\
Matrix & -- & $\mathcal{H}_B{\rightarrow}\mathcal{H}_B$ & $\mathcal{V}{\rightarrow}\mathcal{V}$ & $K{\rightarrow}f$, $f:\mathcal{V}{\rightarrow}\mathcal{V}$ \\
Transf. & $T_B$ & $T_B$ & $T_B$ & $T_R$ \& $T_G$ \\
\end{tabular}
\end{ruledtabular}
\end{table}

CRM is an array of matrices parameterized with continuous $k$. Formally speaking, it is map $K{\rightarrow}f$, which involves two spaces $K$ and $\mathcal{V}$. Thus, there are two dimension-like quantities associated with CRM: its dimension dim($K$)$+{\mathcal{N}}$ and its rank $\mathcal{N}$. Accordingly, Hamiltonian $H$ (originally defined in $\mathcal{H}$) is incarnated by $H_{m,n}(k)$ defined in the ribbon space. Rigorously speaking, $H_{m,n}(k)$ is not a representation of $H$, since spaces $\mathcal{H}$ and $K{\otimes}\mathcal{V}$ are not of the same dimension. 

Although $\hat{r}$ is a linear map in $\mathcal{H}$, its incarnation CRM in $K{\otimes}\mathcal{V}$ loses linearity, reflected by CRM's transformation featuring an inhomogeneous term (Def. 7). That is, the transcribing will not maintain linearity, exactly because the transcribing procedure is different from a basis transformation. Representation of linear operators under different bases are linked by reversible transformations, which should preserve the dimension of spaces. However, the dimension is decreasing from $\infty$ in $\mathcal{H}$ to $N\times\mathcal{N}$ in $\mathcal{H}_B$, and finally to dim($K$)$+\mathcal{N}$ in $K{\times}\mathcal{V}$. Consider two bands in a 1D B.Z., a vector in $\mathcal{H}_B$ is denoted with $2{\times}N$ components $(c_{1,k_1},{\ldots}c_{1,k_N}, c_{2,k_1},{\ldots},c_{2,k_N})$, while in $K{\times}\mathcal{V}$, a point is denoted by $(k,c_1,c_2)$, which is three dimensional. Continuity is crucial in reducing the dimension from $N{\times}\mathcal{N}$ to dim($K$)$+\mathcal{N}$ (Appx. E). One cannot simultaneously maintain the linearity and achieve the convergence. 
Whether DRM or CRM is adopted, it does not form a representation of Weyl algebra for its non-existence mentioned in Sec. 2. 

\vskip 0.5mm

\textbf{The principles of defining $r$-matrix}. Matrices are usually defined as representations of operators, for which the interchangeable terms ``matrix" and ``operator" are often acceptable. To be representations, matrices should reproduce the defining algebra. For example, Pauli matrices are representations of spin (matrix operator) that satisfy Lie algebra. Formally speaking, we seek map from operators to matrices that preserves Lie brackets. Such a structure-preserving map is named isomorphism, \cite{38} which is the principle in defining the matrix for operators (Table~\ref{tab6}). 
\begin{table*}
\caption{\label{tab:table6} Comparison of matrices for $\hat{r}$ operator, spin, and group, in terms of whether the matrices form isomorphism, what structure to be preserved, and the host space. Matrices of spin and group have a common defining principle: isomorphism; but they differ in the structure to be preserved by isomorphism. However, isomorphism is not the principle of defining $r$-matrix, which makes $r$-matrix distinctive from spin and group matrices. \label{tab6}}
\begin{ruledtabular}
\begin{tabular}{c c c c}
 & Matrix of $\hat{r}$ & Matrix of spin & Matrix of group \\
\hline           
Oper. & $\hat{r}$, $k$ & $s_1$, $s_2$, $s_3$ & ${\lbrace}g|g{\in}G{\rbrace}$ \\
Isomorphism  & No & Yes & Yes \\
Str. to preserve & -- & $[s_i,s_j]={\epsilon}_{i,j,k}s_k$ & $M(g_1)M(g_2)=M(g_1{\circ}g_2)$  \\
Space & Ribbon space $V_{\mathscr{R}}$ & Vector space & Vector space \\
\end{tabular}
\end{ruledtabular}
\end{table*}

Another well-known matrix isomorphism is group representation, which describes abstract groups in terms of bijective linear transformations of a vector space to itself (i.e., vector space automorphisms); in particular, they can be used to represent group elements as invertible matrices so that the group operation can be represented by matrix multiplication.

However, matrix for $\hat{r}$ (differential operator) invokes distinct defining principles since matrix isomorphism does not exist for Weyl algebra (Sec. 2). We still build a map from $\hat{r}$ to Hermitian matrices, and this map is not meant to preserve the algebra (that is why we find $r_{m,j}k_{j,n}-k_{m,j}r_{j,n}=0$ in Eq.~\ref{eq78}, while Weyl algebra has $\hat{r}k-k\hat{r}=i$); however, this does not prevent the operator’s information being encoded in the matrix. That is, via the matrix, one can deduce the observable, except for the means of extracting observable is different from matrix operators, since the encoding way is changed.

For $r$-matrix, the transformation rule is inhomogeneous (Eq.~\ref{eq92}), which is impossible to preserve the commutator $[\hat{r},k]=i\hbar$. In other words, the commutator is not invariant under $T_G$. This is a significant difference between Lie algebra and Weyl algebra. Despite the lack of matrix isomorphism, we may still define matrix for differential operators, and it interacts with ribbons in a similar fashion like a “genuine” matrix. 
\vskip 0.5mm

\textbf{Two types of operators}. We shall discriminate two types of operators: matrix operators (e.g., spin) and differential operators (e.g., $\hat{r}$). Their properties and designations are compared in Sec. 5. The two operators are expressed by matrices that are equipped with distinct transformation rules (see Def. 6, 7). Differential operators “appear like a matrix” until ribbon transformation is performed; just like coefficients of a vector are no more than “a column of numbers” until basis transformation is performed when covariance and contravariance manifest. Thus, operator is not just about its matrix elements but also their transformation rules. 

Then why the two matrices follow the rules specified by Def. 6, 7? Why are these particular forms of transformation chosen? This arises from the basic algebras for the two operators. Matrix operator is associative, i.e., $(AB)C=A(BC)$. One is at liberty to act $B$ on the left or the right first and ends up with the same outcome. Then, the matrix elements 
\begin{equation}
\begin{split}
&{\langle}m|\hat{O}|n{\rangle}{\rightarrow}{\langle}m'|\hat{O}|n'{\rangle}={\langle}m|U\hat{O}U^{\dagger}|n{\rangle} \\
&={\langle}m|U|i{\rangle}{\langle}i|\hat{O}|j{\rangle}{\langle}j|U^{\dagger}|n{\rangle}=U_{m,i}O_{i,j}U_{j,n}^{\dagger}
\end{split}
\label{eq161}
\end{equation}
This is exactly the expression for Eq.~\ref{eq91} in Def. 6. The form of Eq.~\ref{eq91} is interpreted as the transformation rule is the operator’s matrix against the updated bases.

However, we shall not take the above procedure for granted. It is only true when $\hat{O}$ is associative. Otherwise, e.g., ${\partial}_k$, which acts on only one side (conventionally on the right), when we try to repeat the procedure as Eq.~\ref{eq161}, we encounter a different situation.
\begin{equation}
\begin{split}
{\langle}m(k)|i{\partial}_kn(k){\rangle}{\rightarrow}&{\langle}m'(k)|i{\partial}_kn'(k){\rangle} \\
=&{\langle}m'(k)|U(k)i{\partial}_kU^{\dagger}(k)n(k){\rangle}.
\end{split}
\label{eq162}
\end{equation}

Obviously, there will be an extra term on the right side, which arises from ${\partial}_k$ being non-associative and obeying ${\partial}_k({\varphi}(k){\psi}(k))={\partial}_k({\varphi}(k)){\psi}(k)+{\varphi}(k){\partial}_k({\psi}(k))$, namely Leibniz rule. If we write the above terms in matrix forms, it is exactly the Def. 7. As such the algebra differences in associative property is incarnated by the matrix’s transformation. In terms of abstract algebra, the particular form of transformation introduced by Def. 7 stems from the Leibniz rule imposed on ${\partial}_k$.

One may wonder what if we can define a generalized differential operator that may act on both sides, such that the associative rule will be recovered for ${\partial}_k$? Unfortunately, this is unachievable due to the intrinsic properties of ${\partial}_k$. Consider two-sided actions: ${\partial}_k$ acting on the right and $\overset{\leftarrow}{\partial}$ acting on the left. Clearly
\begin{equation}
\begin{split}
c{\partial}_k{\varphi}(k){\neq}(c\overset{\leftarrow}{\partial}){\varphi}(k)=0,
\end{split}
\label{eq163}
\end{equation}
where $c$ is a $k$-independent constant. That means ${\partial}_k$ acting left or right sides first will lead to distinct results, and thus associative rule is violated. Thus, differential operator is incompatible with matrix operator in its bottom algebra. It is impossible to turn a differential operator into a matrix. Therefore, one should distinguish the terms “operators” and “matrices of operators”. This distinction is trivial when only matrix operators are involved. 
\vskip 0.5mm

\textbf{Basis-free designation vs matrix designation}. The bra/ket designation could be entirely substituted by matrix designation such as $|{\varphi}{\rangle}{\rightarrow}{\varphi}_n$, ${\langle}m(k)|i{\partial}_kn(k){\rangle}{\rightarrow}A_{m,n}(k)$, etc. Gauge transformation $T_G$ is directly defined with Eq.~\ref{eq132}. about matrix, without reference to first multiplying $|{\varphi}{\rangle}$ with a phase $e^{i{\xi}(k)}$ (or a unitary operator) then deducing $T_G$’s form. Thus, Def. 5-9 are directly established on matrices, not referring to “matrix’s definition” in terms of bra/ket. 

The bra/ket may lead to mistakes when differential operators are present (Sec. 5).  Differential operators commonly exist in transport problems, relativistic quantum mechanics, gauge field theory, etc., wherein matrix could be a better designation. \cite{46} However, bra/ket designation is elegant for matrix operators and broadly used. Thus, in Sec 5. we list the “translation”: matrix’s “definition” in terms of bra/ket inner product. However, it never suggests that one must refer to the “internal structures” of matrices (definition begins with matrix elements, not need to refer to the “origin” of these matrices). Gauge transformation, basis transformation, and extracting observables can all be handled with pure matrix designations. 

Thus, we switch to matrices for a common denotation for both matrix operator and differential operator. Accordingly, the matrix is established on ribbon bases in ribbon space (or bundle space) as a generalization of bases of a vector space. Because of the fundamental algebraic distinctions, the matrices of the two types of operators follow different rules of transformation under different ribbons. 
\vskip 0.5mm

\textbf{Application \& Outlook}. Why can the fundamentals about $r$-matrix give insights for a transport theory? We give two examples, each deserving extensive discussions in a separate work. The first is regarding two arguments established earlier: (i) isomorphism $\mathcal{H}_B{\cong}{\mathcal{V}{\otimes}E}$ (just the $\Pi$ map in Sec. 3); (ii) one-to-one correspondence between $|u_{n,k}{\rangle}$ and vectors in the quotient space $\mathcal{V}$ of $\mathcal{H}_B$ (space spanned by $|u_{n,k}{\rangle}~{\cong}\mathcal{V}$ in Sec. 4). Argument (i) links $\mathcal{V}$ with $\mathcal{H}_B$ as its quotient space, and argument (ii) relates space spanned by $|u_{n,k}{\rangle}$ with $\mathcal{V}$ by isomorphism; such that vector space spanned by $|u_{n,k} {\rangle}$ can be transcribed to the original physical space $\mathcal{H}_B$. Although such a linkage does not influence the evaluation of, for instance, Berry connection or curvature, it provides counting for vectors in $\mathcal{H}_B$ and $\mathcal{V}$ spaces to shed light on phenomenon’s essence. 

To be concrete, consider the adiabatic current $J_d$ evaluated by integration of Berry connection \cite{21,25}
\begin{equation}
\begin{split}
J_d&={\partial}_tP=-\frac{e}{V_{\text{cell}}}{\partial}_t{\langle}\hat{r}{\rangle}_n \\
&=-\frac{e}{V_{\text{cell}}}{\partial}_t\frac{1}{2\pi}{\oint}r_{n,n}(k){\cdot}dk.
\end{split}
\label{eq164}
\end{equation}
What is the input needed for evaluating $J_d$? Eq.~\ref{eq164} relies on $r_{n,n}(k)$ over the entire B.Z. Remember $r_{n,n}(k)$  is the reduced $r$-matrix evaluated with $|u_{n,k}{\rangle}$ in quotient space $\mathcal{V}$. Thus, to understand what $|u_{n,k}{\rangle}$ physically represents, one ought to apply $\Pi$ inversely (called $\Pi_k^*$ map temporally) to yield their pre-images in the original physical space $\mathcal{H}_B$. It can be proved that $|u_{n,k}{\rangle}$ always corresponds to $N$ mutually orthogonal vectors in $\mathcal{H}_B$ (even when $|u_{n,k}{\rangle}$ is constant with $k$). Since the dimension of subspace in $\mathcal{H}_B$ corresponds to particle numbers (orthogonality is due to exclusion principle)
\begin{equation}
\begin{split}
\#(\text{particles})=\text{dim}({\lbrace}{\psi}_k{\rbrace}_{\text{occ}}),
\end{split}
\label{eq165}
\end{equation}
it indicates that the adiabatic current $J_d$ is an $N$-particle phenomenon. Because of that, given full knowledge of a single-particle wave function ${\varphi}(t)$, it is still insufficient to determine $J_d$ with Eq.~\ref{eq164}. In previous formulation, ${\langle}\hat{r}{\rangle}_n$ is evaluated with wavefunction $u_{n,k} (r)$ coordinated by $r$. Thus, one tends to (mistakenly) attribute $J_d$ to a single-particle effect, since $u_{n,k}(r)$ can be extracted from a single-particle Bloch wave ${\psi}_{n,k}(r)$. As such, many-body feature in $J_d$ is concealed.

Note that counting the physical states must be established on vector space (regarding the dimensions of a subspace). It is ill-defined to ask, “how many particles are involved for functions $u_{n,k}(r)$”. That is why arguments (i)(ii) providing counting for vectors is important. In the adiabatic limit, the corresponding of $u_{n,k}(r)$ to $N$ vectors has a simple interpretation: just the $N$ particles occupied by a band ($N$ is the unit cell numbers). In general, the correspondence to $N$ orthogonal states may apply to non-equilibrium, whether inter-band hopping is involved or whether $k$ is preserved. As such, we realize that the $N$ particles have formed a bounded “unit”, even when some of them are excited to a different band or different $k$. This $N$-particle picture gives insight about the description of electronic states in crystals.

Moreover, arguments (i)(ii) concern the stability for adiabatic currents, as $J_d$ will lose gauge invariance even with a single particle missing. In other words, gauge invariance is fragile with particle number variation, e.g., $N{\pm}1$. Then, an intriguing allusion is yielded: the $N$-particle $J_d$ is correlated, although these $N$ particles are non-interacting; that is, correlation still exists given interaction is all removed. In contrast, the traditional wisdom is that correlation exclusively arises from interaction, and free particle is uncorrelated. It is known Berry phase theory of polarization \cite{21} links $J_d$ and its transported charges with global topology. Now with arguments (i)(ii), we reveal that the physical meaning of global topology is the $N$-body correlated effects.

As the second example, we show why the variable extraction from $r$-matrix (functional $\mathfrak{F}$ in Sec. 6) is linked to transport theory. One is often overwhelmed by the diverse transport mechanisms: shift current, \cite{4,8,18} injection current, \cite{4,8,26} adiabatic current \cite{20}; and has to resort to a case study to recognize which mechanism is in effect. Such a situation owns its origin to lacking an unambiguous way of determining current as an observable. In contrast, spin’s expectation value is 
\begin{equation}
\begin{split}
{\langle}\hat{s}{\rangle}={\langle}{\varphi}(t)|\hat{s}|\varphi(t){\rangle},
\end{split}
\label{eq166}
\end{equation}
which is independent of whether ${\varphi}(t)$ changes slowly or fast. However, such a generic definition is lacking for current because of the divergence of $r$-matrix makes $J{\propto}{\partial}_t{\langle}\hat{r}{\rangle}$ ill-defined. Thus, the divergence of $r$-matrix leads to vagueness in extracting observable for $\hat{r}$, leading to ill-defined $J$ with ${\partial}_t{\langle}\hat{r}{\rangle}$, leading to the diverse transport mechanisms that involve different approximations or physical intuitions (e.g., electronic hoping is fast or slow, inter-band or intra-band), leading to the inconsistency among different mechanisms from the point of view of being a single-particle effect or an $N$-particle one, being correlated or not. Thus, finding the converging matrix to recover the original definition $J{\propto}{\partial}_t{\langle}\hat{r}{\rangle}$ is crucial for developing a unified transport theory.

To be concrete, consider adiabatic current $J_d$ and shift current $J_s$. We realize there is a “gulf”. For $J_s$, it is evaluated by Eq.~\ref{eq4}, in which the domain of integration $\int$ is arbitrary, because gauge invariance of $J_s$ remains whether the integration domain is closed or not, connected or not, even on a single $k$ point. In other words, $J_s$ only depends on the initial and final states of hopping, which are two discrete points along the evolution wave function $\varphi(t)$, such that information of $J_s$, within the $J_s$ formulation, is fully encoded in $\varphi(t)$ and thus is a single-particle phenomenon. On the other hand, $J_d$ requires stringently closed $\oint$ and exhibits as an $N$-particle phenomenon, as proceeding discussion suggested, fundamentally different from $J_s$.

Remember there is only “one” current in crystals. Adiabatic current $J_d$, shift current $J_s$, etc. are merely artificial classifications. $J_d$ and $J_s$ are to handle the slow and fast changing Hamiltonians, respectively, characterized by external driving frequencies $\hbar{\omega}$. By tuning $\hbar{\omega}$, the current should gradually switch from one formulation to the other. But how can a single-particle $J_s$ cross the gulf to continuously connect with an $N$-particle $J_d$? Note that both $J_d$ and $J_s$ are non-interacting, irrelevant to interaction causing emergent collective states for electrons. Namely, the gulf is that at high  $\hbar{\omega}$ information of current is encoded in wavefunction of a single particle, while at low $\hbar{\omega}$ one has to know the states of all $N$ particles (without missing any of them) to determine the current -- no known mechanism can realize this. Moreover, it is hard to imagine the transition regime when $\hbar{\omega}$ is intermediate. The current situation is a reminiscent of the inconsistency between quantum and relativity theories in terms of the fundamental description of the space and events: whether local or non-local laws are paramount in universe. The central question is how to reconcile a non-local theory at quantum scales with a local theory at larger scales that describe the same universe. Here, we consider a much modest question: how to reconcile the two transport theories that describe the same non-interacting current, while one is about single-particle and the other is about $N$ particles. Therefore, removing divergence of $r$-matrix and finding the way of extracting observables from CRM will help judge which picture is correct. Our theoretical framework is poised to enhance our understanding of the photocurrent and phonon responses exhibited by topological materials, which are currently the focus of active exploration in THz and ultrafast experiments \cite{51,52,53}.

In short, we should not be just content for an evaluable formulation for currents but also examine (a) whether the formulation is generic and unique? (b) Whether formulations for different limiting situations are compatible and can crossover to each other? (c) Whether the formulation is stable? For example, $J_d$ requires $\oint$; whether the formulation is robust against particle missing in B.Z. In order to address these issues, it boils down to understanding $r$-matrix, the relations between different involved spaces, and also the way of extracting observables.

\section{8. Summary \& Conclusion}
This work surveys the definitions of $\hat{r}$ operator and the $r$-matrix, addressing why although \textit{no} matrix may satisfy the commutation $[r,p]=i\hbar$, a matrix can still be assigned to $\hat{r}$. This involves a fundamental question: what is the defining principle of $r$-matrix? Subject to that principle, we further wonder if one could find CRM to substitute for the well-known diverging DRM, motivated by a belief: the matrix of a physical operator should not diverge. In the CRM to be derived, every element should be finite; the dimensions of CRM are finite and arbitrary. 

In Sec. 2, we first introduce the math involved in defining $\hat{r}$: Weyl algebra, which is characterized by the number of conjugated variable pairs, denoted as $A_N$. Then demonstrate DRM does not satisfy $[r,p]=i\hbar$; indeed, no matrix can satisfy the commutation equation. Thus, a different principle for defining $r$-matrix is needed.
 
In Sec. 3, we first show 1-st Weyl algebra $A_1$ (the familiar substitution $\hat{r}{\rightarrow}i{\partial}_k$) inevitably leads to divergence in $r$-matrix. Then we show the divergence could be resolved by $N$-th Weyl algebra $A_N$ to substitute for $A_1$. A key modification is (Eq.~\ref{eq57})
\begin{equation}
\text{CRM:}~\hat{r}~{\rightarrow}~i{\partial}_{k_1}+i{\partial}_{k_2}+{\cdots}+i{\partial}_{k_N}.
\end{equation}
Note that $A_1{\rightarrow}A_N$ is merely substituting generators of Weyl algebra, not the entire modification, since the space for generators to act upon must adjust too. (Appx. F) For that, we introduce three spaces $\mathcal{H}$, $\mathcal{H}_B$ and $\mathcal{V}$, on top of which a space $\mathcal{V}{\otimes}E$ (\textit{ribbon space} as declared latter) for $A_N$ to act on is constructed before we derive CRM. The constructing is essentially about $\Pi$ map $\mathcal{H}_B{\rightarrow}\mathcal{V}{\otimes}E$, with which we are able to show: (1) Bloch space $\mathcal{H}_B$ proves incomplete for $\hat{r}$ by finding counterexamples (Appx. B); (2) rigorously, inter products such as ${\langle}r|{\psi}_{n,k}{\rangle}$, ${\langle}r|u_{n,k}{\rangle}$ are not accurate, because $|r{\rangle}{\in}{\mathcal{H}}$, $|{\psi}_{n,k}{\rangle}{\in}\mathcal{H}_B$, $|u_{n,k}{\rangle}{\in}\mathcal{V}$ and inner products cannot be defined between vectors belonging to different vector spaces.

In Sec. 4, by acting $N$-th Weyl algebra on the product space $\mathcal{V}{\otimes}E$, we obtain the explicit forms of CRM as Eq.~\ref{eq59}. As two facets of CRM, the “$r$-matrix” and “reduced $r$-matrix” are discriminately introduced linked to Bloch space $\mathcal{H}_B$ and its quotient space $\mathcal{V}$, respectively (Eq.~\ref{eq73},\ref{eq74}). A corollary is achieved that geometric quantities (e.g., Berry connection) are associated with the quotient space $\mathcal{V}$, instead of Bloch space $\mathcal{H}_B$. CRM and DRM are discussed in aspects of what has caused the divergence, how the divergence is resolved, DRM not being a special case of CRM, etc.

In Sec. 5, we show matrices defined through ($N$-th) Weyl algebra and Lie algebra (like spin) will display different properties in transformation and other aspects. As a consequence, two types of operators are recognized: matrix operator (Def. 6) and differential operator (Def. 7), which can be unified under a platform ``ribbon space". The unifying leads to fortuitous discoveries. For example, We show $r_{m,n}=r_{n,m}^*{\nLeftrightarrow}\hat{r}=\hat{r}^{\dagger}$, which subtly affects the well-known Berry curvature formula for polarization (Eq.~\ref{eq94},\ref{eq107}). Designation system must adjust to suit the different transformation properties. We find the ket/bra designations (perfectly workable for matrix operator) might encounter ambiguity for differential operators in certain situations. (Table~\ref{tab3})

In Sec. 6, we extensively discuss the space that hosts CRM: ribbon space, in which ribbon transformation $T_R$ is introduced in analog with basis transformation $T_B$ in a vector space. Particularly, we show how gauge transformation $T_G$ and gauge symmetry make entrance as a natural consequence of $T_R$. We give a formal definition for $T_G$ (Def. 8) associated with a manifold $K$ and gauge symmetry U($N$). Existing gauge-invariant formulations could be classified and distinguished under the present frame. Noteworthy, although labeled with ``gauge invariance", formulations can vary significantly in meanings depending on distinct $K$ and associated gauge symmetries. We further show that $T_G$ owes its origin to differential operators; on the other hand, $T_G$ is only trivially defined for matrix operator. We explain the relationships among ribbon transformation $T_R$, gauge transformation $T_G$ and basis transformation $T_B$. We address why U($1$) gauge symmetry is necessary for an observable, and whether U($N$) gauge symmetry is necessary too? 

In Sec. 7, we review the journey: from generators of Weyl algebra, to the various spaces involved, to the principles of defining CRM on these spaces, to the designation and symbolism, to observable extraction. Remarkably, setting out from the basic definition of $\hat{r}$, a series of concepts (involving geometry, transport, gauge, etc.) will emerge and become intertwined. We reveal two pathways with hinged aspects: 
$$
\mathcal{H}~{\sim}~A_1~{\sim}~\text{DRM}~{\sim}\text{ambiguity in}~{\langle}\hat{r}{\rangle}
$$
$$
\mathcal{V}{\otimes}E~{\sim}~A_N~{\sim}~\text{CRM}~{\sim}\text{unambiguity in}~{\langle}\hat{r}{\rangle}.
$$
Accepting one pathway means one must accept all the related aspects. Accurately speaking, we are not denying the diverging nature of DRM in the original space $\mathcal{H}$, but discover a potentially unique way of resoling the ambiguity (arising from this divergence) in defining $r$-matrix and in obtaining observable $r$. This approach aligns harmoniously with existing arguments regarding the Berry connection, Wannier centers, and other related concepts; and will give more.

Last but never least, we outlook how the CRM, seemingly an abstract notion, is related to concrete applications in transport. The divergence of $r$-matrix necessitates alternative ways of yielding ${\langle}\hat{r}{\rangle}$ based on different approximations, which unfortunately turn out non-unique, leading to diverse transport formalisms. On the other hand, resolving the divergence in $r$-matrix gives a logically unique way of yielding ${\langle}\hat{r}{\rangle}$ thus a unique direction in building transport theory. We shall stress this does not indicate the existing transport formalism is incorrect in view of the success each has achieved. However, they might correspond to different expansion limits of certain unified theory. Moreover, understanding how one transport formalism crossovers to another should bring deeper physical insights such as the description of electronic states in crystals.



\section{Appendix}
\subsection{A. Summary of notation.}
$\hat{r}$~~~~~Position operator

$r$~~~~~Real space coordinate or eigenvalues of $\hat{r}$.

$\mathscr{R}$~~~~Ribbon band (map).

$J_d$~~~~Adiabatic current (or displacement current).

$J_s$~~~~Shift current.

$J_i$~~~~Injection current.
\vskip 1.5mm

$\varphi(t)$~~Generic time-dependent wave function. 

${\psi}_{n,k}(r)$~~Bloch wave functions (function of $r$) associated with band $n$ and crystal momentum $k$.

$u_{n,k}(r)$~~Periodic part of Bloch wave function ${\psi}_{n,k}(r)$. 

$|{\psi}_{n,k}{\rangle}$ Vector in Bloch space $\mathcal{H}_B$.

$|A_{n,k}{\rangle}$ (or $|u_{n,k}{\rangle}$)~~Vector in space $\mathcal{V}$ (different from $u_{n,k}(r)$ which is yet a vector).

$|E_k{\rangle}$ Vector in quotient space $E$
\vskip 1.5mm

$V$~~~~~Vector space (general)

$\mathcal{H}$~~~~~Space spanned by eigenstates of $\hat{r}$.

$\mathcal{H}_B$~~~Bloch space.

$\mathcal{H}_W$~~~Wannier space.

$\mathcal{V}$~~~~~Quotient space of Bloch space $\mathcal{H}_B$ associated with bands.

$E$~~~~~Quotient space of $\mathcal{H}_B$ associated with $k$.

$V_{\mathscr{R}}$~~~Ribbon space.

$V_{O}$~~~Matrix operator space.
\vskip 1.5mm

$K$~~~~~Brillouin Zone (otherwise a generic smooth manifold).

$I_B$~~~~A set of bases of Bloch space $\mathcal{H}_B$.

$\mathcal{I}$~~~~~A set of ribbon bases.

$\Pi$~~~~~Projection map: $I_B{\rightarrow}{\mathcal{V}}{\otimes}E$.

$\Pi_1$~~~~Projection map: $I_B{\rightarrow}{\mathcal{V}}$.

$\Pi_2$~~~~Projection map: $I_B{\rightarrow}E$.
\vskip 1.5mm

$\hat{O}$~~~~~Basis-free designation of matrix operator.

$O_{m,n}$~~Matrix elements of matrix operator.

$M_{m,n}$~~Matrix elements of differential operator.
\vskip 1.5mm

$A_1$~~~~1-th Weyl algebra.

$A_N$~~~$N$-th Weyl algebra.

$\mathcal{A}$~~~~~Linear self-conjugated maps.

$\vartheta$~~~~~Berry phase.

$\mathbb{I}$~~~~~Identity operator.

$I$~~~~~Inversion symmetry (IS) operation.

$\mathcal{T}$~~~~Time-reversal symmetry (TRS) operation.

$\mathfrak{T}$~~~~Time-ordered operator.

$\Re$ ($\Im$)~~Real (imaginary) part.

$k$~~~~~crystal momentum of no particular dimension.

$\textbf{k}$~~~~~Crystal momentum of 2D or 3D.

$\sigma_i$~~~~Pauli matrices representing spin.

$\tau_i$~~~~Pauli matrices representing pseudo-spin.
\vskip 1.5mm

$\mathfrak{r}_{m,n}(k,k')$~~Converging $r$-matrix.

$r_{m,n}^{(\mathcal{N})}(k)$~~~~~~~Reduced $r$-matrix of $\mathcal{N}$-dimension ($r_{m,n}(k)$ is the shorthand notation).

$A_{m,n}(k)$~~~~Berry connection matrix.

$R_{m,n}(k)$~~~~Shift vector ($k$-dependent) from band $n$ and $m$.

$\mathfrak{X}_{m,n}(k)$~~~~Complementary term of shift vector $R_{m,n}$.

$\gamma_{m,n}(k)$~~~~Pumping rate from band $n$ to $m$ at $k$.
\vskip 1.5mm

$:=$~~~Define to be

$\cong$~~~Isomorphic

Aut~~~Automorphism
\vskip 1.5mm

$\zeta_n$~~~~The parity for $n^{th}$ band. 

$T_R$~~~~Ribbon transformation.

$T_B$~~~~Basis transformation.

$T_G$~~~~Gauge transformation.

$F$~~~~~Observable function for matrix operator

$\mathfrak{F}$~~~~~Observable functional for differential operator
\vskip 1.5mm

$\aleph_0$~~~Aleph-null as cardinality of $\mathbb N$.

$\mathbb{N}$~~~~Set of natural numbers (non-negative integers).

$\mathbb{Z}$~~~~Set of integers.

$\mathbb{Q}$~~~~Set of rational numbers.

$\mathbb{R}$~~~~Set of real numbers.

$\mathbb{C}$~~~~Set of complex numbers.

\subsection{B. Proof for the incompleteness of the Bloch space $\mathcal{H}_B$ with respect to $\hat{r}$.}

We aim to prove the incompleteness of $\mathcal{H}_B$ for $\hat{r}$. When Bloch space $\mathcal{H}_B$ is the Hilbert space, it is complete for arbitrary operators defined within it. However, if an operator is defined beyond $\mathcal{H}_B$, it becomes incomplete. Thus, completeness is associated with specific operators, not taken for granted. An “easy” showing of incompleteness is by comparing the dimensions of $\mathcal{H}_B$ with $\hat{r}$: the former is countably infinite, while the latter is uncountably infinite. Although both are infinite, $r$ is even ``more", such that the space required to host $\hat{r}$ is larger than $\mathcal{H}_B$, and $\mathcal{H}_B$ is incomplete. 

To make precise the above statement, we ought to introduce “cardinality”, which extends the measure of the number of elements in a set from finite to infinity. For finite sets, the cardinality could be replaced by “the number”. For infinite set, the cardinality of the natural numbers is denoted as ${\aleph}_0$. A set has cardinality ${\aleph}_0$ if and only if it is countably infinite, that is, there is a bijection between it and the natural numbers, such as integer number $\mathbb{Z}$, rational number $\mathbb{Q}$. Occasionally, it leads to counterintuitive results. For example, although even natural number is a proper subset of $\mathbb{N}$, i.e., even natural number is smaller than $\mathbb{N}$, the cardinalities of the two are equal, because one could build one-to-one map: $n{\mapsto}2n$. Thus, it is possible that a proper subset of an infinite set has the same cardinality as the original set, which is impossible for proper subsets of finite sets. 

The set of all finite ordinals, called $\omega$, which has cardinality ${\aleph}_0$. The label $n,k$ in ${\psi}_{n,k}$ is, at most, with ordinality of ${\omega}^2$ has cardinality ${\aleph}_0$, when both $n,k$ take infinitely many values. On the other hand, if band $n$ is finite, it is $n{\omega}$. Thus, it cannot be equal to the cardinality of real number $\mathbb{R}$. Therefore, the space “shrinks” and cannot be isomorphic to the space spanned by the eigenstates of $\hat{r}$.

To gain more evidence for incompleteness, we prove it from a different angle. That is, if Bloch space $\mathcal{H}_B$ is complete for $\hat{r}$, ${\psi}_{n,k}$ can expand arbitrary functions defined on $\mathbb{R}$ (all eigenvalues of $\hat{r}$). On the other hand, if we construct a set of Bloch bases and show there exists function $f(r)$ that cannot be expanded by this set of bases, $\mathcal{H}_B$ is incomplete for $\hat{r}$.

Firstly, we shall define Bloch functions and Bloch bases. Bloch function refers to function forms ${\psi}(r):{\mathbb{R}}{\rightarrow}{\mathbb{C}}$ that satisfy
\begin{equation}
\begin{split}
{\psi}_k(r)=e^{ikr}u(r),
\end{split}
\label{eq167}
\end{equation}
where $u(r)=u(r+a)$. Then, ${\psi}_k(r)$ is called a Bloch function subject to wavevector $k$ and periodicity $a$. Bloch function specifies a certain form of functions, which has yet involved vector space. 

The Bloch basis is a notion associated with a vector space. Bloch bases are a (finite or infinite) set of mutually orthogonal Bloch functions. Mind it is mistaken to think Bloch bases are formed by all Bloch functions. Because firstly arbitrary two Bloch functions might not be orthogonal; secondly, “all” involves vagueness, and to be a vector space, the dimension needs to be well-specified (whether it is finite or infinite).

We define the following (orthogonal) functions to serve as the set of bases. 
\begin{equation}
\begin{split}
{\psi}_{n,k}(r)=
\begin{cases}
\frac{2}{\sqrt{a}}\sin(\frac{4n\pi}{a}r),~0<r<\frac{1}{2}a, \\
0,~\frac{1}{2}a<r<a,
\end{cases}
\end{split}
\label{eq168}
\end{equation}
where $n=1,2,3{\ldots}$, labelling bands. ${\psi}_{n,k}(r)$ above is defined in a single unit cell $[0,a]$. For other cells, one can find it out by a phase shift: ${\psi}_{n,k}(r+R_i)=e^{ik(r+R_i)}u_{n,k}(r+R_i)=e^{ikR_i}e^{ikr}u(r)=e^{ikR_i}{\psi}_{n,k} (r)$. By such a definition, orthogonality is satisfied. 
\begin{equation}
\begin{split}
&{\int}{\psi}_{m,k_p}^*(r){\psi}_{n,k_q}(r)dr \\
&=\sum_i^Ne^{-i(k_p-k_q)R_i}{\int}_0^a{\psi}_{m,k_p}^*(r){\psi}_{n,k_q}(r)dr \\
&=\sum_i^Ne^{-i(k_p-k_q)R_i}\frac{4}{a}{\int}_0^{\frac{a}{2}}\sin(\frac{4m\pi}{a}r)\sin(\frac{4n\pi}{a}r)dr \\
&=N{\delta}_{p,q}{\delta}_{m,n}.
\end{split}
\label{eq169}
\end{equation}
Thus, the constructed space will form an isomorphic space ${\cong}\mathcal{H}_B$ with dimension $N{\cdot}\mathcal{N}$. For infinite dimension, it is about making $N$ or $\mathcal{N}$ approach to infinity.

Clearly, within each unit cell, as designed, there is a vacuum gap, thus such a bases, which are isomorphic to $\mathcal{H}_B$, will not be able to express any function which is non-vanishing in the vacuum regime. Thus, we have shown that not arbitrary function is expandable with Bloch bases, even though we allow $N$ and $\mathcal{N}$ approach to infinity. 

One might wonder why the counterexample like Eq.~\ref{eq168} could be constructed without referring to Hamiltonian. The answer is we may use the constructed bases Eq.~\ref{eq169} to construct a desired Hamiltonian, i.e., using these bases as Hamiltonian’s eigenstates. Although it is possible to use eigenstates of Hamiltonian to specify Bloch bases, it is not necessary since Bloch bases ultimately is a notion associated with a vector space rather than to Hamiltonian operator. 

One might also wonder the constructed function should be “orbital-like”, rather than “designed”. In fact, in terms of showing the incompleteness or not, whether the basis is an “orbital” yet comes into play. Two ways of rationalizing it. First, we can use this physically weird bases to construct a physically weird Hamiltonian, under which the eigenstates are these bases. That means if we construct the Hamiltonian, these states will be physical, and no longer “weird”. The second way is we may make the vacuum gap a little more physical, not sharply vanish in $[a/2,a]$, but allows certain spread. In that case, we need to reconstruct the ${\psi}_{n,k}(r)$ within. However, the extra complexation in constructing such orbital of more physical comfort will not provide any more fundamentality. 

\subsection{C. Designations for differential operator.}
It is crucial to specify the effectual range of differential operators. We utilize brackets to indicate the range. 
\begin{equation}
\begin{split}
{\partial}_k(\ldots)
\end{split}
\label{eq170}
\end{equation}
The effectual range for ${\partial}_k$ is the expressions contained in the brackets. If there is not $({\ldots})$, it means ${\partial}_k$ will act all the way to the right most. For example,
\begin{equation}
\begin{split}
{\partial}_k({\langle}{\varphi}(k)|{\psi}(k){\rangle})={\langle}{\partial}_k{\varphi}(k)|{\psi}(k){\rangle}+{\langle}{\varphi}(k)|{\partial}_k{\psi}(k){\rangle},
\end{split}
\label{eq171}
\end{equation}
where we have defined
\begin{equation}
\begin{split}
{\langle}{\partial}_k{\varphi}(k)|&:={\partial}_k({\langle}{\varphi}(k)|) \\
|{\partial}_k{\psi}(k){\rangle}&:={\partial}_k(|{\psi}(k){\rangle}).
\end{split}
\label{eq172}
\end{equation}
Note that
\begin{equation}
\begin{split}
&{\partial}_k({\langle}{\varphi}(k)|{\psi}(k){\rangle}){\neq}{\partial}_k{\langle}{\varphi}(k)|{\psi}(k){\rangle} \\
&={\langle}{\partial}_k{\varphi}(k)|{\psi}(k){\rangle}+{\langle}{\varphi}(k)|{\partial}_k{\psi}(k){\rangle}+{\langle}{\varphi}(k)|{\psi}(k){\rangle}{\partial}_k.
\end{split}
\label{eq173}
\end{equation}
That is, without the brackets, one obtains an extra term ${\langle}{\varphi}(k)|{\psi}(k){\rangle}{\partial}_k$.

The Berry connection matrix is defined as
\begin{equation}
\begin{split}
A_{m,n}(k):&=i{\langle}m(k)|{\partial}_kn(k){\rangle}{\neq}i{\langle}m(k)|{\partial}_k|n(k){\rangle} \\ 
&=i{\langle}m(k)|{\partial}_kn(k){\rangle}+i{\langle}m(k)|n(k){\rangle}{\partial}_k.
\end{split}
\label{eq174}
\end{equation}
That means $A_{m,n}(k)$ will be taken as a “normal” matrix in terms of interacting with other matrices (just follow the rules of matrix multiplication) but behaves differently under ribbon transformations compared with a pure matrix operator. In other words, under a fixed ribbon basis, there is no difference between $A_{m,n}(k)$ and a normal matrix. 

Consider a ribbon transformation 
\begin{equation}
\begin{split}
T_R:K{\rightarrow}\text{Aut}(\mathcal{V})
\end{split}
\label{eq175}
\end{equation}
That is
\begin{equation}
\begin{split}
|n(k){\rangle}{\mapsto}U(k)|n(k){\rangle}
\end{split}
\label{eq176}
\end{equation}
We examine the transformation behavior of $r_{m,n}(k)$. The incorrect expression is 
\begin{equation}
\begin{split}
\text{Wrong}:r_{m,n}(k){\mapsto}r_{m,n}'(k)=i{\langle}m(k)|U(k){\partial}_kU^{\dagger}(k)|n(k){\rangle}.
\end{split}
\label{eq177}
\end{equation}
Because as denoted by Eq.~\ref{eq177}, ${\partial}_k$ will act all the way to the right, but the effectual range is only ${\partial}_k(U^{\dagger}(k)n(k))$. Thus, the correct denotation is
\begin{equation}
\begin{split}
r_{m,n}'(k)=i{\langle}m(k)|U(k)|{\partial}_kU^{\dagger}(k)n(k){\rangle}.
\end{split}
\label{eq178}
\end{equation}
Then, we have
\begin{equation}
\begin{split}
&i{\langle}m(k)|U(k)|{\partial}_kU^{\dagger}(k)n(k){\rangle} \\
&=i{\langle}m(k)|U(k)|j(k){\rangle}{\langle}j(k)|{\partial}_k(|l(k){\rangle}{\langle}l(k)|U^{\dagger}(k)|n(k){\rangle}) \\
&=i{\langle}m(k)|U(k)|j(k){\rangle}{\lbrace}{\rangle}j(k)|{\partial}_kl(k){\rangle}{\langle}l(k)|U^{\dagger}(k)|n(k){\rangle} \\
&+{\langle}j(k)|l(k){\rangle}{\partial}_k({\langle}l(k)|U^{\dagger}(k)|n(k){\rangle}){\rbrace} \\
&=U_{m,j}(k)r_{j,l}(k)U_{l,n}^{\dagger}(k)+U_{m,j}(k)i{\partial}_k(U_{j,n}^{\dagger}(k)),
\end{split}
\label{eq179}
\end{equation}
where
\begin{equation}
\begin{split}
&U_{m,j}(k)={\langle}m(k)|U(k)|j(k){\rangle}, \\
&U_{l,n}^{\dagger}(k)={\langle}l(k)|U^{\dagger}(k)|n(k){\rangle}.
\end{split}
\label{eq180}
\end{equation}
We suggest keeping the last brackets in Eq.~\ref{eq179} to indicate the effectual range of ${\partial}_k$ for the same reason as in Eq.~\ref{eq177}. Note that
\begin{equation}
\begin{split}
{\partial}_k(U_{j,n}(k)){\neq}{\langle}l(k)|\dot{U}(k)|n(k){\rangle}:={\langle}l(k)|({\partial}_kU(k))|n(k){\rangle}.
\end{split}
\label{eq181}
\end{equation}
An easy mistake about Eq.~\ref{eq179} is
\begin{equation}
\begin{split}
&i{\langle}m(k)|U(k)|{\partial}_kU^{\dagger}(k)n(k){\rangle} \\
&=i{\langle}m(k)|U(k){\lbrace}U^{\dagger}(k)|{\partial}_kn(k){\rangle}+{\partial}_k(U^{\dagger}(k))|n(k){\rangle}{\rbrace} \\
&=i{\langle}m(k)|{\partial}_kn(k){\rangle}+i{\langle}m(k)|U(k)|j(k){\rangle}{\langle}j(k)|{\partial}_k(U^{\dagger}(k))|n(k){\rangle} \\
&\xrightarrow{\overset{?}{}}r_{m,n}(k)+U_{m,j}(k)i{\partial}_k(U_{j,n}^{\dagger}(k)).
\end{split}
\label{eq182}
\end{equation}
Eq.~\ref{eq182} is inconsistent with the correct expression Eq.~\ref{eq179} for the first term $r_{m,n}(k)$ different from the first term in Eq.~\ref{eq179} $U_{m,j} (k)r_{j,l}(k)U_{l,n}^{\dagger}(k)$. In fact, the inconsistency is due to the mistake of the last step, indicated by $\xrightarrow{\overset{?}{}}$.

\subsection{D. Useful expressions for differential operators.}
For generic ribbons
\begin{equation}
\begin{split}
{\partial}_k|m(k){\rangle}=|{\partial}_km(k){\rangle}+|m(k){\rangle}{\partial}_k, \\
{\partial}_k{\langle}m(k)|={\langle}{\partial}_km(k)|+{\langle}m(k)|{\partial}_k.
\end{split}
\label{eq183}
\end{equation}
The identity expression (Einstein convention) for bases could be generalized to ribbons
\begin{equation}
\begin{split}
1=|m(k){\rangle}{\langle}m(k)|,
\end{split}
\label{eq184}
\end{equation}
and
\begin{equation}
\begin{split}
|{\partial}_k{\psi}(k){\rangle}={\partial}_k(|m(k){\rangle}{\langle}m(k)|{\psi}(k){\rangle}).
\end{split}
\label{eq185}
\end{equation}
For complex conjugation,
\begin{equation}
\begin{split}
{\langle}m(k)|{\partial}_kn(k){\rangle}^*={\langle}{\partial}_kn(k)|m(k){\rangle},
\end{split}
\label{eq186}
\end{equation}
and
\begin{equation}
\begin{split}
{\langle}{\partial}_{\lambda}m(k,{\lambda})|{\partial}_kn(k,{\lambda}){\rangle}^*={\langle}{\partial}_kn(k,{\lambda})|{\partial}_{\lambda}m(k,{\lambda}){\rangle}.
\end{split}
\label{eq187}
\end{equation}
Eq.~\ref{eq186} could be generalized to cases of multiple differential operators.
\begin{equation}
\begin{split}
{\langle}{\partial}_{\lambda}{\partial}_km(k,{\lambda})|n(k,{\lambda}){\rangle}^*={\langle}n(k,{\lambda})|{\partial}_{\lambda}{\partial}_km(k,{\lambda}){\rangle}.
\end{split}
\label{eq188}
\end{equation}
For \textit{orthogonal} ribbons ${\lbrace}|m(k){\rangle}{\rbrace}$, we further have sign reversal properties
\begin{equation}
\begin{split}
{\langle}{\partial}_km(k)|n(k){\rangle}=-{\langle}m(k)|{\partial}_kn(k){\rangle}.
\end{split}
\label{eq189}
\end{equation}
Similarly, (Einstein convention)
\begin{equation}
\begin{split}
|{\partial}_km(k){\rangle}{\langle}m(k)|=-|m(k){\rangle}{\langle}{\partial}_km(k)|.
\end{split}
\label{eq190}
\end{equation}
For generic ribbons, however, Eq.~\ref{eq189} does \textit{not} hold
\begin{equation}
\begin{split}
{\langle}{\partial}_k{\varphi}(k)|{\psi}(k){\rangle}{\neq}-{\langle}{\varphi}(k)|{\partial}_k{\psi}(k){\rangle}.
\end{split}
\label{eq191}
\end{equation}
In addition, the sign reversal properties (Eq.~\ref{eq189}) are invalid for multiple differential operators.
\begin{equation}
\begin{split}
{\langle}{\partial}_{\lambda}m(k,{\lambda})|{\partial}_kn(k,{\lambda}){\rangle}{\neq}-{\langle}m(k,{\lambda})|{\partial}_{\lambda}{\partial}_kn(k,{\lambda}){\rangle}.
\end{split}
\label{eq192}
\end{equation}

\subsection{E. Continuous conditions and existence of the $\Pi$ map}

Given a set of functions ${\lbrace}{\psi}_{n,k_q}(r){\rbrace}_N$ defined on $[0,a]$, 
is it possible to find a set ${\lbrace}a_i^{(n)}(k_q){\rbrace}$ of solutions of the equation
\begin{equation}
\begin{split}
{\int}_0^a{\psi}_{m,k_p}^*(r){\psi}_{n,k_q}(r)dr={\sum}_i^{\mathcal{N}}a_i^{(m)*}(k_p){\cdot}a_i^{(n)}(k_q)\,?
\end{split}
\label{eq193}
\end{equation}
Consider a simple one-band situation, where $m,n$ only take one possible value. Then, the above equation reduces to 
\begin{equation}
\begin{split}
{\int}_0^a{\psi}_{k_p}^*(r){\psi}_{k_q}(r)dr=a^*(k_p){\cdot}a(k_q).
\end{split}
\label{eq194}
\end{equation}
Since $k_p$ takes $N$ possible values, the left side represents $N(N-1)/2$ combinations, which lead to that number of independent constraint equations. On the right side, $a(k_q)$ is an $N$-component vector, which consists of $N$ variables to satisfy the $N(N-1)/2$ constraint equations. Evidently, the overdetermination means that solutions are not guaranteed to exist.

The situation is unchanged for a higher number of bands. There are $\mathcal{N}N(\mathcal{N}N-1)/2$ combinations on the left side, greater than the tunable variable number $\mathcal{N}^2N$ on the right side. The prime message here is that a set of continuous functions ${\psi}_{k_q}(r)$ cannot always be reproduced by inner products of vectors of discrete components. Roughly, this can be rationalized by saying that ${\psi}_{k_q}(r)$ is continuous with $r$, which in principle could be infinite in dimension. Thus, Eq.~\ref{eq193} is equivalent to dimension reduction, which is not always realizable. 

Therefore, the derivative 
\begin{equation}
\begin{split}
{\partial}_k{\int}_0^a{\psi}_k^*(r){\psi}_k(r)dr
\end{split}
\label{eq195}
\end{equation}
is not automatically well-defined globally on $k{\in}K$.
The word “globally” should be stressed, since locally one can take derivatives with the assistance of a series expansion of ${\psi}_k(r)$ with respect to $k$ within its convergence range.

Intuitively speaking, in a local range, the $N(N-1)/2$ constraint equations do not come into play, since the points $k_p$ are distributed all over $K$; locally, we do not have to consider them. On the other hand, if we hope to obtain a global smooth solution, we must go beyond the convergence domain and begin to consider the existence of global solutions, for which the $N(N-1)/2$ constraint equations start to take effect. Technically, connecting local solutions into a global one is known as \emph{sheaf theory} \cite{50}. In this particular case, a global solution might not exist. In practice, the problem can be solved by relaxing equality to obtain an approximate solution. For example, one might introduce an error function and minimize it for a given dimension of $a(k_q)$. 

Note that global solution is of tremendous importance for the global property of Berry phase, and the derivative is meant to be integrated over the entire B.Z. Therefore, an implicit assumption involved is the existence of a global solution. This assumption is equivalent to assuming that the $a(k_q)$ are discrete points on a globally continuous function about $k$. It is an analytic assumption, like requiring the physical wavefunction to be continuous and smooth, with a partial derivative everywhere. In that case, the smoothness involves $r$. Here, it involves $k$. This assumption is indispensable for both the previous method that employed the 1-st Weyl algebra, and the present method with the $N$-th Weyl algebra as outlined in Sec. 3, as long as  $\partial_k$ is involved. 

Returning to the question raised in the main text, since $\mathcal{H}_B{\cong}\mathcal{V}{\otimes}E$, why should $\mathcal{H}_B{\cong}\mathcal{V}{\otimes}E$ be more convenient? And are there any pre-conditions for $\mathcal{H}_B{\cong}\mathcal{V}{\otimes}E$?. The answer is yes: the existence of a global solution for Eq.~\ref{eq32} has been assumed. There is a vague intuition that an exact transformation, without approximation added, cannot reduce the complexity, and thus cannot bring us closer to solving the problem. Why is $\mathcal{H}_B{\cong}\mathcal{V}{\otimes}E$ advantageous over $\mathcal{H}_B$ if the two are isomorphic? It is the assumption of global solutions of Eq.~\ref{eq32} that has neatly removed the complexity arising for continuous $r$.

\subsection{F. Weyl algebra and Ring}
A major result of this paper is convergent $r$-matrix (CRM) by extending $A_1$ to $N$-th Weyl algebra $A_N$. Although Weyl algebra is the math underlying the fundamental quantum operators $\hat{r}$ and $p$, its mathematical identity as a ring is less well known to physicists, as one may simply accept the replacement of $\hat{r}$ by $i{\partial}_k$ and yield the correct result in many situations. However, in others, as shown in this paper, this replacement might lead to inconsistencies; indeed, it is exactly this replacement that has led to the divergent $r$-matrix (DRM). Therefore, if one is aimed to resolve the divergence, the algebra definition in the bottom level cannot be overlooked. In this appendix, we try to bridge the physics-math gap and point out that quantum has already encountered Weyl algebra much more than people had thought. We should cover those aspects most pertinent to physical applications, keeping the math extension in the minimum but an adequate level.

What is a ring? Why Weyl algebra is a ring? A ring is a set of elements, defined with several abstract conditions, namely ring axioms; any set that satisfy them can be called a ring, such as Weyl algebra. A most familiar ring is the set of all integer numbers, usually denoted as $\mathbb{Z}$. The ring is defined from three aspects.

(1) There is an invertible addition operation (a binary operation), which must fulfill closeness. Take $\mathbb{Z}$ as example. Such addition as $1+2=3$ is invertible as subtraction is always well-defined, e.g., $3-2=1$, which means $-2$ is the inverse of $+2$. Such addition clearly fulfills the closeness because addition of two integers will lead to another.

(2) There is a multiplication (not necessarily invertible). In the set of $\mathbb{Z}$, multiplication of two integers will be another. However, different form addition, inverse of multiplication might not exist, like $0^{-1}$ is ill-defined. 

(3) Regarding the interplay of multiplication and addition, they satisfy the distribution rules.

Now we check if $[r, k]=i$ Weyl algebra will be a ring. Accurately, $\hat{r}$ and $k$ are the generators rather than the full set, as the full set should contain infinitely many elements, otherwise the closeness for addition and multiplication would not be satisfied. 

For $A_1$, there are one pair of generators $\hat{r}$ and $k$. For $A_N$ there should be $N$ pairs ${\lbrace}r_i,k_i{\rbrace}_N$. We consider $A_1$ first. The generators should firstly be included into the set: $\hat{r}$ and $k$. Then, we shall apply the addition to $k$, one obtains $2k=k+k$, $3k=2k+k$, etc. Then by multiplying gives $k^n$ should also be included. Then combine the multiplication with addition, we realize the ring should contain polynomials as $c_0+c_1 k+c_2 k^2+$. That is why in Sec. 2, Weyl algebra involves the polynomials. Since a generic wavefunction ${\varphi}(k)$ coordinated with $k$ could be locally expanded, the polynomials are just the wavefunctions. 

Then, we account for $\hat{r}$, which should be $i{\partial}_k$. For the same reason of multiplication, one shall include $i{\partial}^n_k$. Besides, when $i{\partial}^n_k$ interplays with polynomials, one could move the differential to the right most to yield a “standard” form as $f_n(k){\partial}^n_k$, where $f_n (k)$ is a polynomial associated with $n$-th order derivative ${\partial}^n_k$; then with addition one could add different orders together ${\sum}_nf_n(k){\partial}^n_k$. That is how we use the basic commutation rule $[r, k]=i$ to generate a polynomial ring that obeys all the axioms. Note that the distribution rule is valid, which is easy to verify. 

For $N$-th Weyl algebra $A_N$, one simply extends the polynomials variable from $k$ to ${\lbrace}k_i{\rbrace}$; the derivative from ${\partial}_k$ to ${\lbrace}{\partial}_{k_i}{\rbrace}_N$. This is reflected by Eq.~\ref{eq15} in main texts. The $\hat{r}$ and $k$ will contain $N$ variables and the corresponding derivatives. 

A ring is a set equipped with two binary operations satisfying properties analogous to those of addition and multiplication of integers. Ring elements may be numbers such as integers or complex numbers, but they may also be non-numerical objects such as polynomials, square matrices, functions, and power series. A common worry is that ${\partial}_k$ needs to be acting on something; otherwise, it is meaningless. That is incorrect. One should accept the terms are meaningful on their own. Just like numbers 1, 6, etc., are meaningful on their own. We do not require they must multiply with anything or representing anything, such as 1 cat or 6 dogs. 

In front of $\frac{{\partial}^m}{{\partial}^m_k}$, one could multiply with a coefficient, just as one could multiply $a$ with $2$, $3/5$, etc., to yield $2a$, $3a/5$. In addition, one is able to multiply two terms that both contain ${\partial}_k$. As a general property of ring, multiplication should be well-defined between arbitrary pairs between them. This is exactly the same requirement as the closeness of integer multiplication. 

In view of these general aspects above, we see that expressions in quantum mechanics belong to Weyl algebra. One needs to temporally forget the physical meanings of these terms, such as physical observable, wavefunctions. Then merely focus on the abstract multiplication between different parts, for instance, view wavefunction $\varphi$ and physical operator $i{\partial}_k$ on equal status as different elements in a sing set. 

Next, we clarify a few basic terminologies. “Algebra” is a pretty loose and broad concept. Some of them, e.g., Lie algebra, is defined as a vector space, while others, such as Weyl algebra, may not necessarily form vector spaces. Thus, one algebra could be very different from another, although they are both called algebra.

In this paper, we have focused on Lie algebra and Weyl algebra. Many physicists tend to equate Lie algebra as “spin”. This is not too mistaken in many situations; however, it might conceal Lie algebra being a vector space as a formal definition. Because, for spins, physicists tend to imagine there are three of them along $x$, $y$, $z$. While for a vector space, there are infinite elements. Lie algebra, we notice that $s_x$, $s_y$, $s_z$ are merely the bases, one is at the liberty to multiply with a complex number. 

Vector space is usually defined as a set bases (finite or infinite) together with a field such as $\mathbb{R}$ or $\mathbb{C}$. Could we view $\frac{{\partial}^m}{{\partial}^m_k}$ as bases and the polynomial as the coefficients associated with a “abstract” number, such that it will form a vector space? The answer is no. As the coefficients do not form a field (which requires division to be defined). Besides, the multiplication is defined for $\frac{{\partial}^m}{{\partial}^m_k}$ too, which is lacking in a vector space’s definition. Thus, Weyl algebra is usually not viewed as a vector space. 

What is the difference between a vector space and a topological space? One might feel any point, for instance, denoted with two numbers could be viewed as a 2D vector. This is incorrect. The axiom of vector space requires $a+a=2a$, and if $2a=0$ then one must have $a=0$. Consider a point on a sphere surface, represented by $(\theta, \phi)$. Consider a point on equator $a=(\pi/2, \pi)$. If we think $a$ can represent a vector, just like a point in vector space represents a vector $a+a=2a=(0,0)$, which leads to $a=(0,0)$. However, $a{\neq}0$. Thus $S^2$ does not form a vector space, but only a topological space. Topological space requires the neighborhood is well defined, some information about some points being near to another but further from others; without topology, these points will be just like ``grains of sand".


\begin{thebibliography}{99}
\bibitem{12} Introducing a $\delta$-function to yield a formal $r$-matrix attempts to encode $\hat{r}$ into an infinite matrix, which is purely formal because of its inevitable divergence.
\end{thebibliography}
\end{document}